\documentclass[rmp,aps,preprint,nofootinbib]{revtex4}                                          
\usepackage{epsfig}
\usepackage{float}
\usepackage{amsfonts}
\usepackage{color}
\usepackage{threeparttable}

\begin{document}

\title{Jammed Hard-Particle Packings: From Kepler to Bernal and Beyond}

\author{S. Torquato}

\email{torquato@princeton.edu}

\affiliation{\emph{Department of Chemistry,  Department of Physics,
Princeton Center for Theoretical Science,Princeton Institute for the Science and Technology of Materials, and 
Program in Applied and Computational Mathematics, Princeton University}, Princeton
New Jersey, 08544 USA}

\affiliation{\emph{School of Natural Sciences, Institute of Advanced Study}, Princeton
New Jersey, 08540 USA}

\author{F. H. Stillinger}

\email{fhs@princeton.edu}

\affiliation{\emph{Department of Chemistry}, \emph{Princeton University}, Princeton
New Jersey, 08544 USA}

\begin{abstract}

Understanding the characteristics of jammed particle packings provides basic 
insights into the structure and bulk properties of crystals, glasses, and 
granular media, and into selected aspects of biological systems.  This review 
describes the diversity of jammed configurations attainable by frictionless convex 
nonoverlapping (hard) particles in Euclidean spaces and for that purpose it stresses individual-packing
geometric analysis.  A fundamental feature of that 
diversity is the necessity to classify individual jammed configurations according 
to whether they are locally, collectively, or strictly jammed.  Each of these 
categories contains a multitude of jammed configurations spanning a wide and 
(in the large system limit) continuous range of intensive properties, including 
packing fraction $\phi$, mean contact number $Z$, and several scalar order metrics.  
Application of these analytical tools to spheres in three dimensions 
(an analog to the venerable Ising model) covers a myriad of jammed states, 
including maximally dense packings (as Kepler conjectured), low-density strictly-jammed 
tunneled crystals, and a substantial family of amorphous packings.  With 
respect to the last of these, the current approach displaces the historically 
prominent but ambiguous idea of ``random close packing" (RCP) with the precise concept 
of ``maximally random jamming" (MRJ).  Both laboratory procedures and numerical 
simulation protocols can, and frequently have been, used for creation of ensembles 
of jammed states.  But while the resulting distributions of intensive properties may 
individually approach narrow distributions in the large system limit, the distinguishing 
varieties of possible operational details in these procedures and protocols lead
to substantial variability among the resulting distributions, some examples of which 
are presented here.  This review also covers recent advances in understanding 
jammed packings of polydisperse sphere mixtures, as well as convex nonspherical 
particles, e.g., ellipsoids, ``superballs", and polyhedra.  Because of their relevance 
to error-correcting codes and information theory, sphere packings in high-dimensional 
Euclidean spaces have been included as well. We also make some remarks
about packings in (curved) non-Euclidean spaces. In closing this review, several basic 
open questions for future research to consider have been identified.

\end{abstract}

\maketitle
\tableofcontents

\section{Introduction}

\begin{figure}[bthp]
\centerline{\includegraphics[width=3in,keepaspectratio,clip=]{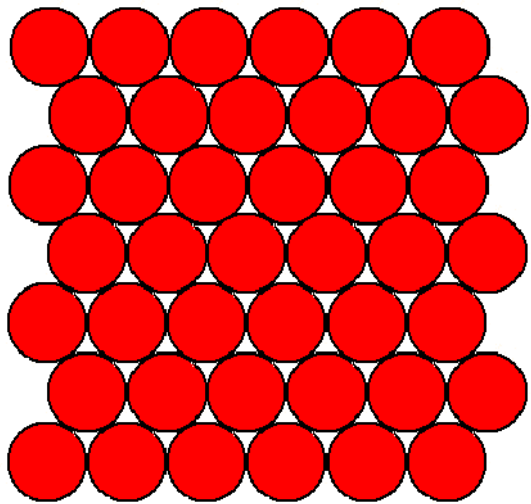}}
\caption{\footnotesize A portion of the densest packing of identical circles in the plane,
with centers lying at the sites of a triangular lattice. The fraction
of $\mathbb{R}^2$ covered by the interior of the circles is 
$\phi=\pi/\sqrt{12}=0.906899\ldots$. The first claim of a
proof was made by Thue in 1892. However, it is generally
believed that the first complete error-free proof was produced only in 1940 by Fejes T{\' o}th; see \onlinecite{Ro64}
and \onlinecite{Co93} for the history of this problem. }
\label{circle}
\end{figure}

The importance of packing hard particles into various kinds of vessels 
and the questions it raises have an ancient history. Bernal has remarked that ``heaps (close-packed arrangements of
particles) were the first things that were ever measured in the form of basketfuls of grain for
the purpose of trading or the collection of taxes" \cite{Ber65}.
Although packing problems are easy to pose, they are notoriously difficult
to solve rigorously. In 1611, Kepler was asked: What is the densest way to stack 
equal-sized cannon balls? His solution, known as ``Kepler's conjecture," 
was the face-centered-cubic (fcc) arrangement (the way your green grocer stacks oranges). 
\onlinecite{Ga31}  proved that this is the densest Bravais lattice packing (defined below). But
almost four centuries passed before Hales proved the general conjecture that there is no other arrangement
of spheres in three-dimensional Euclidean space whose density can exceed
that of the fcc packing \cite{Ha05}; see \onlinecite{As08} for a popular account of the proof. Even the proof
of the densest packing of congruent (identical) circles in the plane, the two-dimensional analog
of Kepler's problem, appeared only 70 years ago \cite{Ro64,Co93}; see Fig. \ref{circle}.

Packing problems are ubiquitous and arise in a variety of applications.
These exist in  the transportation, packaging, agricultural
and communication industries. Furthermore, they have been studied to help understand the 
symmetry, structure and macroscopic physical properties of condensed matter phases, 
 liquids, glasses and crystals \cite{Ma40,Be60,Ber65,St64b,St69,We71,As76,Wo81,Ha86,Sp94,Chaik95}.
Packing problems are also relevant for the analysis of
heterogeneous materials \cite{To02a}, colloids \cite{Ru89,Chaik95,To09a}, and granular media \cite{Ed94}.  
Understanding the symmetries and other mathematical characteristics
of the densest sphere packings in various spaces and dimensions 
is a challenging area of long-standing interest in discrete geometry and number theory 
\cite{Co93,Co03,Ro64} as well as coding theory  \cite{Sh48,Co93,Co07}.

\begin{figure}[bthp]
\centerline{\includegraphics[width=2.5in,keepaspectratio,clip=]{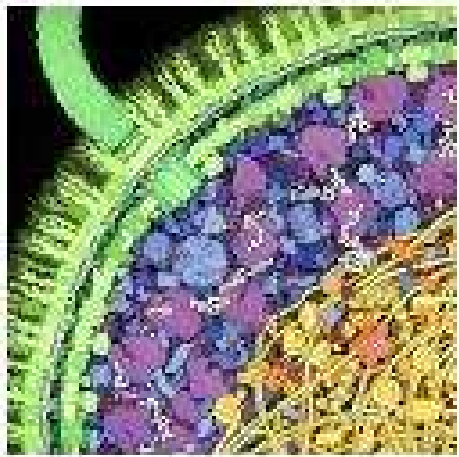}
\hspace{0.25in} \includegraphics[width=3.5in,keepaspectratio,clip=]{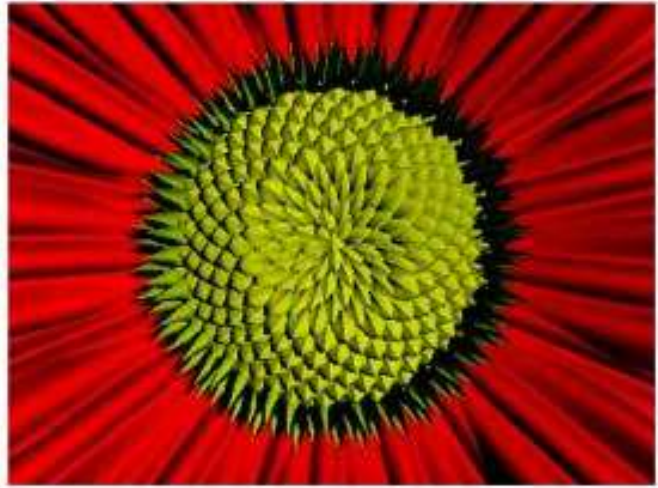}}

\caption{\footnotesize
Biological examples of packing geometries. 
Left panel: Schematic representation of the approximate numbers, shapes and
density of packing of macromolecules inside a cell of Escherichia coli.
(Illustration by David S. Goodsell, the Scripps Research
Institute.)
Molecular crowding occurs because
of excluded volume effects due to the mutual
impenetrability of all of the  macromolecules.
This nonspecific steric repulsion
is always present, regardless of any other attractive
or repulsive interactions that might occur between
the macromolecules. Thus, molecular crowding is essentially
a packing problem \cite{El01}. Right panel: Close-up of a daisy capitulum (flower head),
as given by \onlinecite{Pr90}.
The most prominent feature is two sets
of spirals, one turning clockwise and the other counterclockwise.
There are several models that relate
such phyllotactic patterns to packings problems \cite{Pr90}.
}
\label{bio}
\end{figure}

It is appropriate to mention that packing issues also arise in numerous biological contexts,
spanning a wide spectrum of length scales.
This includes ``crowding" of macromolecules within living cells
\cite{El01}, the  packing of cells to form tissue \cite{To02a,Ge08a},
the fascinating spiral patterns seen in plant shoots  and
flowers (phyllotaxis) \cite{Pr90,Ni10} and the competitive settlement of territories by animals, the patterns
of which can be modeled as random sequential packings \cite{Tan80,To02a}.
Figure \ref{bio} pictorially depicts macromolecular crowding and a familiar phyllotactic pattern.

We will call a packing a large collection of nonoverlapping (i.e., hard) particles
in either a finite-sized container or in $d$-dimensional Euclidean
space $\mathbb{R}^d$. The packing fraction $\phi$ is the fraction
of space covered by (interior to) the hard particles. 
``Jammed" packings are those particle configurations in which each particle
is in contact with its nearest neighbors in  such a way that
mechanical stability of a specific type is conferred to the packing
(see Section \ref{jamming}). Jammed packings and their properties have received considerable attention in the
literature, both experimental and theoretical. 
Within the domains of analytical theory and computer simulations, two conceptual
approaches for their study  have emerged. One is  
the ``ensemble" approach \cite{Be60,Ber65,Ed94,Ed01,Li98,Ma02,Si02,Oh03,Wy05,Si05,Ga06,Ma08,Pa10}, which for
a given packing procedure  aims to understand typical
configurations and their frequency of occurrence. The other, more recently, is the
``geometric-structure" approach \cite{To00b,To01b,Ka02d,To03c,Do04a,Do04b,Do07a,To07}, which emphasizes quantitative characterization of single-packing configurations, without regard
to their occurrence frequency in the algorithmic method used to produce them.
Our primary objective is to review
the latter approach, while at the same time to show that  these two approaches are complementary 
in that they represent different aspects of the larger context of hard-particle
jamming phenomena. In particular, a wide range of jammed packing ensembles 
can be created by the choice of the generating algorithm, and the geometric-structure approach 
analyzes and classifies individual members of those ensembles, whether they be crystalline or
amorphous at any achievable packing fraction $\phi$.

\begin{figure}[bthp]
\centerline{\includegraphics[width=3in,keepaspectratio,clip=]{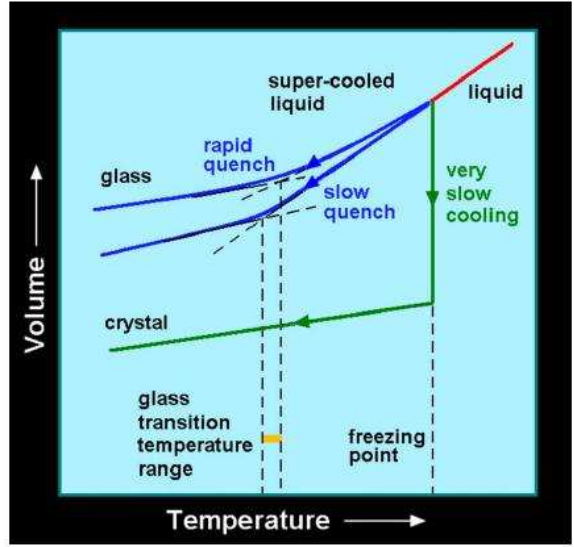}}

\caption{\footnotesize Three isobaric (constant pressure)
 cooling paths by which a typical liquid may solidify, represented in a volume vs. temperature diagram.  An infinitesimal cooling rate from the high-temperature liquid traces out the thermodynamic equilibrium path (shown in green), 
including a discontinuity resulting from the first-order freezing transition. This reversible 
path leads to the ground state defect-free crystalline structure in the $T \rightarrow 0$  limit.  
Very slow but finite cooling rate (not shown) can involve crystal nucleation but typically 
creates defective crystals.  More rapid cooling of the liquid (blue curves) can avoid 
crystal nucleation, passing instead through a glass transition temperature range, and 
resulting in metastable glassy solids at absolute zero.  The volumes, energies, and other 
characteristics of those glasses vary with the specific  cooling rate employed in their production.}
\label{glass}
\end{figure}

     The process of cooling an initially hot liquid ultimately to absolute zero 
temperature provides a close and useful analogy for the subject of hard-particle 
jamming.  Figure \ref{glass} summarizes this analogy in  schematic form, showing typical paths 
for different isobaric cooling rates in the temperature-volume plane.  While these 
paths are essentially reproducible for a given cooling schedule, i.e., giving a 
narrow distribution of results, that distribution depends sensitively on the specific 
cooling schedule, or protocol, that has been used.  A very rapid quench that starts 
with a hot liquid well above its freezing temperature will avoid crystal nucleation, 
producing finally a glassy solid at absolute zero temperature.  A somewhat slower 
quench from the same initial condition can also avoid nucleation, but will yield 
at its  $T=0$ endpoint a glassy solid with lower volume and potential energy.  
An infinitesimal cooling rate in principle will follow a thermodynamically reversible 
path of equilibrium states, will permit nucleation, and will display a volume discontinuity 
due to the first order freezing transition on its way to attaining the structurally 
perfect crystal ground state.  By analogy, for hard-particle systems compression 
qualitatively plays the same role as decreasing the temperature in an atomic or 
molecular system.  Thus it is well known that compressing a monodisperse hard-sphere 
fluid very slowly leads to a first-order freezing transition, and the resulting 
crystal phase corresponds to the closest packing arrangement of those spheres 
\cite{Mau99}.  The resulting hard sphere stacking variants are configurational 
images of mechanically stable structures exhibited for example by the venerable 
Lennard-Jones model system.  By contrast, rapid compression rates applied to a 
hard-sphere fluid will create random amorphous jammed packings \cite{Ri96b}, 
the densities of which can be controlled by the compression rate utilized 
(see Fig. \ref{hsphere} in Sec. IV below).  In both the cases of cooling liquid glass-formers, 
and of compressing monodisperse hard spheres, it is valuable to be able to analyze 
the individual many-particle configurations that emerge from the respective protocols.

We begin this review, after introducing relevant terminology, by specifically considering packings of frictionless, identical spheres in 
the absence of gravity, which represents an idealization
of the laboratory situation for investigations of jammed packings; see Section \ref{lessons}.
This simplification follows that tradition in condensed matter science to exploit
idealized models,
such as the Ising model, which is regarded  as one of the pillars of statistical mechanics \cite{On44,Do60,Ga99}.
In that tradition, this idealization  
offers the opportunity to obtain fundamental as well as practical
insights, and to uncover unifying concepts that describe a broad range of phenomena. 
The stripped-down hard-sphere ``Ising model" for jammed packings
(i.e., jammed, frictionless, identical spheres in 
the absence of gravity) embodies the primary attributes of real packings while simultaneously generating
fascinating mathematical challenges. The geometric-structure
approach to analyzing individual jammed states produced by 
this model covers not only the maximally dense packings (e.g., Kepler's conjecture) 
and amorphous ``Bernal" packings, but an unbounded collection  of other jammed configurations.
This approach naturally
leads to the inevitable conclusion that there is  great
diversity in the types of attainable jammed packings with varying
degrees of order, mechanical stability, and density.

Important insights arise when jammed sphere packings  are placed in a broader context
that includes jammed states of noncongruent spheres as well as nonspherical objects, 
as discussed in Secs. \ref{poly} and \ref{nonspherical}.
These extensions include polydisperse spheres, ellipsoids, ``superballs," and polyhedra
in three dimensions. In addition, this broader context also 
involves sphere packings in  Euclidean space with high dimensions (Sec. \ref{high}), which
is relevant to error correcting codes and information theory \cite{Sh48,Co93},
and packings in non-Euclidean spaces (Sec. \ref{curved}). Finally, in Sec. \ref{open},
we  identify a number of basic open questions for future research.

\section{Preliminaries and Definitions}
\label{defs}

Some basic definitions concerning packings are given here. A {\it packing} $P$ is 
a collection of nonoverlapping solid objects or particles in $d$-dimensional
Euclidean space $\mathbb{R}^d$. Packings can be defined in other spaces
(e.g., hyperbolic spaces and compact spaces, such as the surface
of a $d$-dimensional sphere), but our primary focus in this review
is $\mathbb{R}^d$. A {\it saturated} packing
is one in which there is no space available to add another particle 
of the same kind to the packing.

\begin{figure}[bthp]
\centerline{\includegraphics[width=1.6in,keepaspectratio,clip=]{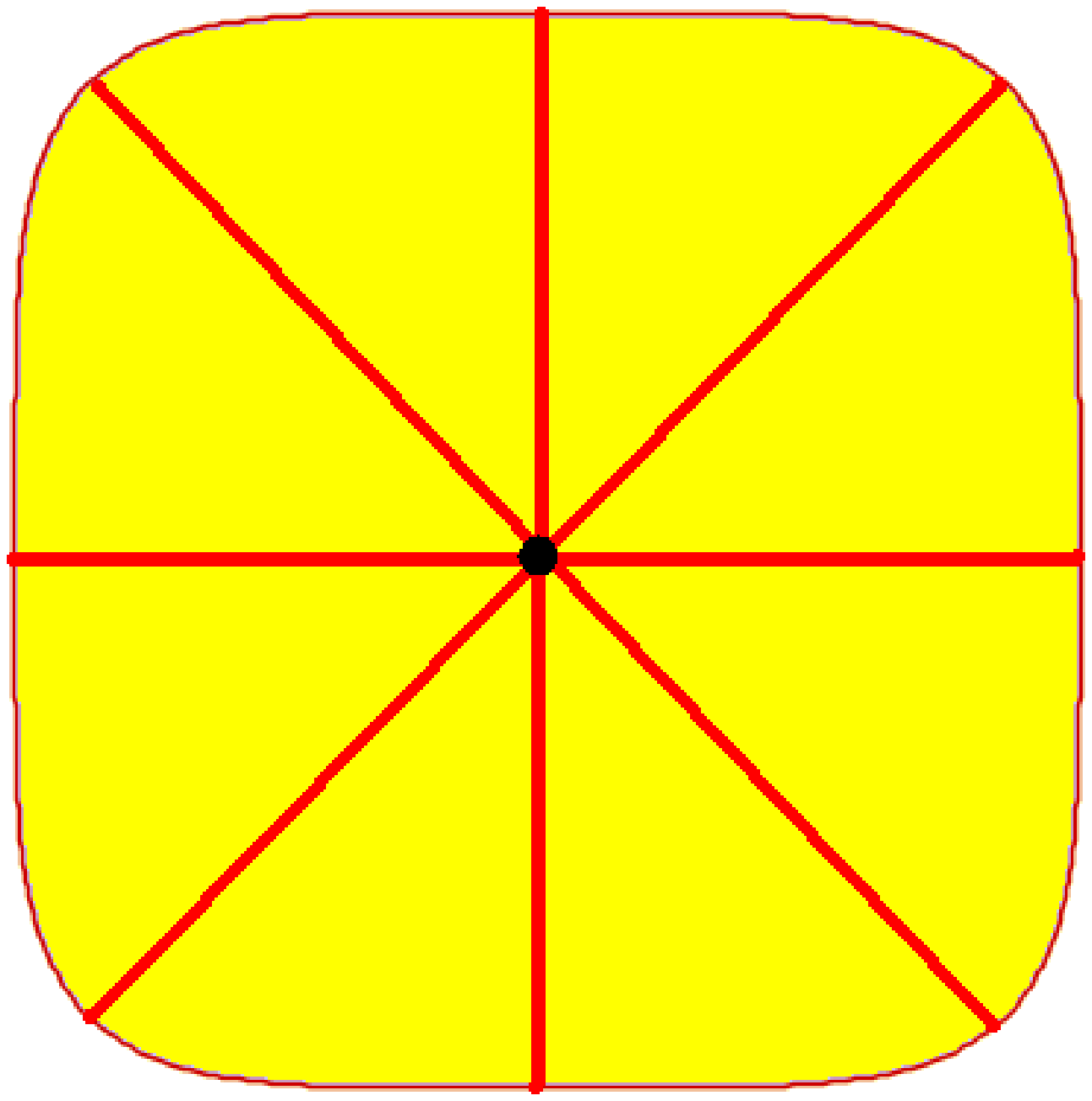}
\hspace{0.4in}\includegraphics[width=1.6in,keepaspectratio,clip=]{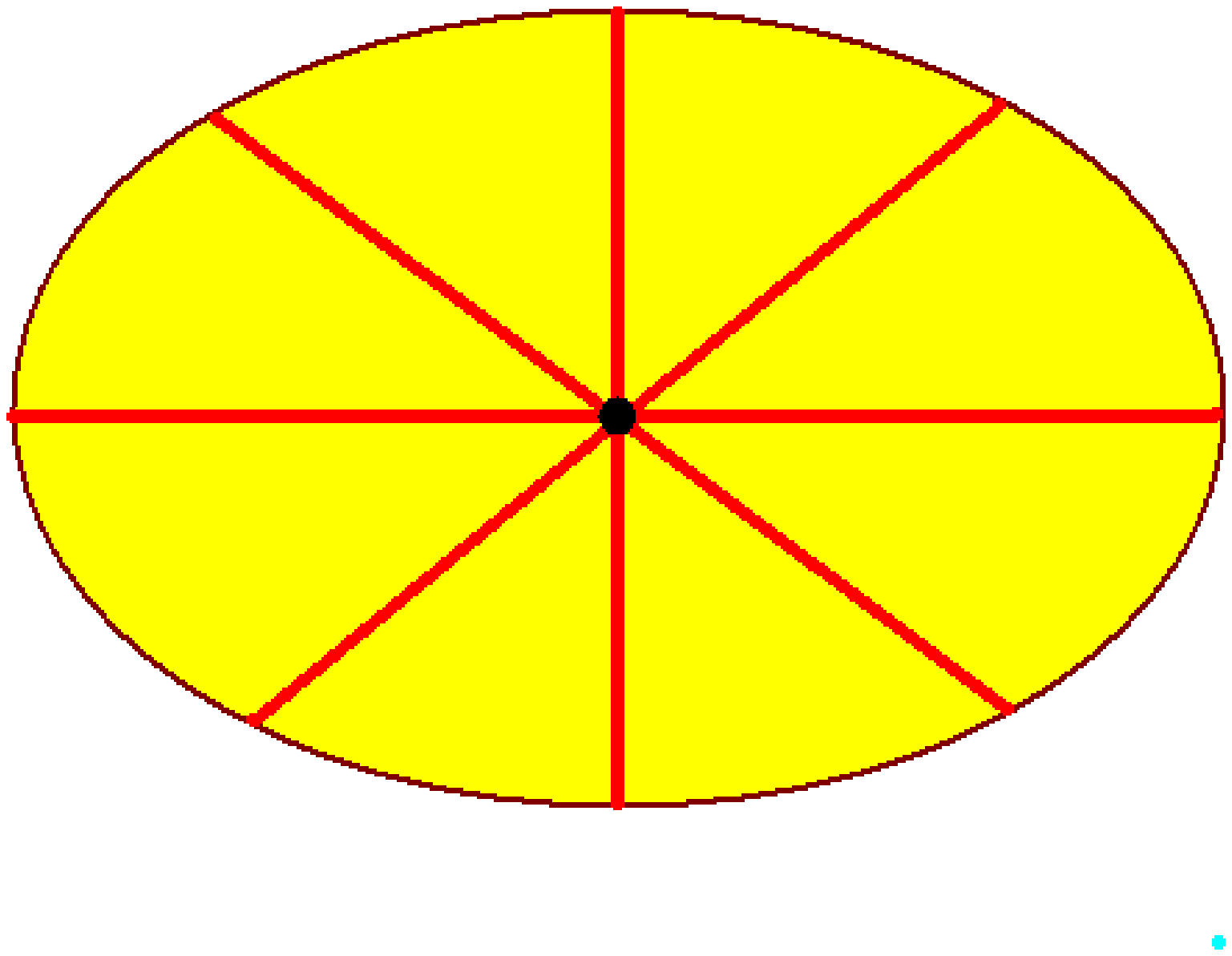}
\hspace{0.4in}\includegraphics[width=1.6in,keepaspectratio,clip=]{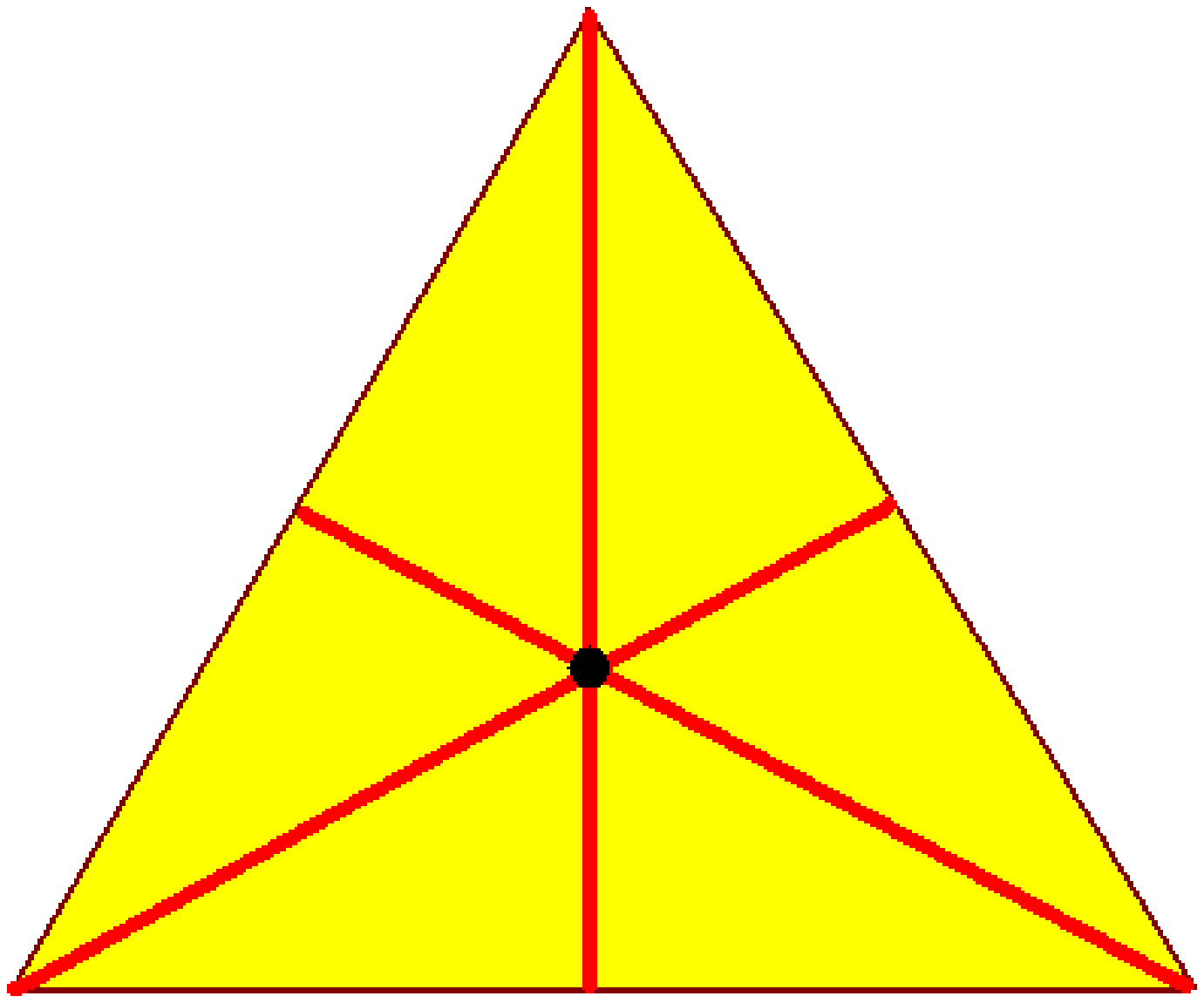}}
\caption{ Symmetries associated with three particle shapes.
Chords pass through each particle centroid. 
Left panel: A ``superdisk" is centrally symmetric and possesses
two equivalent principal axes. Middle panel: An ellipse  is centrally symmetric but does
not possess two equivalent principal axes. Right panel: A triangle is not
centrally symmetric. }
\label{central}
\end{figure}

We will see subsequently that whether a particle possesses  central symmetry
plays a fundamental role in determining its dense packing characteristics.
A $d$-dimensional particle is {\it centrally symmetric} if it has a center $C$ that
bisects every chord through $C$ connecting any two boundary points
of the particle, i.e., the center is a point of inversion symmetry.
 Examples of centrally symmetric particles in  $\mathbb{R}^d$ are spheres, ellipsoids and
superballs (defined in Section \ref{nonspherical}). A triangle and tetrahedron are
examples of non-centrally symmetric two- and three-dimensional particles, respectively.
Figure \ref{central} depicts examples of centrally and non-centrally symmetric
two-dimensional particles. A $d$-dimensional centrally symmetric particle for $d \ge 2$
is said to possess
$d$ equivalent principal (orthogonal) axes (directions) associated
with the moment of inertia tensor if those
directions are two-fold rotational symmetry axes such
that the $d$ chords along those directions and connecting
the respective pair of particle-boundary points are equal. 
(For $d=2$, the two-fold (out-of-plane) rotation along an orthogonal axis
brings the shape to itself, implying the rotation axis is a ``mirror image" axis.)
Whereas a $d$-dimensional superball has $d$
equivalent directions, a $d$-dimensional ellipsoid generally does not
(see Fig. \ref{central}).

\begin{figure}[bthp]
\centerline{\includegraphics[width=2.5in,keepaspectratio,clip=]{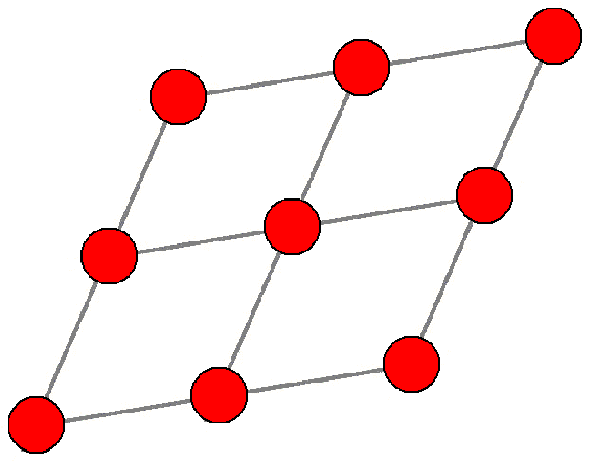}
\hspace{0.3in} \includegraphics[width=2.5in,keepaspectratio,clip=]{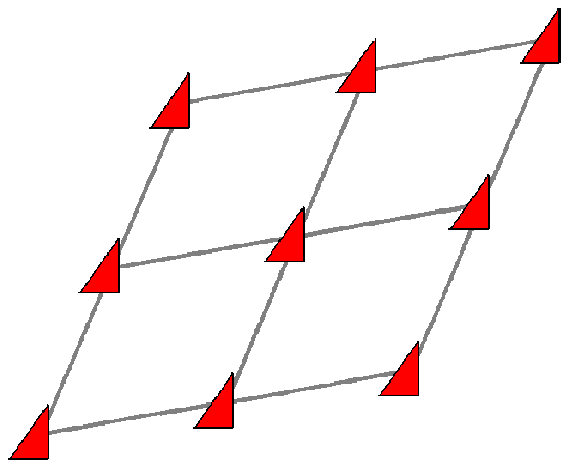}}
\caption{ Examples of lattice packings (i.e., Bravais lattices)
depicted in two dimensions.
Left panel: A portion of a lattice packing of congruent spheres.
Each fundamental cell (depicted as a rhombus here) has
exactly one assigned sphere center. Right panel: A portion of a lattice packing of 
congruent nonspherical particles. Each fundamental cell  has
exactly one particle centroid. Each particle in the packing must have the same orientation. }
\label{lattice}
\end{figure}

A {\it lattice} $\Lambda$ in $\mathbb{R}^d$ is a subgroup
consisting of the integer linear combinations of vectors that constitute
a basis for $\mathbb{R}^d$. In the physical sciences and engineering, this is referred to as a {\it Bravais}
lattice. Unless otherwise stated, the term ``lattice" will refer here
to a Bravais lattice only.
A {\it lattice packing} $P_L$ is one in which  the centroids of the nonoverlapping identical
particles are located at the points of $\Lambda$, and all particles have a common orientation.
The set of lattice packings is a subset of all possible packings in $\mathbb{R}^d$.
In a lattice packing, the space $\mathbb{R}^d$ can be geometrically divided into identical
regions $F$ called {\it fundamental cells}, each of which contains
the centroid of just one particle. Thus, the density of a lattice packing is given by
\begin{equation}
\phi= \frac{v_1}{\mbox{Vol}(F)},
\end{equation}
where $v_1$ is the volume of a single $d$-dimensional particle and
$\mbox{Vol}(F)$ is the $d$-dimensional volume of the fundamental cell.
For example, the volume $v_1(R)$ of  a $d$-dimensional spherical particle of radius $R$
is given explicitly by
\begin{equation}
v_1(R)=\frac{\pi^{d/2} R^d}{\Gamma(1+d/2)},
\label{vol-sph}
\end{equation}
where $\Gamma(x)$ is the Euler gamma function. Figure \ref{lattice}
depicts lattice packings of congruent spheres and congruent nonspherical
particles. 

\begin{figure}[bthp]
\centerline{\includegraphics[width=2.5in,keepaspectratio,clip=]{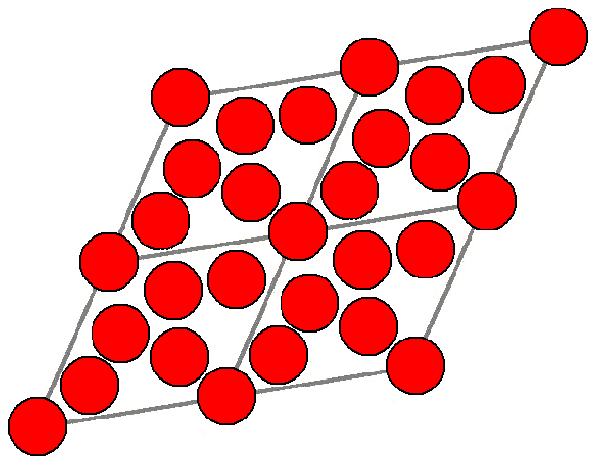}
\hspace{0.3in} \includegraphics[width=2.5in,keepaspectratio,clip=]{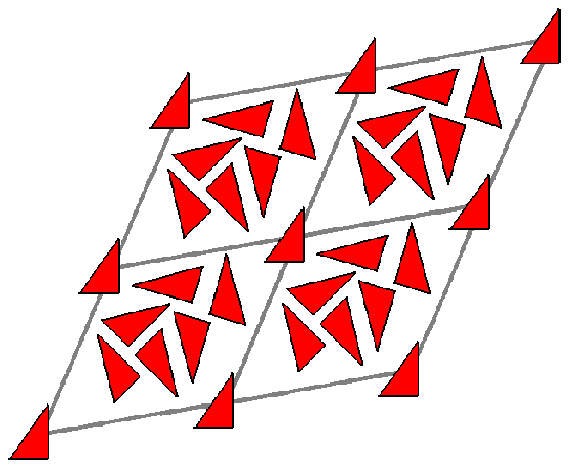}}
\caption{ Examples of periodic non-lattice packings
depicted in two dimensions.
Left panel: A portion of a periodic non-lattice packing
of congruent spheres. The fundamental cell contains multiple 
 spheres located anywhere within the cell subject to the
nonoverlap constraint. 
Right panel: A portion of a periodic non-lattice packing
of congruent nonspherical particles. The fundamental cell contains  multiple
nonspherical particles with arbitrary positions and orientations within the cell subject to the
nonoverlap constraint.}
\label{periodic}
\end{figure}

A more general notion than a lattice packing is a periodic
packing. A {\it periodic} packing of congruent particles
is obtained by placing a fixed configuration of $N$ particles (where $N\ge 1$)
with {\it arbitrary nonoverlapping orientations} in one 
fundamental cell of a lattice $\Lambda$, which is then periodically replicated
without overlaps.
 Thus, the packing is still
periodic under translations by $\Lambda$, but the $N$ particles can occur
anywhere in the chosen fundamental cell subject to the overall nonoverlap condition.
The packing density of a  periodic packing is given by
\begin{equation}
\phi=\frac{N v_1}{\mbox{Vol}(F)}=\rho v_1,
\end{equation}
where $\rho=N/\mbox{Vol}(F)$ is the number density, i.e., the number of particles per unit volume.
Figure \ref{periodic}
depicts a periodic non-lattice packing of congruent spheres and congruent nonspherical
particles. Note that the particle orientations within a fundamental cell in the latter
case are generally not identical to one another.

\begin{figure}[bthp]
\centerline{\includegraphics[height=2in,keepaspectratio,clip=]{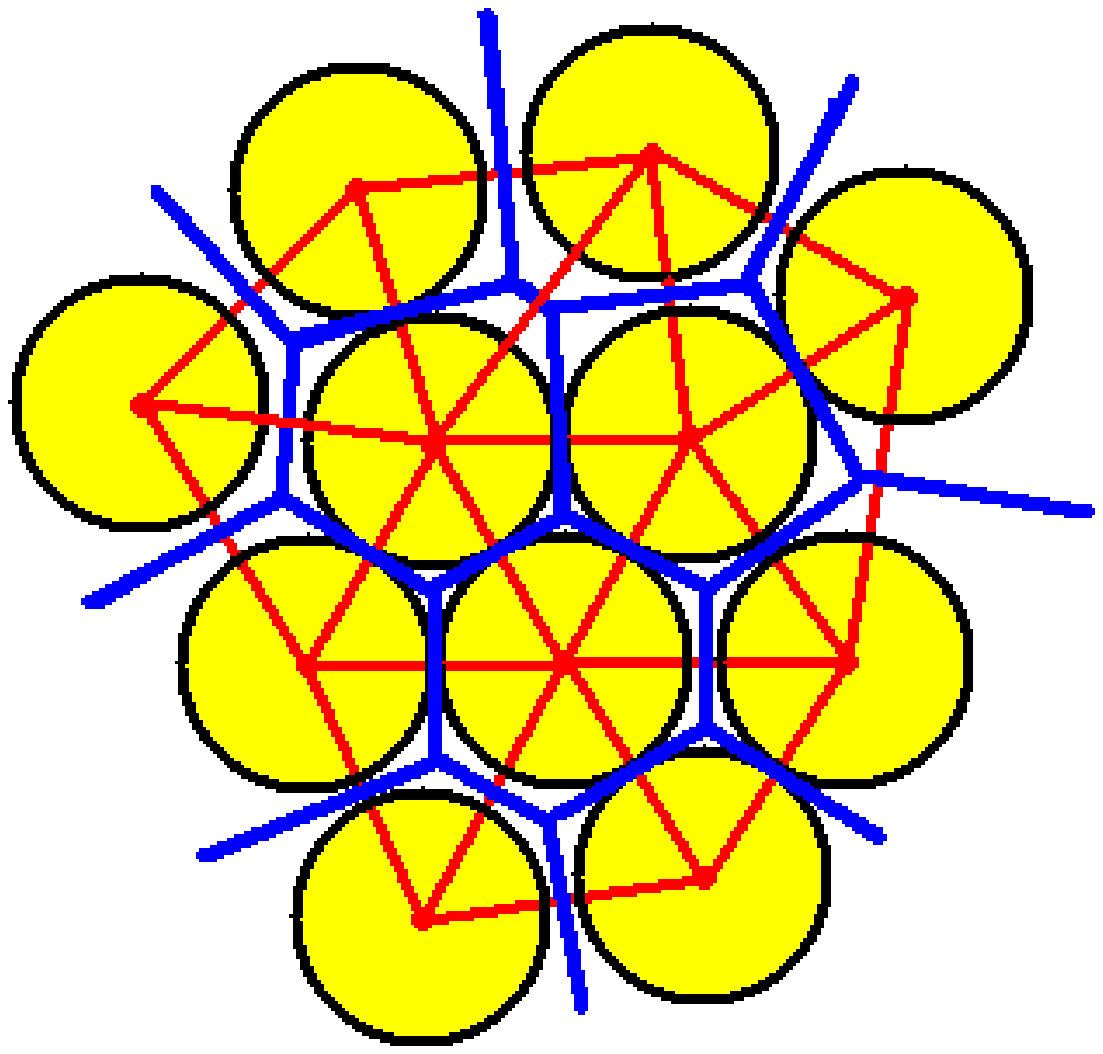}
\hspace{0.25in} \includegraphics[height=2in,keepaspectratio,clip=]{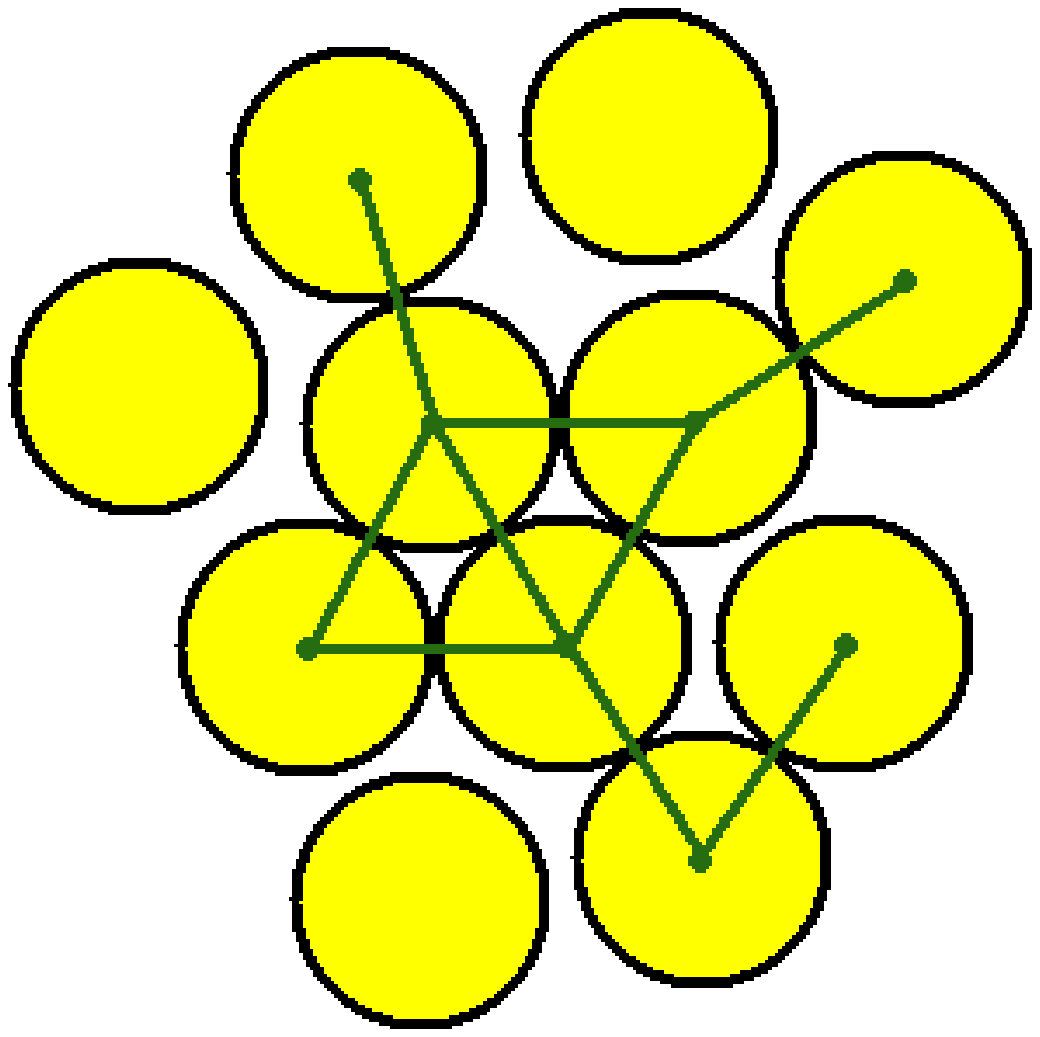}}
\caption{ \footnotesize Geometric characterization of packings 
via bond networks.
Left panel: Illustrations of the Voronoi and Delaunay tessellations
for a portion of a packing of congruent circular disks. The blue and red lines are edges
of the Voronoi and Delaunay cells, respectively. Right panel:  The corresponding
contact network shown as green lines.
}
\label{cells}
\end{figure}

Consider any discrete set of  points with position vectors 
$X\equiv\{{\bf r}_1, {\bf r}_2,\ldots\}$ in $\mathbb{R}^d$. 
Associated with each point ${\bf r}_i \in X$
is its {\em Voronoi cell}, $\mbox{Vor}({\bf r}_i)$,
which is defined to be the region of space no farther from
the point at ${\bf r}_i$ than to any other point ${\bf r}_j$ in the set, i.e.,
\begin{equation}
\mbox{Vor}({\bf r}_i)=\{{\bf r}: |{\bf r}-{\bf r}_i| \le |{\bf r}-{\bf r}_j|
\mbox{for all}\; {\bf r}_j \in X\}.
\end{equation}
The Voronoi cells are convex polyhedra whose interiors
are disjoint, but share common
faces, and therefore the union of all of the polyhedra
is the whole of $\mathbb{R}^d$. This partition of space is called the {\em Voronoi} tessellation.
While the Voronoi polyhedra of a lattice are congruent (identical) to one another,
the Voronoi polyhedra of a non-Bravais lattice are not identical 
to one another. Attached to each vertex of a Voronoi polyhedron  is a {\em Delaunay cell}, 
which can be defined as the convex hull of the Voronoi-cell 
centroids nearest to it, and these Delaunay cells
also tile space.  Very often, the Delaunay tessellation is a {\it triangulation} of space, i.e.,
it is a partitioning of $\mathbb{R}^d$ into $d$-dimensional simplices \cite{To02a}.  Geometrically the Voronoi and Delaunay tessellations are dual
to each other. The {\it contact network} is only defined for a packing
in which a subset of the particles form interparticle contacts. For example,
when the set of points $X$ defines the centers of 
spheres in a sphere packing, the network of interparticle contacts forms the contact network
of the packing by associating with every sphere a ``node" for
each contact point and edges that connect all of the nodes.
As we will see in Sec.~\ref{jamming}, the contact network is crucial
to determining the rigidity properties of the packing
and corresponds to a subclass of the class of fascinating objects called {\it tensegrity
frameworks}, namely {\it strut frameworks}; see  \onlinecite{Con96} for details.
Figure \ref{cells} illustrates the Voronoi, Delaunay and contact networks
for a portion of a packing of congruent circular disks.

Some of the infinite packings that we will be considering in this review
can only be characterized spatially via statistical correlation functions.
For simplicity, consider a nonoverlapping configuration of $N$ identical $d$-dimensional
spheres centered at the positions ${\bf r}^N \equiv \{{\bf r}_1, {\bf r}_2, \cdots, {\bf r}_N$\} in
a region of volume $V$ in $d$-dimensional Euclidean space $\mathbb{R}^d$.
Ultimately, we will pass to the {\it thermodynamic limit}, i.e.,
$N \rightarrow \infty$, $V \rightarrow \infty$ such that 
the {\it number density} $\rho=N/V$ is a fixed positive constant.
For statistically homogeneous sphere packings in $\mathbb{R}^d$, the quantity
$\rho^n g_{n}({\bf r}^n)$  is proportional to
the probability density for simultaneously finding $n$ sphere centers at
locations ${\bf r}^n\equiv \{{\bf r}_1,{\bf r}_2,\dots,{\bf r}_n$\} in $\mathbb{R}^d$ \cite{Ha86}.
 With this convention, each {\it $n$-particle correlation function} $g_n$ approaches
unity when all particle positions become widely separated from one another.
Statistical homogeneity implies that $g_n$ is translationally
invariant and therefore only depends on the relative displacements 
of the positions with respect to some arbitrarily chosen origin of the system, {\it i.e.},
\begin{equation}
g_n=g_n({\bf r}_{12}, {\bf r}_{13}, \ldots, {\bf r}_{1n}),
\end{equation}
where ${\bf r}_{ij}={\bf r}_j - {\bf r}_i$.

The {\it pair correlation} function $g_2({\bf r})$ is the one
of primary interest in this review. If the
system is also rotationally invariant (statistically
isotropic), then $g_2$ depends on the radial distance $r \equiv |{\bf r}|$ only, {\it i.e.},
$g_2({\bf r}) = g_2(r)$. It is important to introduce
the {\it total correlation} function $h({\bf r})\equiv g_2({\bf r})-1$, which,
for a {\it disordered} packing,  decays to zero for large $|{\bf r}|$ sufficiently rapidly
\cite{To06b}. We define the {\it structure factor} $S({\bf k})$ for
a statistically homogeneous packing via the relation
\begin{equation}
S({\bf k})=1+\rho {\tilde h}({\bf k}),
\end{equation}
where ${\tilde h}({\bf k})$ is the Fourier transform of the {\it total correlation function}
$h({\bf r})\equiv g_2({\bf r})-1$ and $\bf k$ is the wave vector.
Since the structure factor is the Fourier transform of an autocovariance
function (involving the ``microscopic" density) \cite{Ha86,To06b}, 
then it follows it is a nonnegative quantity for all $\bf k$, {\it i.e.},
\begin{equation}
S({\bf k}) \ge 0 \qquad \mbox{for all} \quad {\bf k}.
\label{factor}
\end{equation}
The nonnegativity condition follows physically from the fact
that $S({\bf k})$ is proportional to the intensity of the scattering of incident radiation
on a many-particle system \cite{Ha86}.
The structure factor $S({\bf k})$ provides a measure of the density fluctuations in the packing
at a particular wave vector $\bf k$.

\section{Lessons From Disordered Jammed Packings of Spheres}
\label{lessons}

     The classical statistical mechanics of hard-sphere systems 
has generated a huge  collection of scientific publications, stretching back at least to 
\onlinecite{Bo64}.  That collection includes examinations of equilibrium, transport, and jammed packing phenomena.   With respect to the last of these, the concept of a unique ``random close packing" (RCP) state, pioneered by \onlinecite{Be60,Ber65} to model 
the structure of liquids, 
has been one of the more persistent themes with a venerable history \cite{Sc69,An72,Vi72,Go74b,Be83,Jo85,To88,Zi94,Ju97,Po97,Ka07}.   
Until about a decade ago, the prevailing notion of the RCP state was that it 
is the {\it maximum} density that a large, random collection of congruent (identical) spheres can attain and that this density is a well-defined  quantity.
This traditional view has been summarized as
follows: ``Ball bearings and similar objects have been shaken,
settled in oil, stuck with paint, kneaded inside rubber
balloons--and all with no better result than (a packing fraction of) $\ldots$ $0.636$'' 
\cite{An72}. \onlinecite{To00b} have argued that this 
RCP-state concept is actually ill-defined and thus should be abandoned in favor
of a more precise alternative.

It is instructive  to review briefly these developments 
because they will point to the need for a geometric-structure approach
generally to understand jammed packings, whether disordered or not.
It has been observed \cite{To00b}  that 
there has existed ample evidence in the literature, in the
form of actual and computer-simulation  experiments, to 
suggest strongly that the RCP state is indeed ill-defined and, in particular, dependent
on the protocol used to produce the packings 
and on other system characteristics.
In a classic experiment,  \onlinecite{Sc69} obtained  the 
``RCP" packing fraction value $\phi \approx 0.637$ by pouring
ball bearings into a large container, vertically vibrating
the system for  sufficiently long times to
achieve a putative maximum densification, and extrapolating
the measured volume fractions to eliminate finite-size effects. 
Important dynamical parameters for this kind of experiment
include the pouring rate as well as the amplitude, frequency 
and direction of the vibrations. The shape, smoothness and rigidity of 
the container boundary are other crucial characteristics.
For example, containers with curved or flat boundaries
could frustrate or induce crystallization, respectively, in the packings,
and hence the choice of container shape can limit the portion of configuration space that can be sampled.
The key interactions are interparticle forces, including (ideally)
repulsive hard-sphere interactions, friction between the particles (which
inhibits densification), and gravity. The 
final packing fraction will inevitably be sensitive to these
system characteristics. Indeed, one can achieve denser
(partially and imperfectly crystalline) packings when the particles are poured at low 
rates into horizontally shaken containers with flat boundaries \cite{Po97}.

It is tempting to compare experimentally
observed statistics of so-called RCP configurations
(packing fraction, correlation functions, Voronoi
statistics) to those generated on a computer.
One must be careful in making such comparisons
since it is difficult to simulate the features
of real systems, such as the method of preparation
and system characteristics (shaking, friction, gravity, etc.).
Nonetheless, computer algorithms are valuable because they can be used to
generate and study idealized random packings, but the final states are clearly
{\it protocol-dependent}. For example, a popular rate-dependent densification 
algorithm~\cite{Jo85,Ju97} achieves $\phi$
between $0.642$ and $0.649$,  a Monte Carlo scheme~\cite{To88}
gives $\phi \approx 0.68$, a differential-equation densification scheme
produces $\phi \approx 0.64$ \cite{Zi94}, and a ``drop and roll''
procedure~\cite{Vi72} yields $\phi \approx 0.60$; and each of these protocols
yields different sphere contact statistics.

As noted earlier, it has been argued that these variabilities 
of RCP arise because it is an ambiguous concept, explaining why
there is no rigorous prediction of the RCP density, in spite of
attempts to estimate it  \cite{Go74b,Be83,Ma08}.
The phrase ``close packed'' implies that the spheres are in contact with one
another with the highest possible average contact number $Z$.
This would be consistent with the aforementioned traditional view that RCP presents the highest  
density that a random packing of close-packed spheres can possess.
However, the terms ``random'' and ``close packed'' are
at odds with one another. Increasing the degree of coordination (nearest-neighbor
contacts),
and thus the bulk system density, comes at the expense of disorder.
The precise proportion of each of these competing effects
is {\em arbitrary}, and therein lies a fundamental problem. 
Moreover, since ``randomness" of selected jammed packings has  never been quantified,
the proportion of these competing effects 
could not be specified. To remedy these serious flaws, \onlinecite{To00b}
replaced the notion of ``close packing" with
``jamming" categories (defined precisely in Sec. \ref{jamming}),
which requires that each particle 
of a particular packing has a minimal number of properly arranged contacting particles. Furthermore, they introduced the notion
of an ``order metric" to quantify the degree
of order (or disorder) of a single packing configuration.

\begin{figure}[bthp]
\centerline{\includegraphics[height=2in,keepaspectratio,clip=]{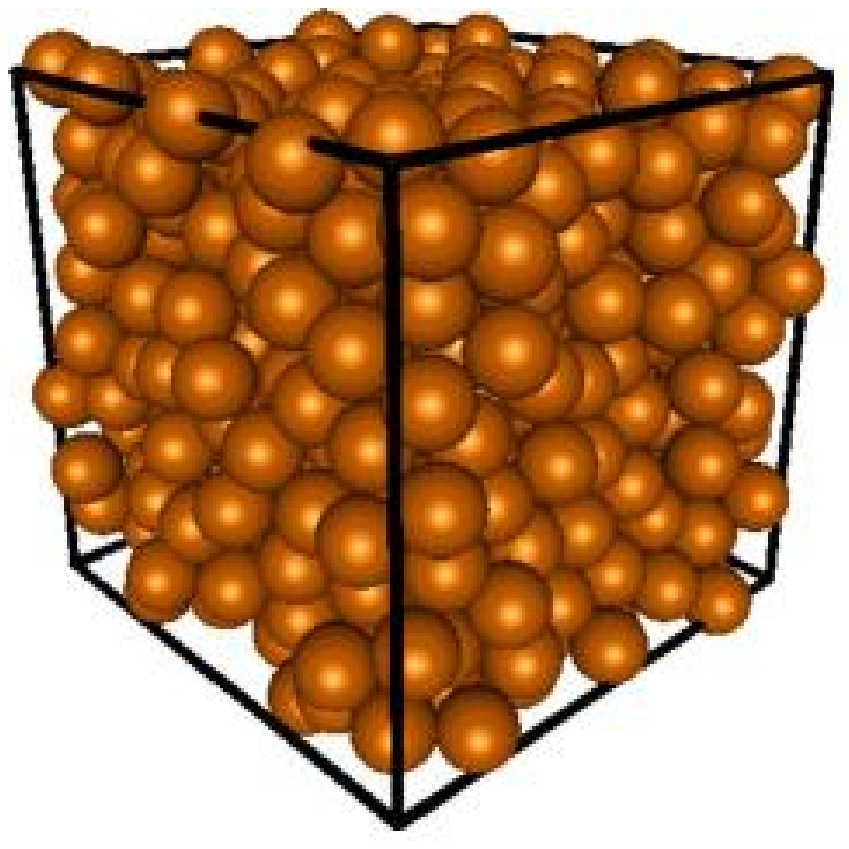}
\hspace{0.25in} \includegraphics[height=2in,keepaspectratio,clip=]{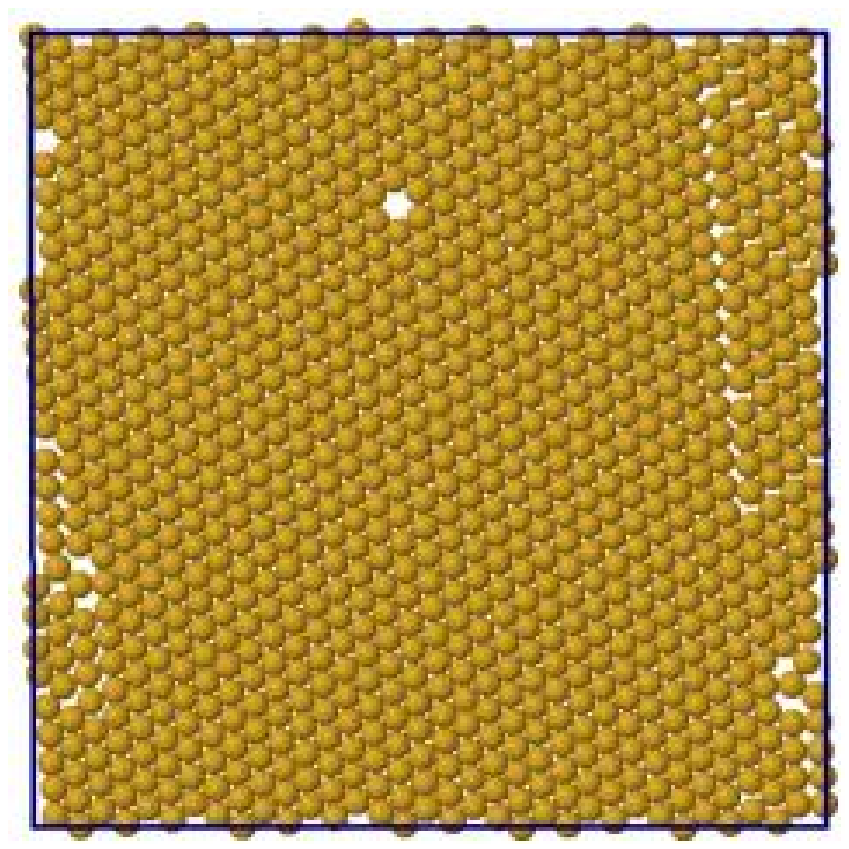}}
\caption{\footnotesize Typical protocols used to generate disordered sphere packings
in three dimensions produce highly crystalline packings in two dimensions. 
Left panel: A three-dimensional MRJ-like configuration
of 500 spheres with $\phi \approx 0.64$ produced
using the Lubachevsky-Stillinger (LS) algorithm with a fast expansion rate \cite{To00b}.
Right panel: A crystalline collectively jammed configuration 
(Sec. \ref{jamming-cat}) of 1000 disks with $\phi \approx 0.88$ produced
using the LS algorithm with a fast expansion rate \cite{Do04a}.
}
\label{MRJ}
\end{figure}

Using the Lubachevsky-Stillinger (LS) (1990) 
molecular dynamics growth algorithm  to generate jammed packings,
it was shown \cite{To00b} that fastest particle growth rates generated the most disordered sphere (MRJ) packings (with $\phi \approx 0.64$; see the left panel
of Fig. \ref{MRJ}), but that by slowing
the growth rates larger packing fractions could be {\it continuously}
achieved
up to the densest value $\pi/\sqrt{18} \approx 0.74048\ldots$
such that the degree of order increased monotonically with $\phi$. 
Those results demonstrated that the notion of RCP
as the highest possible density that a random sphere packing can attain
is ill-defined, since one can achieve packings with arbitrarily
small increases in density at the expense of correspondingly small increases
in order. This led  \onlinecite{To00b} 
to supplant the concept of RCP with the maximally random jammed (MRJ) state,
which is defined to be that jammed state 
with a minimal value of an order metric (see Sec. \ref{order}).
This work pointed the way toward a quantitative means
of characterizing all packings, namely, the geometric-structure approach.

 We note that the same LS packing protocol that leads to a uniformly disordered jammed state in
three dimensions typically yields a highly crystalline ``collectively" jammed packing in two
dimensions. Figure \ref{MRJ}  illustrates the vivid visual difference between the textures produced in
three and in two dimensions (see Section VII for further remarks). 
The low-concentration occurrence of crystal defects in the latter
is evidence for the notion that there are far fewer collectively jammed states for $N$ hard disks
in two dimensions compared to $N$ hard spheres in three dimensions. This distinction can be placed
in a wider context by recalling that there is only one type of jammed state for hard rods in one
dimension, and it is a defect-free perfect one-dimensional crystal. 
These cases for $d=1,2$, and 3,
numerical results for MRJ packing for $d=4,5$, and 6, and theoretical
results \cite{To06b}, indicating that packings in large dimensions
are highly degenerate,
suggest that the number of distinct collectively jammed packings (defined
in Sec. \ref{jamming-cat}) for a fixed 
large number $N$ of identical hard spheres rises monotonically 
with Euclidean dimension $d$. The questions and issues raised by these 
differences in the degree of disorder across dimensions emphasizes the need 
for a geometric-structure approach, to be elaborated in the following.

\section{Jamming Categories, Isostaticity, and Polytopes}
\label{jamming}

\subsection{Jamming Categories}
\label{jamming-cat}

In much of the ensuing discussion, we will treat packings of frictionless, congruent
spheres of diameter $D$ in $\mathbb{R}^d$ in the absence of gravity, i.e.,
the ``Ising model" of jammed sphere packings. 
Packing  spheres is inherently a geometrical problem 
due to exclusion-volume effects.
Indeed, the singular nature of the hard-sphere
pair potential (plus infinity or zero
for $r <D$ or $r \ge D$, respectively, where $r$ is the pair separation)
is crucial because it enables one to be precise about the concept of jamming. 
Analyzing this model directly is clearly  preferable to methods that begin 
with particle systems having ``soft" interactions, 
which are then intended to mimic packings upon passing to the 
hard-sphere limit \cite{Do04e}.
 
Three broad and mathematically precise
``jamming" categories of sphere packings can be distinguished depending on the nature of their 
mechanical stability \cite{To01b,To03a}.  In order of increasing stringency (stability), for
a finite sphere packing, these are the following:
(1) {\it Local~jamming}: Each particle in the packing is locally trapped by
its neighbors (at least $d+1$ contacting particles,
not all in the same hemisphere), i.e., it cannot be translated while fixing the positions
of all other particles;
(2) {\it Collective~jamming:} Any locally jammed configuration is collectively jammed if no
subset of particles can simultaneously be displaced so that its members
move out of contact with one another and with the remainder set; and
(3) {\it Strict~jamming:} Any collectively jammed configuration that disallows
all  uniform volume-nonincreasing strains of the system
boundary is strictly jammed.

\begin{figure}[bthp]
\centerline{\includegraphics[width=1.95in,keepaspectratio,clip=]{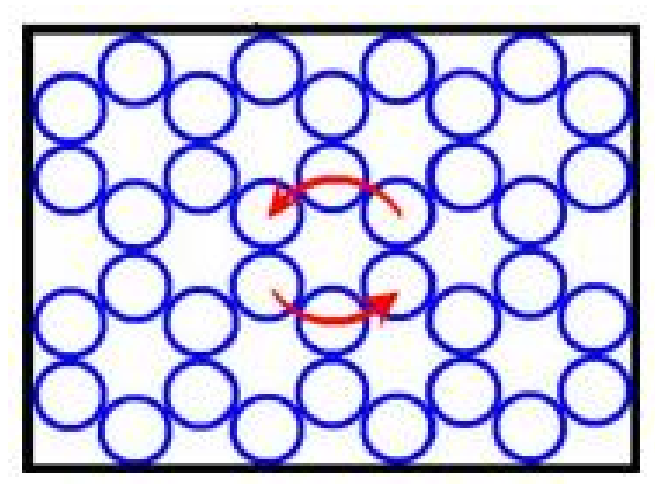}
\hspace{0.3in} \includegraphics[width=1.6in,keepaspectratio,clip=]{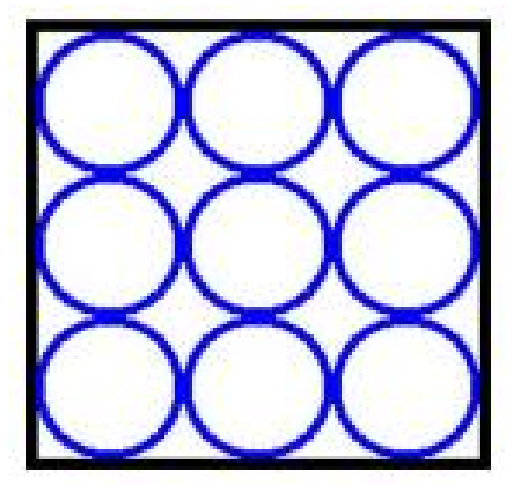}
\hspace{0.3in} \includegraphics[width=1.6in,keepaspectratio,clip=]{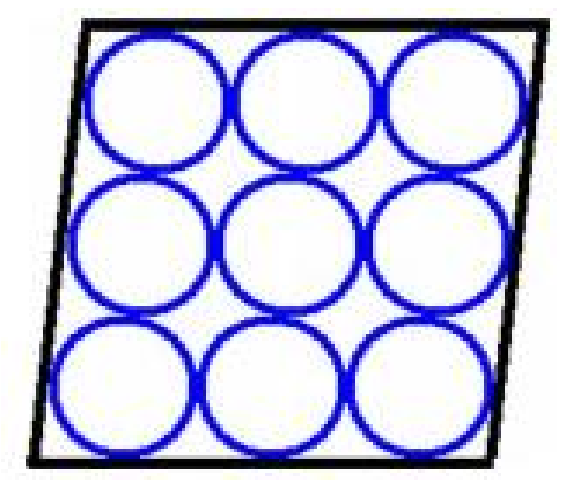}}

\caption{\footnotesize Illustrations of jamming categories.
Leftmost panel: Honeycomb-lattice packing within a rectangular hard-wall container
is locally jammed, but is not collectively jammed (e.g., a collective rotation
of a hexagonal particle cluster, as shown, will unjam the packing). Middle panel:
Square-lattice packing within a square hard-wall container is collectively jammed.
Rightmost panel: 
The square-lattice packing shown in the middle panel can be sheared
and hence is not strictly jammed. Thus, we see that
the square-lattice packing with square hard-wall boundary conditions 
can only be collectively jammed even in the infinite-volume limit.
Thus, the distinction between collective and strict jamming for such
packings remains in the thermodynamic limit.  }
\label{examples}
\end{figure}

We stress that these hierarchical jamming categories do not exhaust the universe
of possible distinctions \cite{Co98,To01b,Do04a,Do04c}, but they  span a reasonable spectrum of possibilities.
Importantly, the jamming category of a given sphere 
configuration depends on the boundary conditions employed. For example, hard-wall boundary
conditions \cite{To01b} generally yield different
jamming classifications from periodic boundary conditions \cite{Do04a}.
These jamming categories, which are closely related to the concepts of ``rigid" and ``stable"
packings found in the mathematics literature \cite{Co98},
mean that there can
be no ``rattlers" (i.e., movable but caged particles) in the packing.  Nevertheless, it
is the significant majority of spheres that compose the
underlying jammed network that confers rigidity to the
packing, and in any case, the rattlers could be removed
(in computer simulations) without disrupting the jammed remainder.
Figure \ref{examples} shows examples of ordered locally and collectively jammed
packings of disks in two dimensions 
within hard-wall containers. Observe 
the square-lattice packing with square hard-wall boundary conditions
can only be collectively jammed (not strictly jammed) even in the infinite-volume limit.
This is to be contrasted with MRJ  packings, where the
distinction between collective and strict jamming vanishes
in the infinite-volume limit, as discussed in Sec. \ref{iso}.
Note that the configurations shown in Fig. \ref{MRJ},
generated under periodic boundary conditions,
are at least collectively jammed.

To emphasize the fact that the jamming category depends
on the boundary conditions of a packing, we tabulate 
whether common periodic structures are locally, collectively
or strictly jammed. Table \ref{class1} gives the jamming classification
for such packings with hard-wall boundary conditions. These
results are compared to corresponding jamming categories
for periodic boundary conditions in Table \ref{class2}. The latter results will
depend on the choice of the number of particles $N$ within
the fundamental cell.

\begin{threeparttable}
\caption{Classification of
some of the common jammed periodic (crystal) packings of identical spheres
in two and three dimensions, where
$Z$ denotes the contact number per particle and $\phi$ is
the packing fraction for the infinite packing \cite{To01b}.
Here hard boundaries are applicable: in two dimensions
we use commensurate rectangular boundaries
and in three dimensions we use a cubical
boundary, with the exception of the hexagonal close-packed
crystal in which the natural choice is a hexagonal prism.}
\small
\begin{tabular}{c|c|c|c} \hline \hline
Periodic (crystal) structures  &Locally jammed & Collectively jammed &  Strictly
jammed \\ \hline \hline
Honeycomb ($Z=3$, $\phi= 0.605\ldots$) &yes & no&no\\ \hline
Kagom{\'e} ($Z=4$, $\phi= 0.680\ldots$)&no\tnote{a} &no\tnote{a} & no\tnote{a}\\ \hline
Square ($Z=4$, $\phi = 0.785\ldots$)&yes &yes & no\\ \hline
Triangular ($Z=6$, $\phi = 0.907\ldots$) &yes& yes&yes\\ \hline \hline
Diamond ($Z=4$, $\phi = 0.340\ldots$)&yes&no&no\\ \hline
Simple cubic ($Z=6$, $\phi = 0.524\ldots $)&yes& yes&no \\ \hline
Body-centered cubic ($Z=8$, $\phi = 0.680\ldots$)&yes&yes&no\\ \hline
Face-centered cubic ($Z=12$, $\phi=  0.740 \ldots$) &yes&yes&yes\\ \hline
Hexagonal close-packed ($Z=12$, $\phi= 0.740\ldots$) &yes&yes&yes\\ \hline \hline
\end{tabular}
\begin{tablenotes}
\item[a] With appropriately placed regular triangular- or
hexagonal-shaped boundaries, the Kagom{\'e} structure
is locally, collectively and strictly jammed.
\end{tablenotes}
\label{class1}
\end{threeparttable}

\begin{threeparttable}
\caption{Classification of
the same periodic sphere packings described in Table \ref{class1}
but for periodic boundary conditions \cite{Do04a}.  We chose the smallest fundamental cells 
with $N$ particles for which an unjamming motion exists, if there is one.
The lattice vectors for each packing are the standard ones \cite{Do04a}}.
\small
\begin{tabular}{c|c|c|c|c} \hline \hline
Periodic (crystal) structures  & ~$N$~ & Locally jammed & Collectively jammed &  Strictly
jammed \\ \hline \hline
Honeycomb & 4   &yes & no&no\\ \hline
Kagom{\'e}& 3   &yes &no & no\\ \hline
Square    & 2  &yes &no & no\\ \hline
Triangular&  1 &yes& yes&yes\\ \hline \hline
Diamond & 4   &yes&no&no\\ \hline
Simple cubic &2  &yes& no&no \\ \hline
Body-centered cubic &2  &yes&no&no\\ \hline
Face-centered cubic & 1 &yes&yes&yes\\ \hline
Hexagonal close-packed & 2 &yes&yes&yes\\ \hline \hline
\end{tabular}
\label{class2}
\end{threeparttable}
\bigskip

Rigorous and efficient linear-programming algorithms have been devised
to assess whether a particular sphere packing is locally, collectively, 
or strictly jammed \cite{Do04a,Do04c}. It is noteworthy that the jamming categories can
now be ascertained in real-system experiments using imaging techniques 
that enable one to determine configurational coordinates of a packing 
[e.g., tomography \cite{As06}, confocal microscopy  \cite{Bru03} and magnetic resonance 
imaging \cite{Ma05}]. Given these coordinates (with
high precision), one can rigorously test the ``jamming" category 
of the experimentally generated packing using the aforementioned linear programming techniques.

\subsection{Isostaticity}
\label{iso}

A packing of $N$ hard spheres of diameter $D$ in a {\it jammed} framework
in $d$-dimensional
Euclidean space is characterized by the $(Nd)$-dimensional configuration
vector of centroid positions $\bf{R}={\bf r}^N \equiv \{{\bf r}_{1},\ldots,{\bf r}_{N}\}$.
 Assume that a configuration ${\bf R}_{\bf J}$ represents a
collectively jammed {\it framework} of a packing (i.e., excluding
rattlers)  with packing
fraction $\phi_{J}$, where there are $M$ interparticle contacts.

\emph{Isostatic} packings are jammed packings that possess the minimal
number of contacts for a jamming category; namely, under periodic 
boundary conditions, for collective jamming,
$M=2N-1$ and $3N-2$ for $d=2$ and $d=3$, respectively, 
and for strict jamming, $M=2N+1$ and $3N+3$ for $d=2$ and $d=3$, respectively \cite{Do05c}.
Thus, we see that the relative differences between isostatic collective and strict jammed packings
diminish as $N$ becomes large, and since the number
of degrees of freedom is essentially equal to $N d$ (depending
on the jamming category and boundary conditions \cite{Do05c}), 
an isostatic packing has a mean contact number per particle, $Z$,
equal to  $2d$.  Collectively/strictly jammed MRJ packings
in the infinite-volume limit are isostatic. Note that packings in which $Z=2d$
are not necessarily collectively or strictly jammed; for example,
we see from Table \ref{class2} that the square and Kagom{\'e} lattices
(with $Z=4$) and the simple-cubic lattice (with $Z=6$) are neither
collectively nor strictly jammed.
Isostaticity  has attained a special status in the field, and 
has been closely linked to ``generic"
or ``random" packings \cite{Mouk98}. In fact, 
as we will show, isostatic packings can be perfectly ordered;
see also  \onlinecite{Ma09}.
Packings having more contacts than
isostatic ones are \emph{hyperstatic} and 
those having fewer contacts
than isostatic packings are \emph{hypostatic}; 
for sphere packings, these latter packings cannot be 
collectively or strictly jammed in the above sense 
\cite{Do07a}. The terms {\it overconstrained}
and {\it underconstrained} (or hypoconstrained), respectively, were used 
by \onlinecite{Do07a} to describe such packings.

\subsection{Polytope Picture of Configuration Space}
\label{jam-poly}

     Full understanding of the many-body properties of $N$ particles 
in a $d$-dimensional container of content (volume) $V$ is facilitated 
by viewing the system in its $dN$-dimensional configuration space.  
This is an especially useful approach for hard-particle models, and 
helps to understand the full range of issues concerning the approach 
to a jammed state.  For the moment, we restrict attention specifically 
the cases of $d$-dimensional hard spheres.  When container content 
$V$ is very large for fixed $N$, i.e., when packing fraction $\phi \approx 0$, 
the hard spheres are free to move virtually independently.  
Consequently, the measure (content) $\cal C$ of the available 
multidimensional configuration space is simply ${\cal C} \approx V^N$.  
But decreasing $V$ enhances the chance for sphere collisions and 
correspondingly reduces $\cal C$, the remaining fraction of configuration 
space that is free of sphere overlaps.  The amount of reduction is 
related exponentially to the excess entropy $S^{(e)}(N,V)$  for the $N$-sphere system:
\begin{equation}
{\cal C}(N,V) \approx V^N\exp[S^{(e)}(N,V) /k_B],
\end{equation}
where $k_B$  is Boltzmann's constant.

In the low-density regime, the excess entropy admits of a power series 
expansion in covering fraction  $\phi$:
\begin{equation}
\frac{S^{(e)}(N,V)}{N k_B} = N \sum_{n \ge 1} \left(\frac{\beta_n}{n+1}\right)
\left(\frac{\phi}{v_1}\right)^n \;.
\end{equation}
Here $v_1$  is the volume of a particle, as indicated earlier in Eq. (\ref{vol-sph}).  
The $\beta_n$  are the {\it irreducible} Mayer cluster integral 
sums for $n+1$  particles that determine the virial coefficient of order 
$n+1$ \cite{Ma40}.   For hard spheres in dimensions $1 \le d \le 8$,
these coefficients for low orders $1 \le n \le 3$ are known exactly, 
and accurate numerical estimates are available for $n+1 \le 10$ \cite{Cl06}. 
This power series represents a  function of $\phi$  
obtainable by analytic continuation along the positive real axis
to represent the thermodynamic behavior for the fluid phase from $\phi=0$  
up to the freezing transition, which occurs at $\phi \approx 0.4911$  
for hard spheres in three dimensions \cite{No08}.
This value is slightly below the minimum density $\phi \approx 0.4937$ at which collective 
jamming of $d=3$   hard spheres is suspected first to occur \cite{To07}.  
Consequently, the available configuration space measured by    
${\cal C}(N,V)$ remains connected in this density range, i.e., 
any nonoverlap configuration 
of the $N$ spheres can be connected to any other one by a continuous displacement 
of the spheres that does not violate the nonoverlap condition.

     A general argument has been advanced that thermodynamic functions must 
experience a subtle, but distinctive, essential singularity at first-order 
phase transition points \cite{An64,Fish70}. In particular, this applies to the 
hard-sphere freezing transition, and implies that attempts to analytically continue 
fluid  behavior into a metastable over-compressed state are dubious.  
Aside from any other arguments that might be brought to bear, this indicates that 
such extrapolations are fundamentally incapable of identifying unique random jammed 
states of the hard-sphere system.  Nevertheless, it is clear that increasing $\phi$  
beyond its value at the thermodynamic freezing point soon initiates partial fragmentation 
of the previously connected nonoverlap configuration space in finite
systems.  That is, locally disconnected portions 
are shed, each to become an individual jammed state displaying its own geometric 
characteristics. (The jammed tunneled crystals 
mentioned in Sec.~\ref{map} are examples of such localized regions near the freezing point.) We shall elaborate on this point within this subsection
after discussing the polytope picture of configuration space near
jamming points.

Consider decreasing the packing fraction slightly 
in a {\it sphere} packing that is at least collectively 
jammed by reducing the particle diameter
by $\Delta{D}$, $\delta=\Delta{D}/D\ll1$, so that the packing fraction is
lowered to $\phi=\phi_{J}\left(1-\delta\right)^{d}$. 
We call $\delta$ the \emph{jamming gap} or distance to jamming.
It can be shown that there is a sufficiently small $\delta$ that
does not destroy the jamming confinement property, in the sense that the configuration
point $\bf{R}=\bf{R}_{J}+\bf{\Delta{R}}$ remains trapped in a small neighborhood
$\mathcal{J}_{\bf{\Delta{R}}}$ around $\bf{R}_{J}$ \cite{Co82}.
Indeed, there exists a range of positive values of $\delta$ that depends
on $N$ and the particle arrangements that maintains the jamming confinement property.
Let us call $\delta_*$ the threshold value at which jamming is lost. How does
$\delta_*$ scale with $N$ for a particular $d$?  An elementary analysis based on
the idea that in order for a neighbor pair (or some larger local group) of
particles to change places, the surrounding $N-2$ (or $N-3,\ldots$) particles
must be radially displaced and compressed outward so as to concentrate the
requisite free volume around that local interchangeable group concludes
that $\delta_* \sim C N^{-1/d}$, where the constant $C$ depends on the dimension
$d$ and the original jammed particle configuration.

\begin{figure}[bthp]
\centerline{\includegraphics[height=1.9in,keepaspectratio,clip=]{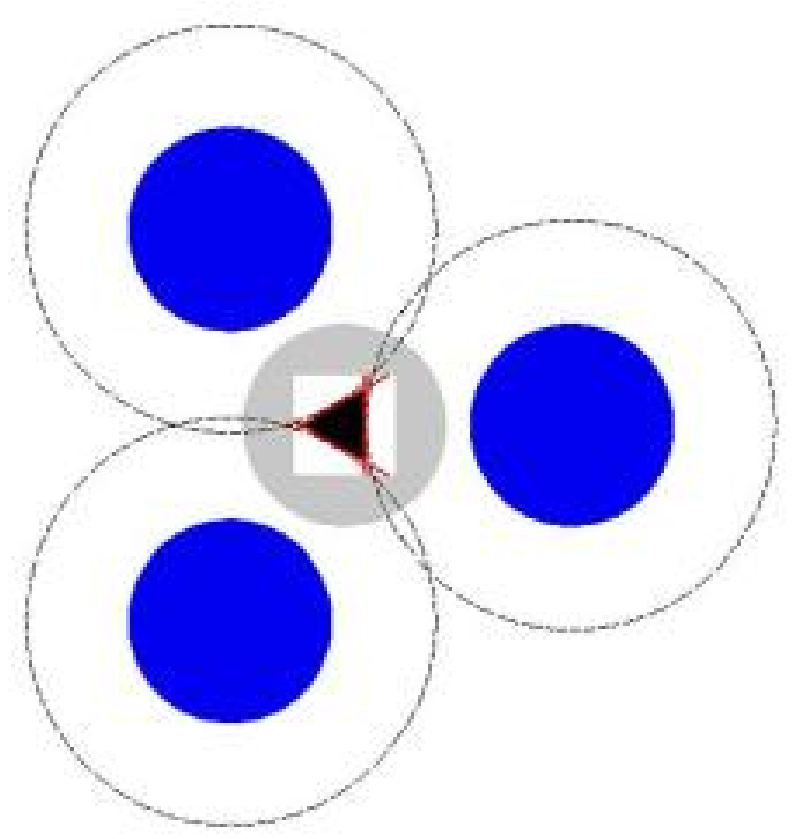}
\hspace{0.35in}\includegraphics[height=1.9in,keepaspectratio,clip=]{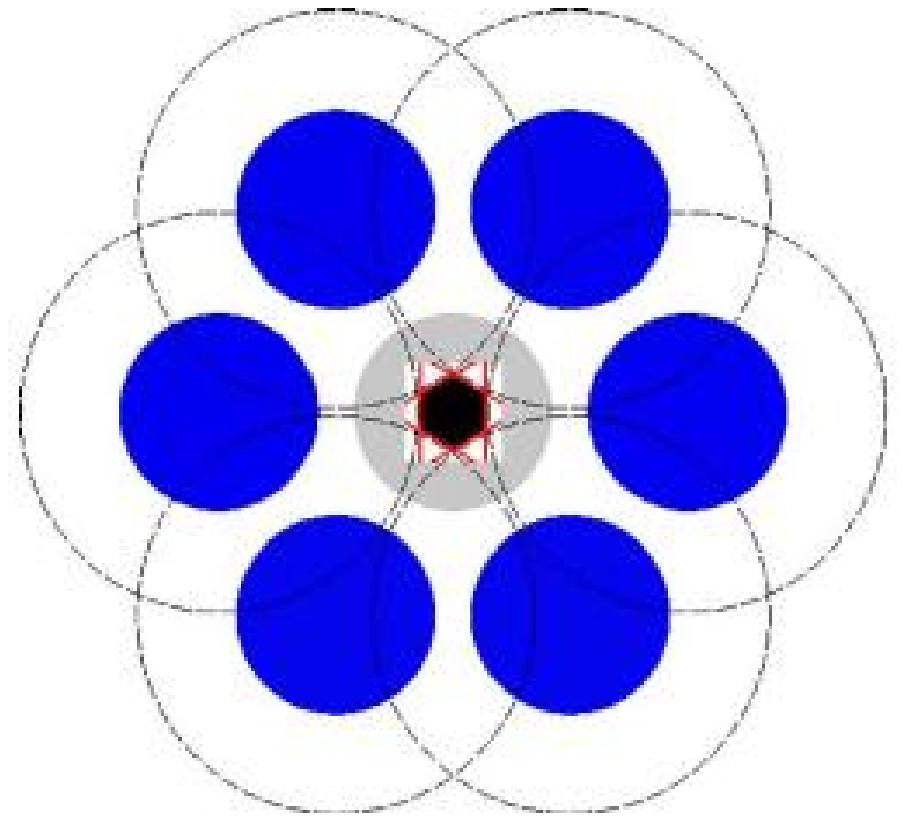}}
\caption{\footnotesize The polytope of allowed displacements  $\mathcal{P}_{\bf{\Delta{R}}}$ for a {\it locally} jammed disk (light shade) trapped among three
(left) or six (right, as in the triangular lattice) fixed disks (blue).
The exclusion disks (dashed lines) of diameter twice the
disk diameter are drawn around each of the fixed disks, along with
their tangents (red lines) and the polytope $\mathcal{P}_{\bf{\Delta{R}}}$  they bound (black).
The polytope for the isostatic (left) and overconstrained (right) case
is a triangle and  hexagon, respectively.
}
\label{single-jam}
\end{figure}

It is noteworthy that for fixed $N$ and sufficiently small $\delta$, it can be shown that asymptotically
(through first order in $\delta$) the set of displacements that are accessible to the packing approaches
a convex \emph{limiting polytope} (a closed polyhedron in high
dimension) $\mathcal{P}_{\bf{\Delta{R}}}\subseteq\mathcal{J}_{\bf{\Delta{R}}}$
\cite{Sa62,St69}. This polytope $\mathcal{P}_{\bf{\Delta{R}}}$
is determined from the linearized impenetrability equations \cite{Do04a,Do04c}
and, for a fixed system center of mass, is necessarily bounded for a jammed
configuration. This implies that 
the number of interparticle contacts $M$ is at least one larger
than the dimensionality $d_{CS}$ of the relevant configuration space.
Examples of such low-dimensional polytopes for a single locally jammed disk are shown
in Fig. \ref{single-jam}.

Importantly, for an isostatic contact
network, $\mathcal{P}_{\bf{\Delta{R}}}$ is a \emph{simplex} \cite{Do05c}.
A $d$-dimensional simplex in $\mathbb{R}^d$ is a closed convex polytope 
whose $d+1$ vertices ($0$-dimensional points) do not all lie in a ($d-1$)-dimensional 
flat sub-space
or, alternatively, it is a finite region of $\mathbb{R}^d$
enclosed by $d+1$ hyperplanes [($d-1$)-dimensional ``faces"] (e.g., a triangle for $d=2$,
a tetrahedron for $d=3$ or a pentatope for $d=4$). 
For overconstrained jammed packings (e.g., ordered maximally dense states), the limiting high-dimensional polytopes
have more faces than simplices do and can be geometrically very complex \cite{Sa62,St69}.
The fact that $\mathcal{P}_{\bf{\Delta{R}}}$ is a \emph{simplex}
for an isostatic packing enables one to derive rigorous
results, as we will now describe.

Consider adding thermal kinetic energy to a nearly jammed
sphere packing in the {\it absence of rattlers}. While the system will not be 
globally ergodic over the full system configuration space and thus
not in thermodynamic equilibrium,  one can still define a 
macroscopic pressure $p$ for the trapped but locally ergodic system
by considering time averages as the system
executes a tightly confined motion around the \emph{particular} configuration
$\bf{R}_{J}$. The probability distribution $P_{f}(f)$ of the {\it time-averaged} interparticle forces $f$ has been rigorously linked to the contact 
value $r=D$ of the pair correlation function $g_2(r)$ defined
in Section II, and this in turn can be related to the distribution of
simplex face hyperareas for the limiting polytope  \cite{Do05c}. 
Moreover, since the available (free)
configuration volume scales in a predictable way with the jamming
gap $\delta$, one can show that the reduced pressure is asymptotically given by
the free-volume equation of state \cite{Sa62,St69,Do05c},
\begin{equation}
\frac{p}{\rho k_{B}T} \sim \frac{1}{\delta}=\frac{d}{1-(\phi/\phi_{J})},
\label{free}
\end{equation}
where $T$ is the absolute temperature and $\rho$ is the number density.
So far as the limiting polytope picture is concerned, the extremely
narrow connecting filaments that in principle connect the jamming
neighborhoods have so little measure that they do not overturn the
free-volume leading behavior of the pressure, even as system size is allowed
to go to infinity.  Although there is no rigorous proof yet 
for this claim, all numerical evidence strongly suggests that it is correct.
 Relation (\ref{free}) is remarkable, since it enables
one to determine  accurately the true jamming density of a given packing,
even if the actual jamming point has not quite yet been reached, just by
measuring the pressure and extrapolating to $p=+\infty$.

\begin{figure}[bthp]
\centerline{\includegraphics[width=3in,keepaspectratio,clip=]{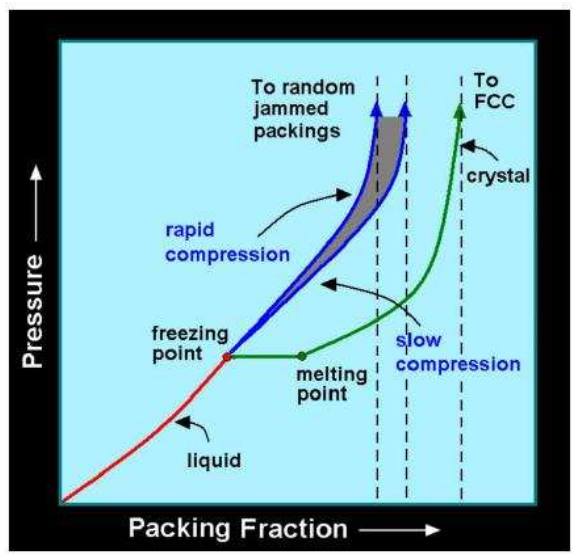}}
\caption{\footnotesize The isothermal phase behavior of
three-dimensional hard-sphere model in the pressure-packing fraction plane, adapted from \onlinecite{To02a}.
Increasing the density plays the same role as decreasing temperature of a molecular liquid;
see Fig. \ref{glass}. Three different isothermal 
densification paths by which a hard-sphere liquid 
may jam are shown.  An infinitesimal compression rate of the liquid traces out the thermodynamic equilibrium path (shown in green), 
including a discontinuity resulting from the first-order freezing transition
to a crystal branch.  Rapid compressions of the liquid while suppressing 
some degree of local order (blue curves) can avoid 
crystal nucleation (on short time scales) and produce a range
of amorphous metastable extensions of the liquid branch that jam
only at the their density maxima. }
\label{hsphere}
\end{figure}

This free-volume form has been used to estimate
the equation of state along ``metastable" extensions of the hard-sphere
fluid up to the  infinite-pressure endpoint, 
assumed to be  random jammed states \cite{To95b,To02a}. 
To understand this further, it is useful to recall the hard-sphere phase
behavior in three dimensions; see Fig. \ref{hsphere}.
For densities between zero and the ``freezing" point
($\phi \approx 0.49$), the thermodynamically stable phase is a liquid. Increasing the density
beyond the freezing point results in a first-order phase transition
to a crystal branch that begins at the melting point ($\phi \approx 0.55$)
 and whose ending point is the maximally dense fcc packing ($\phi \approx 0.74$),
which is a jammed packing in which each particle contacts 12 others \cite{Mau99}.
However, compressing a hard-sphere liquid rapidly, under the
constraint that significant crystal nucleation is suppressed,
can produce a range of metastable branches whose density
end points are random ``jammed" packings \cite{Ri96b,To02a}, which can be regarded to be glasses.
A rapid compression leads to a lower random jammed
density than that for a slow compression. 
The most rapid compression presumably leads to the MRJ state with $\phi \approx 0.64$ \cite{To02a}.
\onlinecite{To95a,To95b} reasoned that the functional form of the pressure of the stable liquid branch 
(which appears to be dominated by an unphysical pole at $\phi=1$) must be fundamentally 
different from the free-volume form (\ref{free}) that applies near jammed 
states, implying that the equation
of state is nonanalytic at the freezing point and proposed the
following expression along any constrained metastable branch:
\begin{equation}
\frac{p}{\rho k_{B}T}= 1+ 4\phi g_F \frac{1-\phi_F/\phi_J}{1-\phi/\phi_J} \qquad \mbox{for}\quad \phi_F \le \phi \le \phi_J,
\end{equation}
where $\phi_F \approx 0.491$ is the packing fraction at the freezing point,
$g_F \approx 5.72$ is the corresponding value of the pair correlation
function at contact, and $\phi_J$ is the jamming density, whose
value will depend on which metastable path is chosen. 
[\onlinecite{To95a,To95b} actually considered the more general
problem of nearest-neighbor statistics of hard-sphere systems, which required
an expression for the equation of state.] Unfortunately,
there is no unique metastable branch (see Fig. \ref{hsphere}) because 
it depends on the particular constraints used to generate the metastable
states or, in other words, the protocol employed, which again emphasizes
one of the themes of this review. Moreover, in practice, 
metastable states of identical spheres in $\mathbb{R}^3$ 
have an inevitable tendency to crystallize \cite{Ri96b}, but even
in binary mixtures of hard spheres chosen to avoid crystallization
the dispersion of results and ultimate nonuniqueness of the jammed states still apply.
We note that \onlinecite{Ka07} assumed the same free-volume 
form to fit the pressure of ``metastable" states for monodisperse hard spheres as obtained
from both numerical and experimental data to determine
$\phi_J$. Their best fit yielded $\phi_J=0.6465$.   

We note that density of states (vibrational modes) in packings of
soft spheres has been the subject of recent interest \cite{Wy05,Si05}.
Collective jamming in hard-sphere packings corresponds
to having  no ``soft modes" in soft-sphere systems, i.e.,  
no unconstrained local or global particle translations are
allowed, except those corresponding to rattlers.
Observe that it immediately follows that if a hard-sphere
packing is collectively jammed
to first order in $\delta$,  a corresponding configuration
of purely soft repelling particles will possess
quadratic modes in the vibrational energy spectrum for such 
a system of soft spheres.

\section{Order Metrics}
\label{order}

The enumeration and classification of both ordered and disordered jammed 
sphere packings for the various jamming categories is an outstanding problem.
Since the difficulty of the complete enumeration of  jammed packing configurations rises
exponentially with the number of particles,  it is 
desirable to devise a small set of intensive parameters that can characterize packings well. 
One  obvious property of a sphere packing is the packing fraction $\phi$.
Another important characteristic of a packing is some
measure of its ``randomness" or degree of disorder.
We have stressed that one ambiguity of the old RCP concept was that 
``randomness" was never quantified.
To do so is a  nontrivial challenge, but even the tentative solutions
that have been put forth during the last decade
have been profitable not only to characterize
sphere packings \cite{To00b,Tr00,Ka02d,To03a} but also glasses, simple liquids, and water \cite{Tr00,Er01,Er02,Er03}.

One might argue that the maximum of an appropriate ``entropic"
metric would be a potentially useful way to characterize the
randomness of a packing and therefore the MRJ state.
However, as pointed out by \onlinecite{Ka02d}, a substantial hurdle to
overcome in implementing such an order metric is the necessity to generate all possible jammed
states or, at least, a representative sample of such states in an
unbiased fashion using a ``universal" protocol in the large-system
limit, which is an intractable problem. Even if such a universal protocol could be developed,
however, the issue of what weights to assign the resulting
configurations remains. Moreover, there are other fundamental
problems with entropic measures, as we will discuss in Sec.~\ref{mrj},
including its significance for two-dimensional monodisperse
hard-disk packings as well as polydisperse hard-disk packings 
with a sufficiently narrow size distribution. It is for this reason that
we seek to devise order metrics that can be applied to single
jammed configurations, as  prescribed by the geometric-structure point of view.

A many-body system of $N$ particles is completely characterized statistically 
by its $N$-body probability density function $P({\bf R};t)$ that is associated
with finding the $N$-particle system with configuration ${\bf R}$ at
some time $t$. Such complete
information is virtually never  available for large $N$ and, in practice, one
must settle for reduced information, such
as a scalar order metric $\psi$.
Any order metric $\psi$ conventionally possesses the following
three properties: (1) it is a well-defined scalar function of
a configuration ${\bf R}$; (2) it is subject 
typically to the normalization $0\le \psi \le 1$;
and, (3) for any two configurations ${\bf R}_A$ and ${\bf R}_B$, $\psi({\bf R}_A) > \psi({\bf R}_B)$ implies that 
configuration  ${\bf R}_A$ is to be
considered as more ordered than configuration ${\bf R}_B$. 
The set of order parameters that one selects is unavoidably subjective, given that there 
appears to be no single universal scalar
measure of order. However, one can construct order metrics that lead to consistent
results (e.g., common minima for jammed packings), as we will discuss after considering specific examples.

\subsection{Specific Order Metrics}

\begin{figure}[bthp]
\centerline{\includegraphics[height=2.3in,keepaspectratio,clip=]{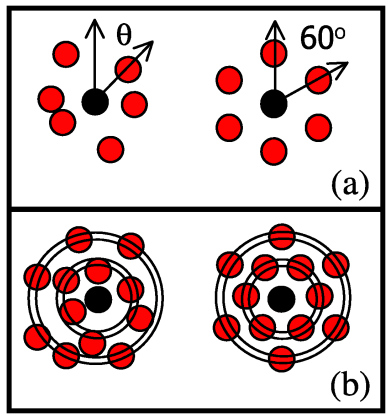}}

\caption{\footnotesize Two types of order. (a) Bond-orientational order contains information about
the orientation of the vectors connecting neighboring particles in
the packing of interest (left). If these orientations persist throughout the packing,
as they do in a triangular lattice (right), the packing is considered
to be perfectly bond-orientationally ordered. (b) Translational order
contains information about the relative spacing of
particles in the packing of interest (left) relative to that of the
densest packing at the same density (right).}
\label{order-Q-T}
\end{figure}

Many relevant order metrics have been devised, but
here we briefly describe only some of them.
The bond-orientational order $Q_{\ell}$ \cite{St83} in three dimensions is defined in terms of the spherical
harmonics $Y_{{\ell}m}(\theta_i, \varphi_i)$, where $\theta_i$ and $\varphi_i$ are the
polar and azimuthal angles (relative to a fixed coordinate
system) 
 of the near-neighbor bond for particle pair $i$.
  A near neighbor could be defined as any sphere
within a specified local radius (e.g., set by the
first minimum of the pair correlation function beyond contact)  
or by a sphere within
a face-sharing Voronoi polyhedron \cite{To02a}.
The average of $Q_{\ell}$ over all of the near-neighbor bonds $N_b$
provides a global measure of symmetries in many-particle systems.
Of particular interest is the average when $\ell=6$, i.e.,
\begin{equation}
Q_6 \equiv 
\left(\frac{4\pi}{13}\sum_{m=-6}^{6}\left|\frac{1}{N_b}
\sum_{i=1}^{N_b}Y_{6m}(\theta_i, \varphi_i)\right|^2\right)^{1/2},
\end{equation}
since it reaches its maximum value for the perfect fcc lattice  and is zero for a Poisson (uncorrelated) point distribution in the infinite-volume limit \cite{Ri96c}
(see Fig. \ref{order-Q-T} for the two-dimensional analog). It is seen that
$Q \equiv Q_6/Q_6^{fcc}$ lies in the closed interval [0,1] and therefore qualifies as
an order metric \cite{To00b}. 

A more local measure of bond-orientational order, $Q_{6,\text{local}}$,
can be obtained by evaluating the bond order at each
sphere individually, and then averaging over all spheres \cite{Ka02d}, i.e.,
\begin{equation}
Q_{6,\text{local}} \equiv
\frac{1}{N}\sum_{j=1}^{N}
\left(\frac{4\pi}{13}\sum_{m=-6}^{6}\left|\frac{1}{n_j}
\sum_{i=1}^{n_j}Y_{6m}(\theta_i, \varphi_i)\right|^2\right)^{1/2},
\end{equation}
where $n_j$ is the number of nearest-neighbors of
sphere~$j$. This is analogous to the two-dimensional
definition of local bond-orientational order studied 
by \onlinecite{Ka00}. As noted in that work, such a local measure of order
is more sensitive to small crystalline regions within a packing
than is its global counterpart $Q_6$, and thus
avoids the possibility of ``destructive" interference between
differently oriented crystalline regions.

Using the fcc lattice radial coordination structure as a reference system,
one can define a translational order metric $T$
based upon the mean occupation of sphere centers  within thin concentric shells 
around each sphere in the packing as compared to the mean occupation
of the same shells in the fcc lattice and the ideal gas at the
same packing fraction \cite{To00b,Tr00}, i.e.,
\begin{equation}
T \equiv \left| \frac{\displaystyle \sum_{i=1}^{N_{\text{shells}}}
\left(n_{i} - n_{i}^{\text{ideal}}\right)}
{\displaystyle \sum_{i=1}^{N_{\text{shells}}}\left(n_{i}^{\text{fcc}} -
 n_{i}^{\text{ideal}}\right)} \right|.
\end{equation}
Here $n_{i}$ is the average occupancy of the $i^{th}$
shell, with superscripts having the obvious meaning, and
$N_{\text{shells}}$ is the total number of shells employed.
Figure \ref{order-Q-T} illustrates the two-dimensional analog
of this order metric.

All of the order metrics defined above were constructed
to yield their maximum values of unity (when appropriately
normalized) for the densest and most symmetrical 
(close-packed crystal) packing. Order metrics that are not
based on any specific crystal structure have also been devised. For example,
two different translational order metrics
have been constructed that are based on functionals
of the pair correlation function \cite{Tr00}.
It has been suggested that local density
fluctuations within a ``window" of a
given size also can be a useful order metric \cite{To03a}.
In particular, calculation of the local number variance 
for a variety of crystal, quasicrystal
and ``hyperuniform" disordered point patterns reveals that
it provides a useful rank-order of 
these hyperuniform spatial patterns at {\it large length scales} \cite{To03a,Za09}.
A hyperuniform point pattern is one which the
infinite-wavelength density fluctuations
vanish or, equivalently, possesses a
structure factor $S({\bf k})$ (defined in Sec. II) 
that tends to zero in the limit ${\bf k}\rightarrow {\bf 0}$  \cite{To03a}.

\subsection{Characteristics of a Good Order Metric}

The specific order metrics have both strengths and weaknesses.
This raises the question of what are the characteristics
of a good order metric?
There is clearly an enormous family of scalar functions
that possess the aforementioned three generic properties of an order metric
$\psi$, but they may not necessarily be useful ones. It has been suggested that
a  good order metric should have the following additional properties \cite{Ka02d}:
(1) sensitivity to any type of
ordering without  bias toward any reference system; (2) ability to reflect the hierarchy of
ordering between prototypical systems given by common
physical intuition (e.g., perfect crystals with high symmetry
should be highly ordered, followed by quasicrystals, correlated disordered packings
without long-range order, and finally spatially uncorrelated or Poisson
distributed particles); (3) capacity to detect order at any length scale;
and (4) incorporation of both the variety of local coordination patterns 
as well as the spatial distribution of such patterns should be included.  
Moreover, any useful set of order metrics
should consistently produce results that are positively correlated with one another
\cite{To00b,To02a}. The development of improved order
metrics deserves continued research attention.

\section{Order Maps and Optimal Packings}
\label{map}

The geometric-structure classification naturally emphasizes that  
there is a great diversity in the types of attainable
    jammed packings with varying magnitudes of overall order, density,
and other intensive parameters. The notions
    of ``order maps" in combination with the mathematically precise ``jamming
    categories" enable one to view and characterize well-known packing
    states, such as the densest sphere packing (Kepler's conjecture) and maximally random jammed (MRJ) packings as extremal states in the order map for a
    given jamming category. Indeed, this picture encompasses not only these
    special jammed states, but an uncountably infinite number of other
    packings, some of which have only recently been identified as physically
    significant, e.g., the jamming-threshold states (least dense 
jammed packings) as well as states between these and MRJ.

The  so-called {\it order} map \cite{To00b} provides
a useful means to classify packings, jammed or not. 
It represents any attainable hard-sphere configuration as a point in the $\phi$-$\psi$ plane. 
This two-parameter description is but a very small subset
of the relevant parameters that are necessary to fully characterize a 
configuration, but it  nonetheless enables one to draw important
conclusions. For collective jamming, a highly schematic order map has previously been proposed
\cite{To00b}.

\begin{figure}[bthp]
\centerline{\includegraphics[height=1.8in,keepaspectratio,clip=]{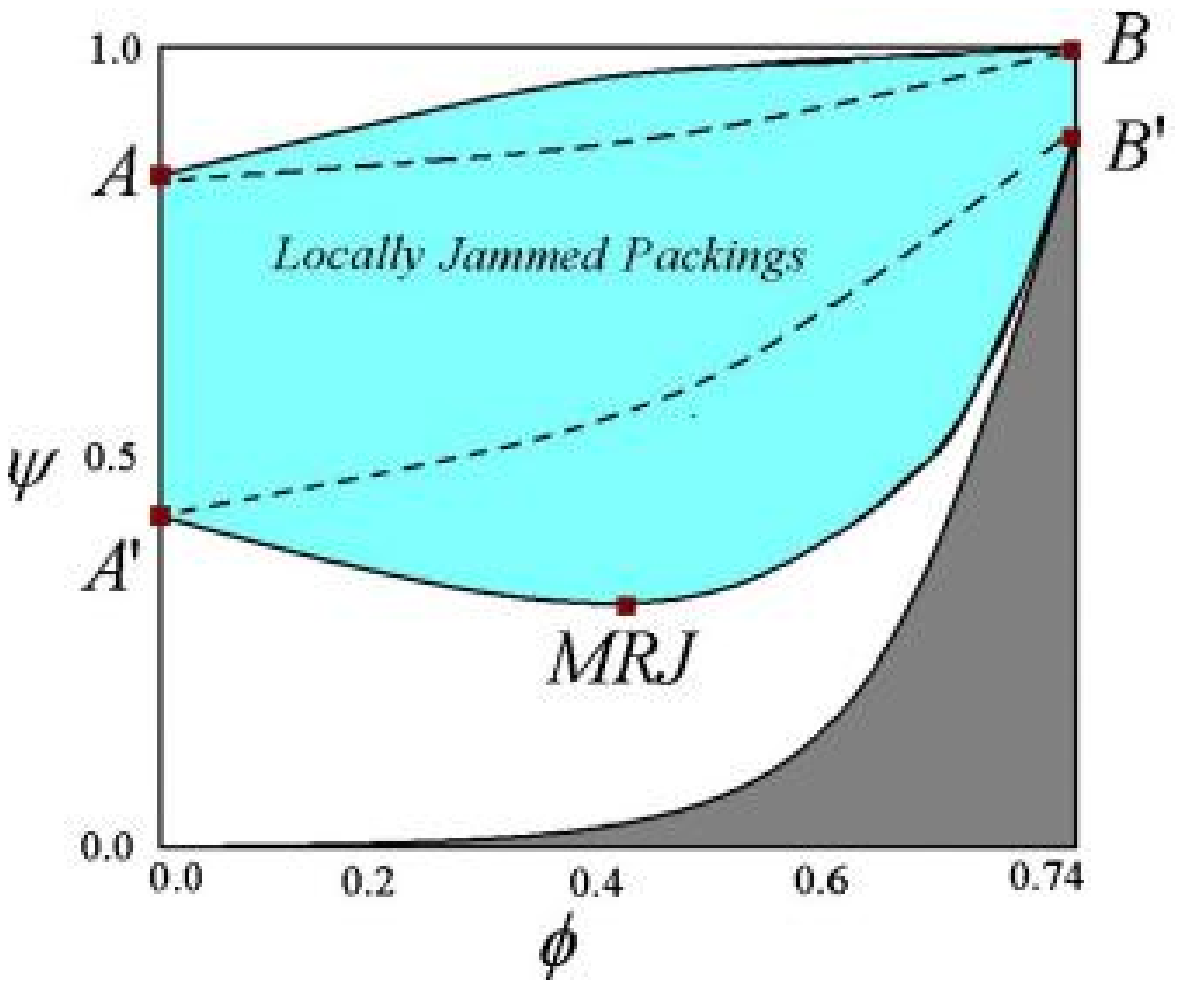}
\includegraphics[height=1.8in,keepaspectratio,clip=]{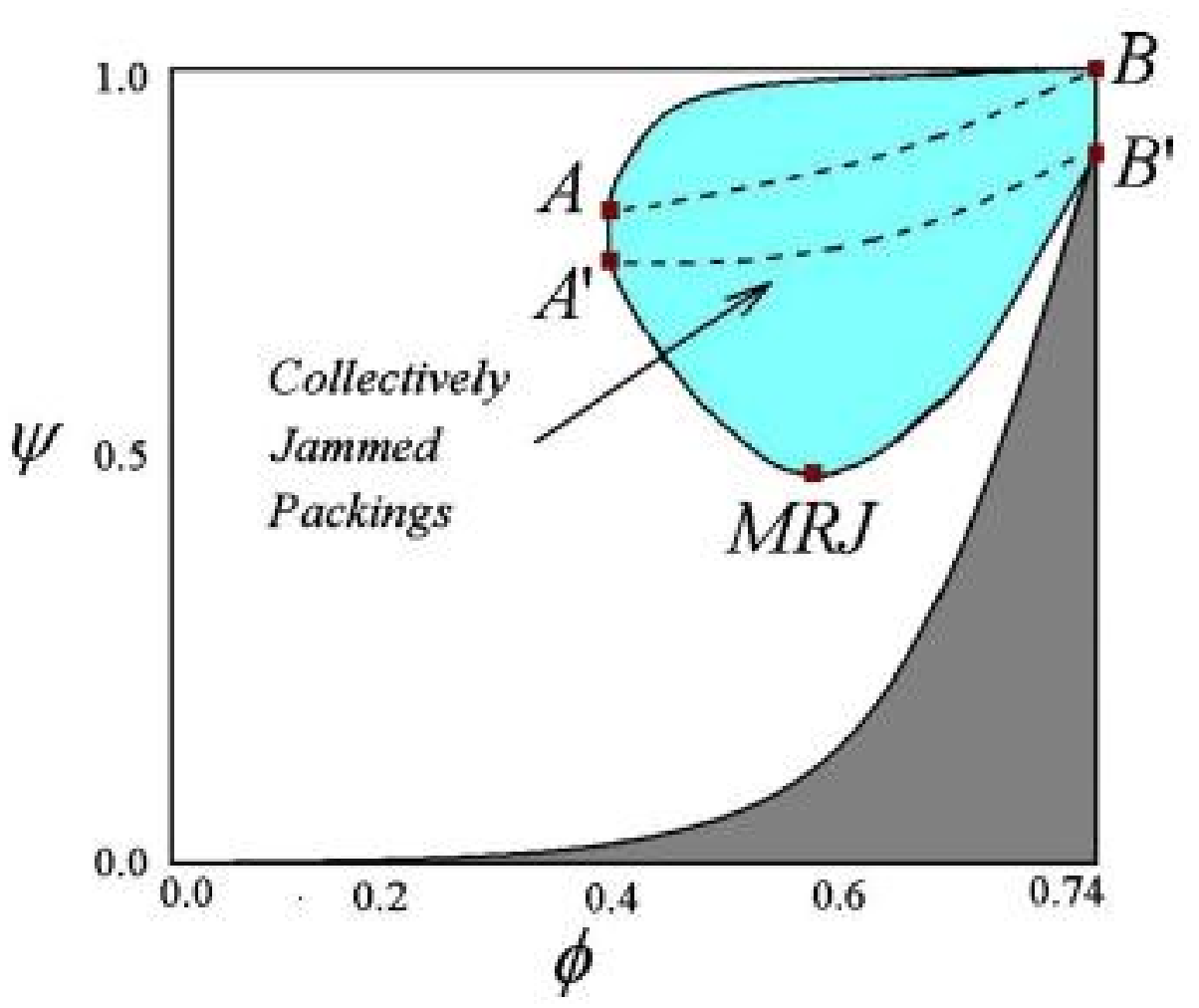}
\includegraphics[height=1.8in,keepaspectratio,clip=]{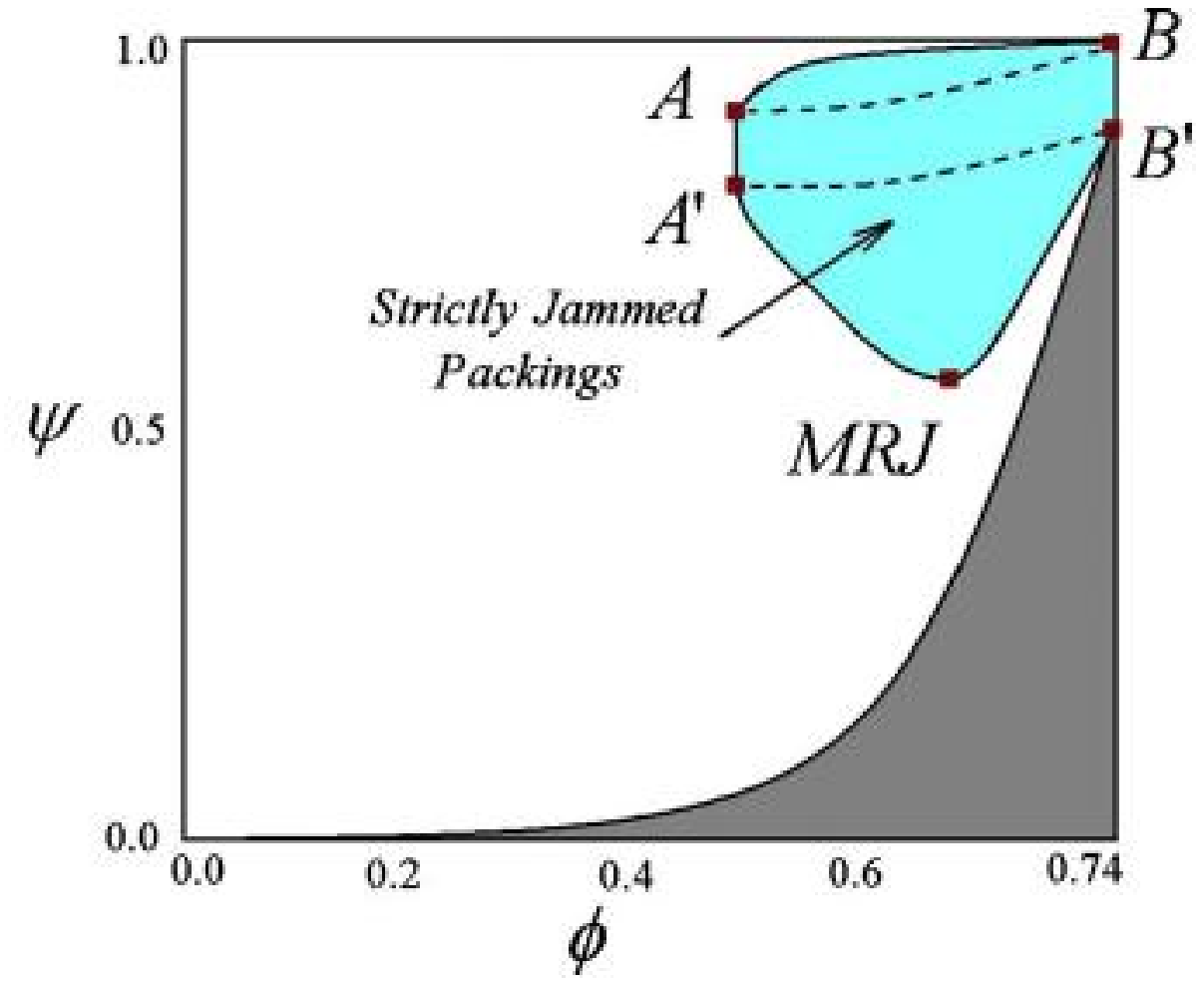}}
\caption{\footnotesize Schematic order maps in the density-order
($\phi$-$\psi$) plane for the three different jamming categories in $\mathbb{R}^3$
under periodic boundary conditions. White and blue regions contain the attainable
packings, blue regions represent the jammed subspaces, and  dark shaded regions
contain no packings.
The locus of points $A$-$A^{\prime}$ correspond to the lowest-density jammed
packings. The locus of points $B$-$B^\prime$ correspond
to the densest jammed packings. Points MRJ represent
the maximally random jammed states, i.e., the most disordered
states subject to the jamming constraint. It should be noted that the packings
represented are not subject to rattler exclusion.}
\label{maps}
\end{figure}

\begin{figure}[bthp]
\centerline{\includegraphics[height=1.8in,keepaspectratio,clip=]{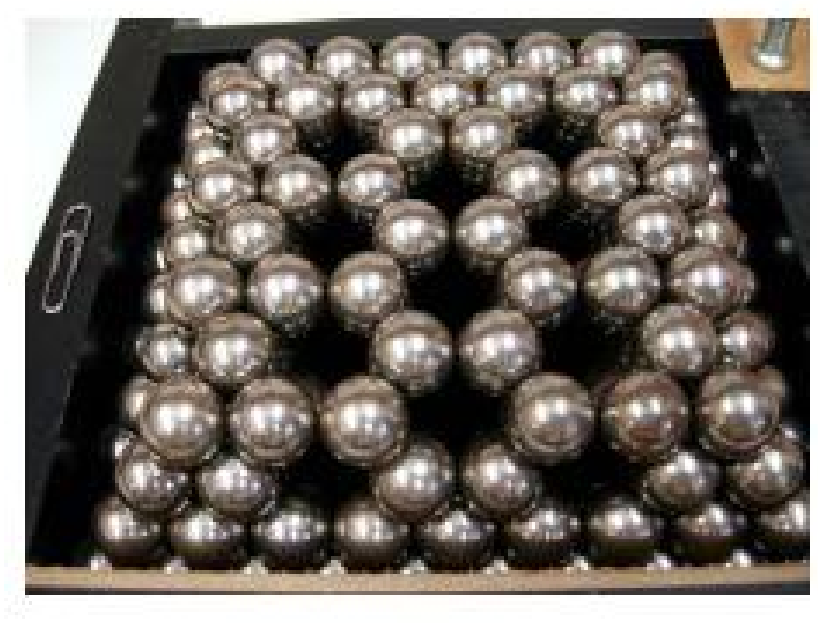}
\hspace{0.2in}\includegraphics[height=1.9in,keepaspectratio,clip=]{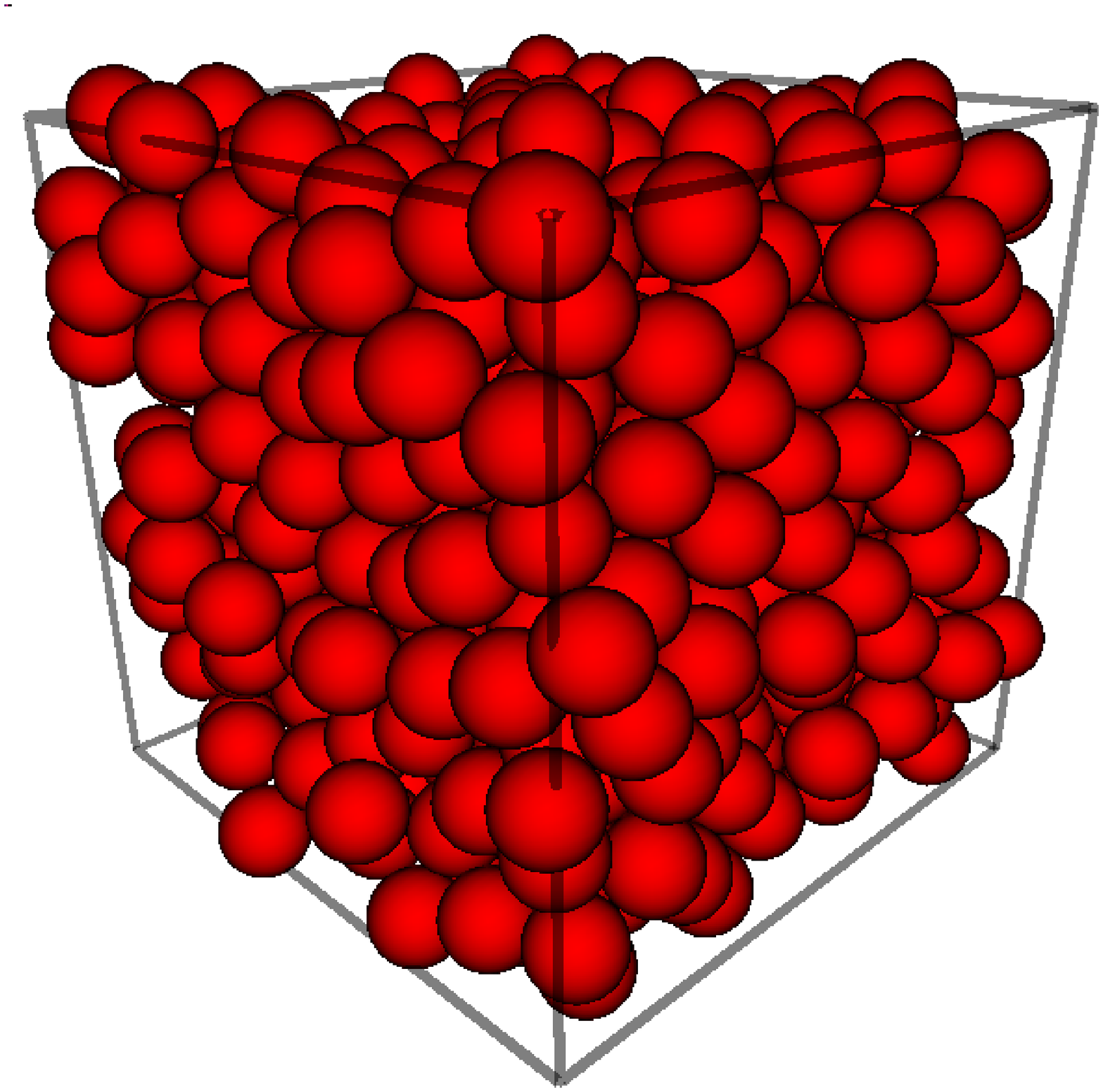}
\hspace{0.2in}\includegraphics[height=1.8in,keepaspectratio,clip=]{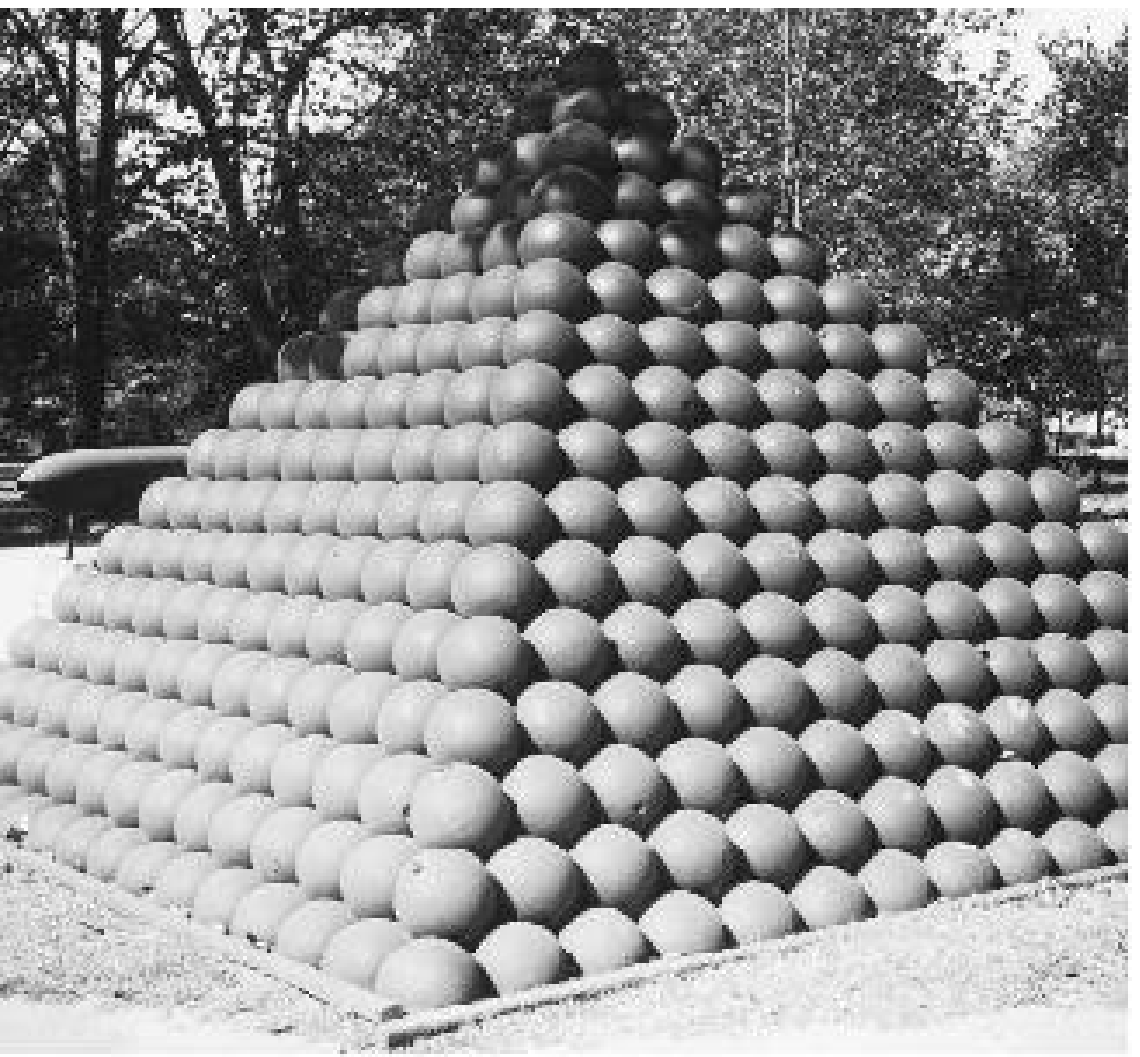}}
\caption{Three different optimal strictly jammed packings identified in the right
most graph of Fig. \ref{maps}.
Left panel: A: $Z=7$. Middle panel: MRJ: $Z=6$ (isostatic). Right panel: B: $Z=12$.}
\label{three}
\end{figure}

Here we present  a set
of refined order maps for each of three jamming categories in $\mathbb{R}^3$ 
(see Fig. \ref{maps}) based both upon early work \cite{To00b,Ka02d} and the most recent investigations \cite{Do04a,To07}.
Crucially, the order maps shown in Fig. \ref{maps}
are generally different across jamming categories and
independent of the protocols used to generate hard-sphere
configurations, and for present purposes include {\it rattlers}.
In practice, one needs to use a variety of protocols to
produce jammed configurations in order to populate the interior 
and to delineate the boundary of the
jammed regions shown in the Fig. \cite{Ka02d}. Moreover, the frequency of
occurrence of a particular configuration is irrelevant insofar as the
order map is concerned. In other words, the order map emphasizes a
geometric-structure approach to packing by characterizing single
configurations, regardless of how they were generated or their occurrence probability.
In Fig. \ref{maps}, the white and blue regions represent geometrically
possible configurations, while 
the dark shaded regions are devoid of packings
(e.g., maximally dense packings with very low order-metric values
do not exist). Clearly, an appreciably reduced region of attainable packings will be occupied
by jammed packings and, for any finite packing, 
its size must decrease as the stringency of the jamming category increases. In the infinite-size limit (not 
depicted in Fig. \ref{maps}), the regions occupied by collectively and strictly
jammed sets become  identical.
The following extremal points or loci in each jammed region are 
particularly interesting (Fig.~\ref{three}):
\vspace{-0.2in}

\begin{enumerate}
\itemsep -0.12in 
\item The locus of points $A$-$A^{\prime}$ corresponds to the lowest-density jammed
packings. We denote by $\phi_{\mbox{\scriptsize min}}$ the corresponding
{\it jamming-threshold} packing fraction. These packings are expected to be characterized
by a relatively high degree of order \cite{To07}.

\item The locus of points $B$-$B^\prime$ correspond
to the densest jammed packings with packing fraction  $\phi_{\mbox{\scriptsize max}}$.
\item The  MRJ point represents the {\it maximally random jammed}
state. Exclusion of rattlers from the MRJ state compromises
its maximal irregularity; the corresponding displaced
position in the order map involves a small reduction in packing
fraction from $\phi \approx 0.64$
and a slight increase in order measure. 

\item More generally, any point along the boundary of the blue
region is an {\it extremal} point, residing at the limit of attainability
for the jamming category under consideration.

\end{enumerate}
\vspace{-0.35in}

\subsection{Strict Jamming} 
\vspace{-0.15in}

We will first discuss the strict-jamming order map.
The densest sphere packings in three dimensions, which
lie along the locus $B$-$B^\prime$
are strictly jammed \cite{To01b,Do04a}, implying that their
shear moduli are infinitely large \cite{To03c}. 
We take point B to correspond to the fcc packing, i.e., it is 
the {\it most ordered} and {\it symmetric} densest packing. The other points
along the line $B$-$B^\prime$  represent the stacking variants of the fcc packing.
All can be conveniently viewed as stacks of planar
triangular arrays of spheres, within which each sphere contacts six neighbors.  These triangular
layers can be stacked on one another, fitting spheres of one layer into ``pockets" formed by
nearest-neighbor triangles in the layer below.  At each such layer addition there are two choices of
which set of pockets in the layer below are to be filled. Thus, all of the stacking variants
can individually be encoded by an infinite binary sequence
and therefore constitute an uncountably infinite number
of maximally dense packings called the Barlow packings \cite{Ba83}.
The most disordered subset of these is denoted by point $B^{\prime}$.
A rigorous proof  that $\phi_{\mbox{\scriptsize max}}= \pi/\sqrt{18}=0.74048\ldots$
has only recently appeared \cite{Ha05}. In two dimensions, the 
strictly jammed triangular lattice is the unique
densest packing \cite{Fe64} and so for $d=2$ the line $B$-$B^\prime$ collapses to a single point $B$.

The MRJ state is a well-defined minimum in an order map in that for a particular choice of jamming
category and order metric it can be identified unambiguously. 
The MRJ concept is automatically compromised
by passing either to the maximal packing density (fcc and its stacking variants) or the minimal
possible density for strict jamming (tunneled crystals), thereby
causing any reasonable order metric to rise on either side. This eliminates
the possibility of a flat horizontal portion of the lower boundary
of the jammed accessible region in the $\phi-\psi$ plane in Fig.~\ref{maps}
(multiple MRJ states with different densities) and therefore indicates the
uniqueness of the MRJ state in density for a particular order metric.
Indeed, at least for collective
and strict jamming in three dimensions, a variety of
sensible order metrics produce an MRJ state with a packing
fraction approximately equal to 0.64 \cite{Ka02d} (see Fig. \ref{MRJ}), 
close to the traditionally  advocated density of the RCP state, and with
an isostatic mean contact number $Z=6$. This consistency among
the different order metrics speaks to the utility of the
order-metric concept, even if a perfect order metric has not yet been identified.
However, the packing fraction of the MRJ state should not be confused
with the MRJ state itself. It is possible to have a rather
ordered strictly jammed packing at this very same density \cite{Ka02d}, as indicated in 
Fig. \ref{maps}; for example, a
jammed but vacancy-diluted fcc lattice packing. This is one reason
why the two-parameter order map description of packings is not only
useful, but  necessary. In other words, density
alone is far from sufficient in characterizing a jammed packing.

The packings corresponding to the locus of points $A$-$A^{\prime}$ have received little attention
until recently.
Although it has not yet been rigorously established as such, a candidate for the lower limiting packing fraction  
$\phi_{\mbox{\scriptsize min}}$ for strictly jammed packings is the subset of ``tunneled crystals" that contain linear arrays of vacancies \cite{To07}.  
These relatively sparse structures are generated by stacking planar ``honeycomb'' layers one upon another, and they all amount to removal of one-third of the spheres from 
the maximally dense structures with packing fraction $\phi_{\mbox{\scriptsize max}}$.  
Consequently, $\phi_{\mbox{\scriptsize min}}=2\phi_{\mbox{\scriptsize max}}/3=
0.49365\ldots$.  Every sphere in a tunneled crystal 
contacts 7 immediate neighbors in one of two possible coordination geometries, and all of the stacking variants exhibit  some form of long-range order. 
It is appropriate to view the two families 
of maximum-density and of minimum-density strictly jammed packings as structural siblings of one another.  
Note that jammed 
packings can trivially be created whose densities span the entire range between these
extremal cases   simply by filling an 
arbitrary fraction of the vacant sites in any one of the tunneled structures.  
The dashed lines joining the points $A$ to $B$ and points $A^\prime$ to $B^{\prime}$  shown in Fig. \ref{maps} are 
the result of sequentially filling the most ordered and disordered tunneled crystals
with spheres until the filling process
ends with the most ordered and disordered densest Barlow packings, respectively. 
Interestingly, the tunneled crystals exist at the edge of mechanical stability, since removal of 
any one sphere from the interior would cause the entire packing to collapse.  
It is noteworthy that \onlinecite{Bu08} have shown that an infinite
subclass of the tunneled crystals has an underlying topology that greatly
simplifies the determination of their magnetic phase structure for
nearest-neighbor antiferromagnetic interactions and $O(N)$ spins.

It should come as no surprise that ensemble methods that produce ``most probable"
configurations typically miss interesting extremal points in the order map, such
as the locus of points $A$-$A^{\prime}$ and the rest of the
jamming-region boundary, including remarkably enough
the line $B$-$B^\prime$. However, numerical protocols
can be devised to yield unusual extremal jammed states,
as discussed in Sec. \ref{protocol}, for example.

Observe that irregular jammed packings can be created
in the entire 
non-trivial range of packing fraction $0.64 < \phi < 0.74048\ldots$ \cite{To00b,Ka02d}
using the LS algorithm. Thus, in the rightmost plot  in Fig. \ref{maps},
the MRJ-$B^\prime$ portion of the boundary of the jammed set, possessing
the lowest order metric, is demonstrably achievable. Until recently, no algorithms
have produced disordered strictly jammed packings to the left of the MRJ point.
A new algorithm described elsewhere \cite{To10c} has indeed 
yielded such packings with $\phi \approx 0.60$, which
are overconstrained with $Z \approx 6.4$, implying that
they are more ordered than the MRJ state (see 
Sec. \ref{protocol} for additional details). The existence of disordered strictly jammed packings
with such anomalously low densities expands conventional
thinking about the nature and diversity of disordered packings
and places in a broader context those protocols that 
produce ``typical" configurations.

Indeed, there is no fundamental reason
why the entire lower  boundary of the jammed set between
the low-density jamming threshold and MRJ point cannot also be realized.
Note that such low-density disordered packings
are not so-called ``random loose" packings, which are even less well-defined
than RCP states. For example, it is not clear that
the former are even collectively jammed. A necessary first step
would be to classify the jamming category of a random loose
packing (RLP), which has yet to be done.  Therefore, in our view,
the current tendency in the literature to put so-called RCP 
and RLP on the same footing as far as jamming is concerned \cite{Ma08}
is premature at best.

In $\mathbb{R}^2$, the so-called ``reinforced" Kagom{\' e}
packing with precisely 4 contacts per particle (in the infinite-packing limit) is evidently 
the lowest density strictly jammed subpacking of the triangular lattice
packing \cite{Do04a} with $\phi_{\mbox{\scriptsize min}} = \sqrt{3} \pi/8 = .68017\ldots$. 
Note that this packing has the isostatic contact number $Z=4$  and yet
is an ordered packing, which runs counter to the prevalent
notion that isostaticity is a consequence of ``genericity"
or randomness \cite{Mouk98}.

\subsection{Collective and Local Jamming}

Observe that the locus of points $B$-$B^{\prime}$ is invariant under
change of the jamming category, as shown in Fig. \ref{maps}. This is not true of the MRJ
state, which will generally have a different location in
the local-jamming and collective-jamming order maps.
Another important distinction  is that it is possible to pack spheres
subject only to the weak locally-jammed criterion, so that the resulting packing fraction
is arbitrarily close to zero \cite{Bor64,St03}.  But demanding either collective jamming or
strict jamming evidently forces $\phi$ to equal or exceed a lower 
limit $\phi_{\mbox{\scriptsize min}}$ that is well above zero.

\subsection{Broader Applications to Other Condensed States of Matter}
   
Although methods for characterizing structural order in regular crystalline
solids are well established \cite{As76,Chaik95}, similar techniques for
noncrystalline condensed states of matter are not nearly as advanced.
The notions of order metrics and order maps have been fruitfully extended
to characterize the degree of structural order in condensed phases of matter
in which the constituent molecules (jammed or not) possess both attractive and repulsive interactions. This includes the determination of the order maps of models of simple liquids, glasses and crystals with isotropic interactions \cite{Tr00,Er03},  
models of water \cite{Er01,Er02}, and models of amorphous polymers \cite{Stach03}.

\section{Protocol Bias, Loss of Ergodicity, and Nonuniqueness of Jammed States}     
\label{protocol}

     A dilute system of $N$ disks or spheres is free to reconfigure largely 
independently of particle-pair  nonoverlap constraints.  However, as $\phi$  
increases either as a result of compression or of particle size growth, 
those constraints consume larger and larger portions of the $dN$-dimensional  
configuration space, making reconfiguring more and more difficult. 
Indeed, the available subspace begins to fracture, producing isolated 
``islands" that each eventually  collapse with increasing $\phi$ into jammed states 
\cite{Sa62}.  This fracturing, or disconnection, implies dynamical {\it non-ergodicity}.  
Owing to permutation possibilities for $N$ identical objects, each disconnected region belongs to a large family 
of essentially $N!$  equivalent regions.  But the configuration space fracturing has even greater complexity in that 
the number of inequivalent such families and their jamming limits rises exponentially with $N$ in the large-system 
asymptotic limit.  The jamming-limit $\phi$  values for the families vary over the ranges indicated in Fig. \ref{maps} 
for collective and strict jamming.

\begin{figure}[bthp]
\centerline{\includegraphics[height=2.4in,keepaspectratio,clip=]{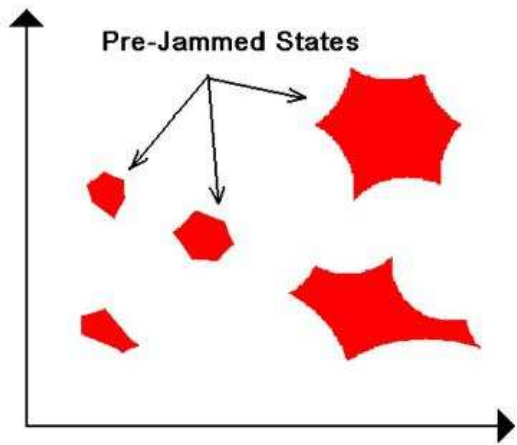}}
\caption{\footnotesize A schematic of the disjoint set of nearly jammed packings that develop in multidimensional configuration space as covering fraction $\phi$ increases. 
The two axes represent the collection of configurational coordinates.
Note that each individual region approaches a polytope in the jamming
limit, as discussed in Sec. \ref{jam-poly}.}
\label{config}
\end{figure}

     Figure \ref{config} offers a simple schematic cartoon to illustrate this $dN$-dimensional disconnection feature.  
Several allowed regions with different sizes and shapes are shown.  Their boundaries consist of sets of  slightly curved hypersurfaces, each of which corresponds to a particle pair contact, or contact with a hard wall if present.  Particle growth or system compression causes hypersurfaces (numbering at least $dN+1$ for hard
walls) to move inward, reducing region content toward zero.  
The larger the region shown, the larger should be understood its jamming $\phi$  value.
The basic issue involved in either laboratory or computer experiments  is how and why the 
various jamming protocols used populate the disconnected regions.  Presumably any given algorithm has associated with 
it a characteristic set of occupation weights, leading in turn to well-defined averages for any property of interest, 
including packing fraction $\phi$  and any chosen order metric  $\psi$.  
The fact that these averages indeed vary with algorithm is a major point of the present review.

Ensemble methods have been invoked to attach special significance to so-called
``typical" or ``unique" packings because of their frequency of
occurrence in the specific method employed. In particular, significance has been attached to the so-called
    unique J (jammed) point, which is suggested to correspond
to the onset of collective jamming in soft sphere systems \cite{Oh03}. The order maps described in Sec. \ref{order} as well as the ensuing discussion
demonstrate that claims of such uniqueness overlook the wide variability
of packing algorithms and the distribution of configurations that they generate. Individual
 packing protocols (numerical or experimental) produce jammed
    packings that are strongly concentrated  in isolated pockets of configuration space
that are individually selected by those protocols. Therefore,
    conclusions drawn from any particular protocol are highly specific
    rather than general in our view.

\begin{figure}[bthp]
\centerline{
\includegraphics[height=2.3in,keepaspectratio,clip=]{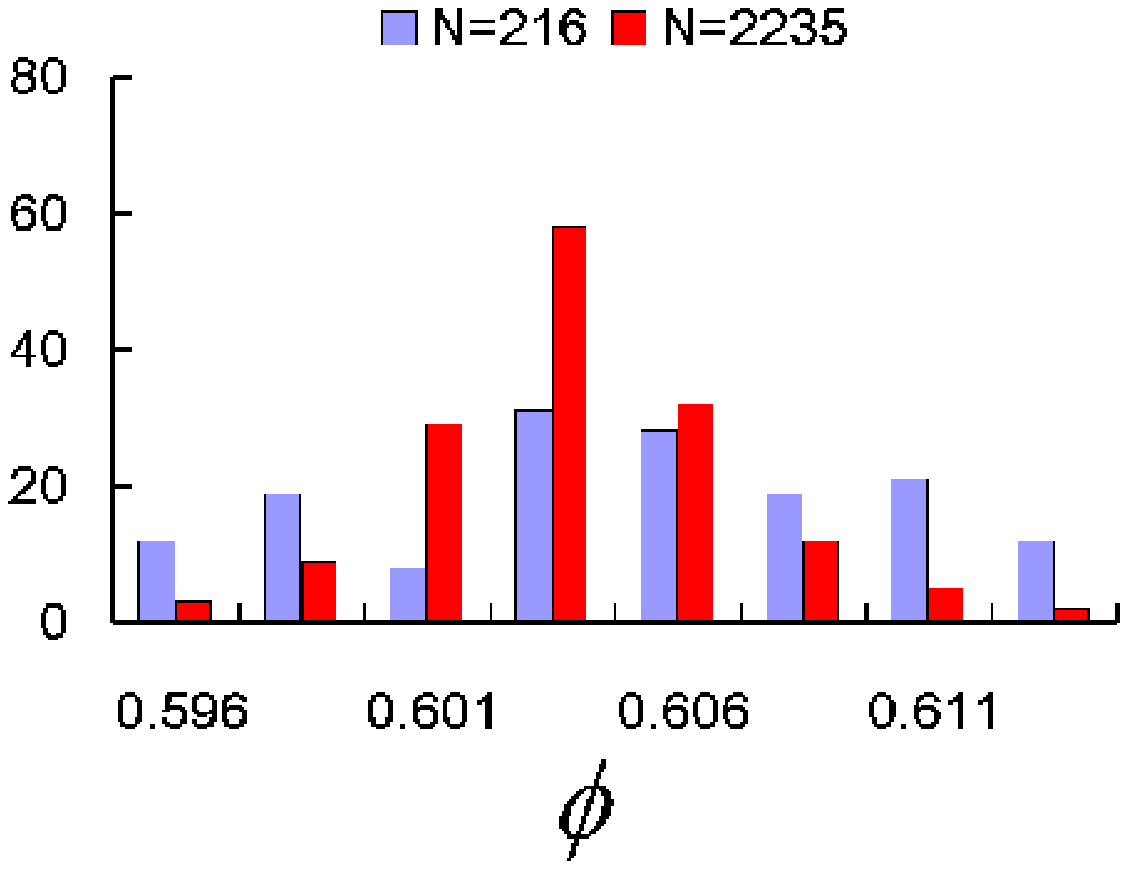}
\hspace{0.2in}\includegraphics[height=2.3in,keepaspectratio,clip=]{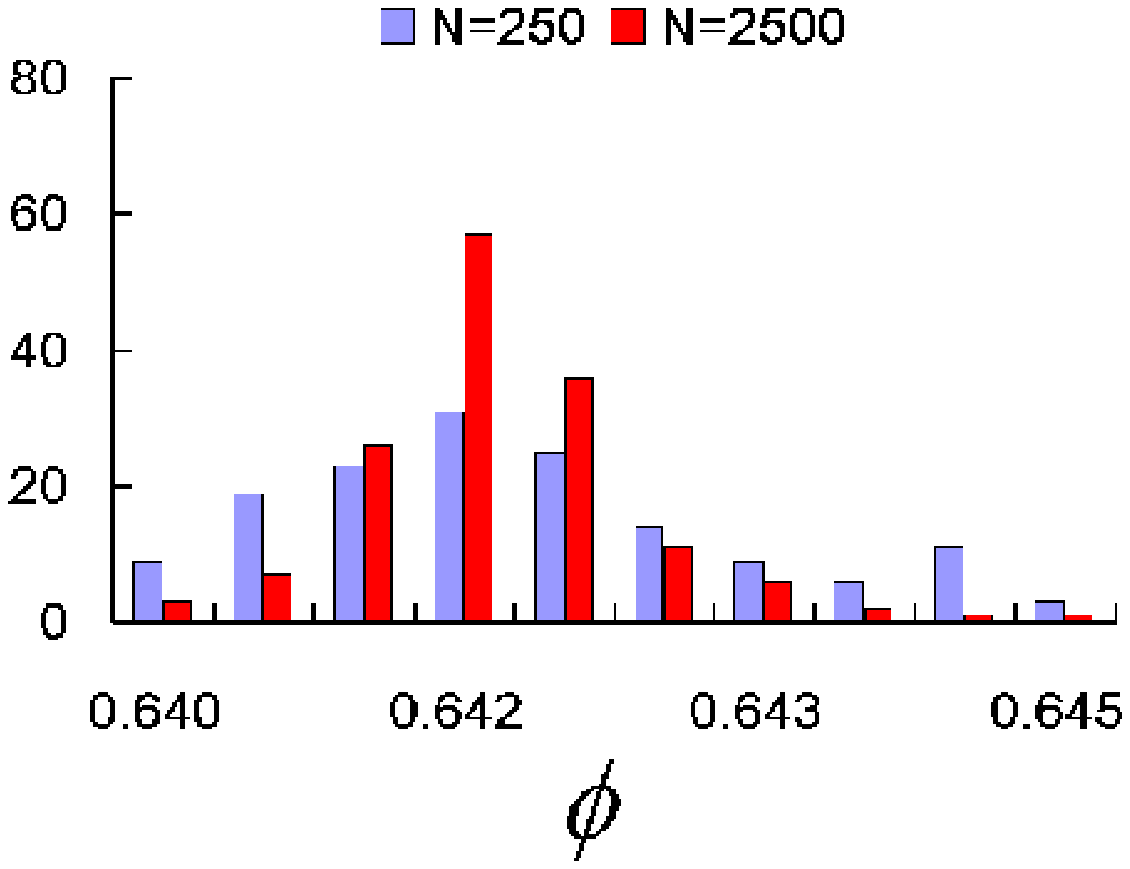}}
\centerline{
\includegraphics[height=2.3in,keepaspectratio,clip=]{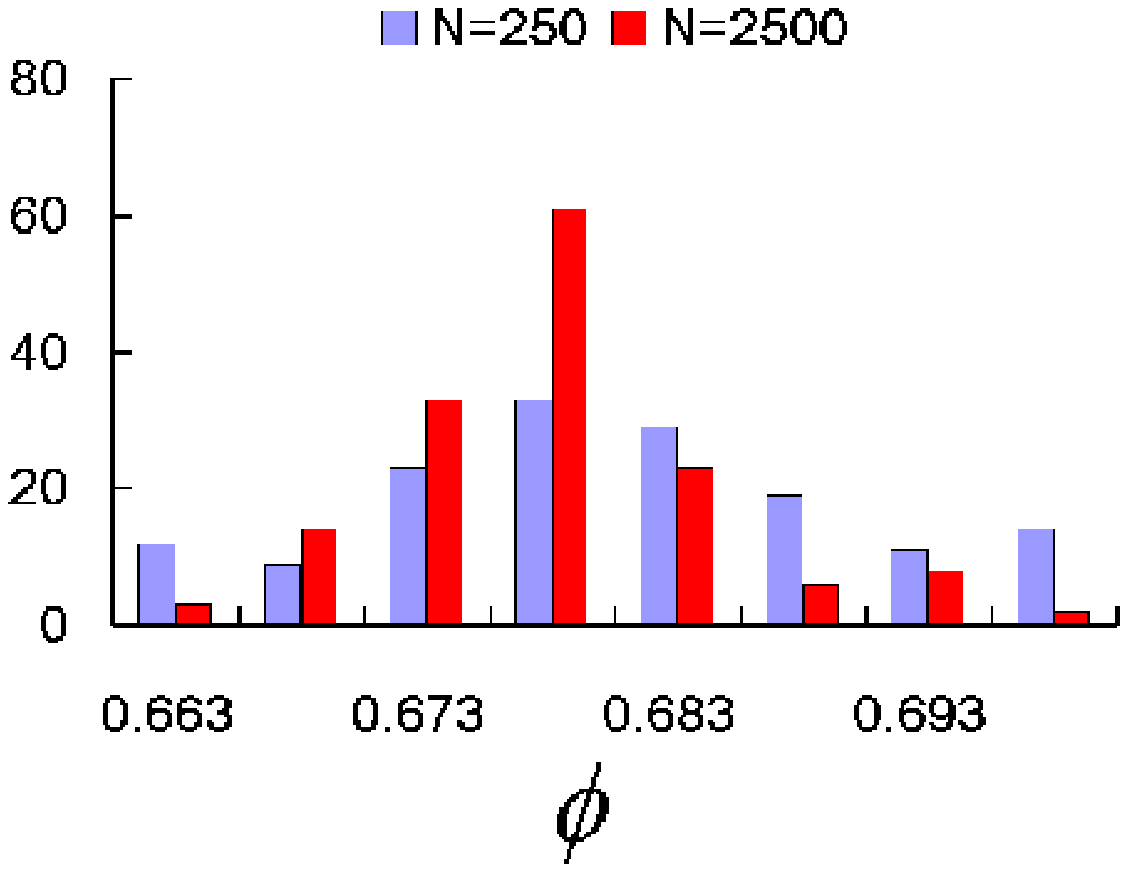}
\hspace{0.2in}
\includegraphics[height=2.3in,keepaspectratio,clip=]{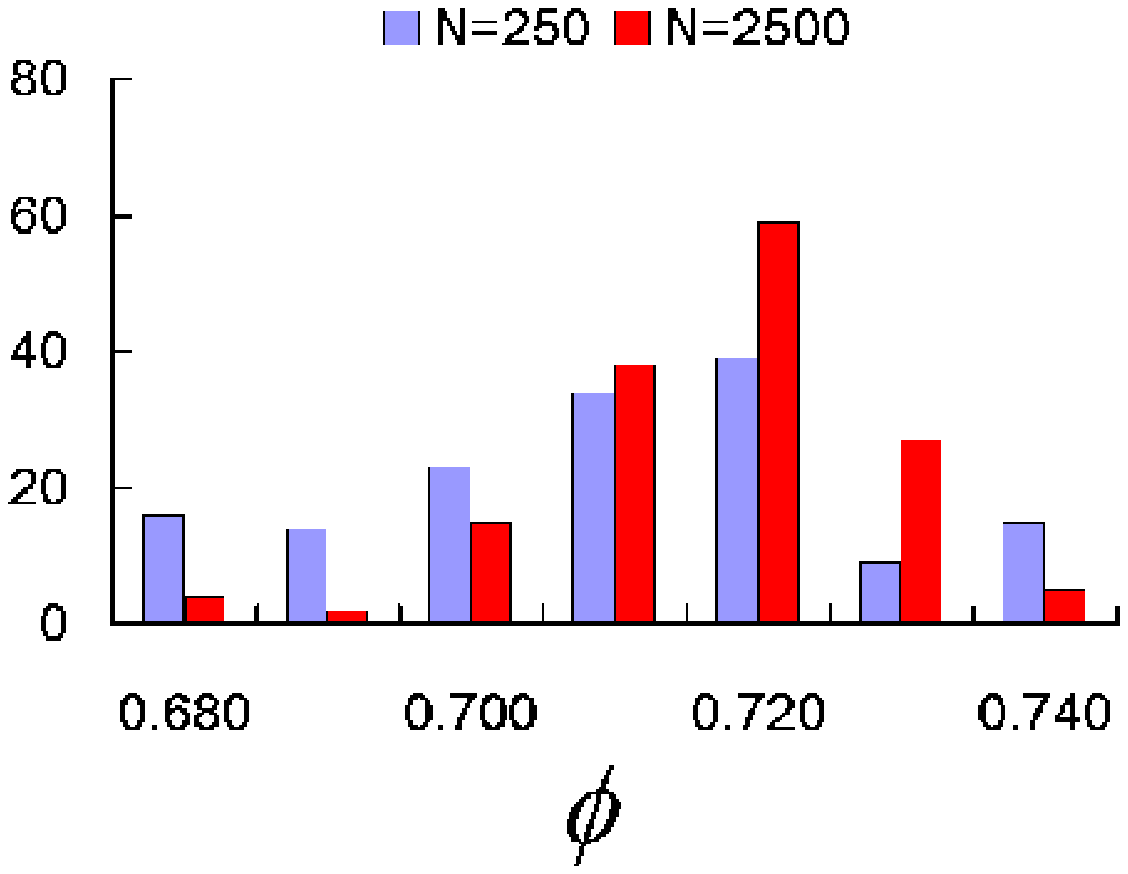}}
\caption{\footnotesize Packing protocols can be devised that lead to strictly jammed states
at any specific density with a high probability of occurrence
anywhere over a wide density range. Shown are  histograms of jammed
packings that are centered around four different packing fractions:
$\phi \approx 0.60, 0.64, 0.68$ and 0.72, as obtained by \onlinecite{Ji10c}.
 The distributions
become narrower as the system size increases.}
\label{histo}
\end{figure}

Indeed, one can create protocols that can
lead to jammed packings at any preselected density with a high probability of occurrence
anywhere over a wide density range.
Unless it were chosen to be highly restrictive, a typical disk or sphere jamming algorithm applied to a 
large number $N$ of particles would be capable of producing a large number of geometrically distinguishable results.  
In particular, these distinguishable jammed configurations from a given algorithm would show some dispersion in their  
$\phi$ and  $\psi$ values.  However, upon comparing the distributions of obtained results for a substantial range of 
particle numbers $N$ (with fixed boundary conditions), one must expect a narrowing of those distributions with increasing 
$N$ owing to operation of a central limit theorem.  Indeed, this narrowing would converge individually onto values 
that are algorithm-specific, i.e., different from one another.  Figure \ref{histo} provides a clear illustration of 
such narrowing with respect to $\phi$  distributions, with evident variation over algorithms, as obtained by \onlinecite{Ji10c}  The examples shown contrast results for two distinctly different sphere system sizes ($\sim 250$ and $\sim 2500$ particles), and for two different algorithms that have results for disordered jammed packings 
converging, respectively,  onto packing fractions of about 0.60, 0.64, 0.68 and 0.72 
The histogram for the lowest density was produced
using the new algorithm \cite{To10c} noted in Section \ref{map}, while the other two histograms were
generated using the LS algorithm.
We stress once again that any temptation to select 
a specific $\phi$  value as uniquely significant (e.g., 0.64) is primarily based on inadequate sampling of the full range of algorithmic richness and diversity that is available at least in the underlying mathematical  theory of sphere jamming.

\section{Attributes of the Maximally Random Jammed State}
\label{mrj}

\begin{figure}
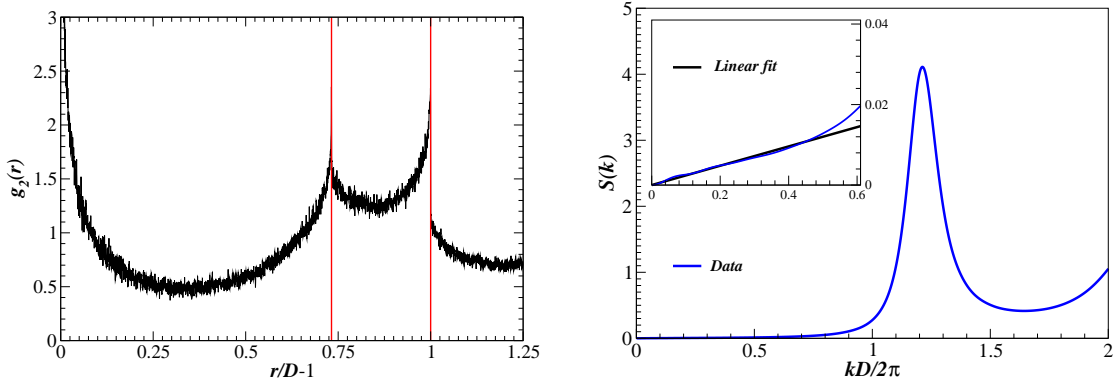

\begin{center}
{\includegraphics[width=2.8in,keepaspectratio,clip=]{fig17a.eps}\hspace{0.25in}
\includegraphics[width=2.7in,keepaspectratio,clip=]{fig17b.eps}}
\end{center}
\caption{\footnotesize Pair statistics for packings in the immediate neighborhood of the 
three-dimensional
MRJ state \cite{Ka02d} with $\phi \approx 0.64$. Left panel: Computational data on 
the pair correlation function  $g_{2}(r)$ versus $r/D-1$ averaged 
over $5$ packings of $10,000$ spheres \cite{Do05c} of diameter $D$.
It is normalized so that at large distances it tends to unity, indicating
no long-range order. The split second peak, the discontinuity at twice 
the sphere diameter, and the divergence near contact
are clearly visible. Right panel: The corresponding structure factor $S(k)$ as a function
of the dimensionless wavenumber $kD/(2\pi)$ for a million particle packing \cite{Do05d}. The inset shows the linear ( $|\bf k|$) nonanalytic behavior at $k=0$.}
\label{g2-S}
\end{figure}

The MRJ state under the strict-jamming constraint is a prototypical glass 
\cite{To07} in that it is maximally disordered without any long-range order 
and perfectly rigid (the elastic moduli are indeed unbounded \cite{To03c}). This endows such packings, which are isostatic
($Z=2d$),
    with special attributes \cite{Do05c,Do05d}. For example, 
the pair correlation function $g_2(r)$ (which provides
the distribution of pair distances) of three-dimensional MRJ packings
possesses a split
second peak \cite{Za83}, with a prominent discontinuity at twice the sphere diameter, 
as shown in the left panel of Fig. \ref{g2-S},
which is a well-known characteristic of disordered jammed packings.
The values $r=\sqrt{3}D$ and $r=2D$ are highlighted in the figure, and match the
two observed singularities. Interestingly, an integrable power-law divergence 
($1/(r/D-1)^{\alpha}$ with $\alpha \approx 0.4$) exists for near contacts \cite{Do05c}. No peaks
are observed at $r=\sqrt{2}D$ or $r=\sqrt{5}D$, which are typical
of crystal packings, indicating that there is no detectable undistorted crystal
ordering in the packing. We note that in a computational
study of stiff ``soft" spheres \cite{Si02}, a nearly square-root divergence
for near contacts was found.

The MRJ state possesses unusual spatial density fluctuations. It was conjectured \cite{To03a} 
that all strictly jammed {\it saturated} packings
of congruent spheres (disordered or not) are {\it hyperuniform},  i.e., 
infinite-wavelength density fluctuations vanish or, equivalently, the structure
factor $S(k)$ vanishes in the limit $k \rightarrow  0$. (Recall
that a  saturated packing is one in which no space exists to insert additional particles.)  Disordered hyperuniform point distributions are uncommon.
Not only was this conjecture verified numerically 
for an MRJ-like state using a million-particle packing 
of monodisperse spheres, 
but it was shown that the structure factor has an unusual nonanalytic linear dependence 
near the origin \cite{Do05d}, namely, $S(k) \sim |k|$ for $k \rightarrow 0$,
or equivalently, a quasi-long-ranged tail of the total pair
correlation function $h(r) \sim - r^{-4}$. This same linear nonanalytic
behavior of $S(k)$ near the origin is also found in such diverse
three-dimensional systems as the early Universe \cite{Pe93}, 
ground state of liquid helium \cite{Fe56,Re67} and nonintercting spin-polaried
 fermionic ground states \cite{To08c,Sc09}.
The generalization of the aforementioned conjecture 
that describes when strictly jammed saturated packings 
of noncongruent sphere packings as well as other particle
shapes has been given by \onlinecite{Za10a}.
Specifically, the void spaces of general MRJ packings
are highly constrained by the underlying contact network,
which induce hyperuniformity and quasi-long-range behavior of the
two-point probability function for the \emph{void phase}.

The quasi-long-range behavior of $g_2(r)$  as well
as the aforementioned  pair-correlation features 
distinguish the MRJ state strongly from that of the
equilibrium hard-sphere fluid \cite{Ha86}, which is characterized
by a structure factor that is analytic at $k=0$ and thus
has a pair correlation function that decays exponentially fast to unity
for large $r$. Consequently, early attempts \cite{Be60} to
use disordered jammed packings to model liquid structure were imprecise.

It should be recognized that MRJ-like  sphere packings created in
practice via computer algorithms \cite{To00b,Oh02,Do05d} or actual experiments may
contain a small concentration of rattlers, the average
concentration of which is protocol-dependent.
The packings leading to the data depicted in Fig. \ref{g2-S} contain
between 2-3 percent rattlers.  Thus, the hyperuniformity
property of the MRJ state requires that the rattlers
be retained in the packing.

It is well known that lack of ``frustration" \cite{Ju97,To02a} in two-dimensional
analogs of three-dimensional computational and experimental
protocols that lead to putative RCP states result in packings of identical disks
that are highly crystalline, forming rather large triangular
coordination domains (grains). Such a 1000-particle  packing with $\phi \approx 0.88$ is depicted 
 in the right panel of Fig. \ref{MRJ}, and is only collectively
jammed at this high density.  Because such highly ordered
packings are the most probable outcomes for these typical protocols, 
``entropic measures" of disorder would identify these as the most disordered,
a misleading conclusion.  An appropriate order metric, on
the other hand, is capable of identifying a particular configuration
(not an ensemble of configurations) of considerably
lower density (e.g., a jammed vacancy-diluted triangular lattice
or its multi-domain variant) that
is consistent with our intuitive notions of maximal disorder.
However, typical packing protocols would almost never generate
such disordered disk configurations because of their
inherent implicit bias toward undiluted crystallization.
Note that the same problems persist even for 
polydisperse disk packings provided that 
the size distribution is sufficiently narrow.

Importantly, previously reported  low packing
fractions of  0.82 -- 0.84 for so-called RCP disk arrangements \cite{Be83} 
were found not even to be collectively jammed \cite{Do04a}.
This conclusion clearly demonstrates that the distinctions between
the different jamming categories are crucial. Moreover, the
geometric-structure approach to jamming reveals the basic
importance of collective motions potentially involving an arbitrarily large
number of particles. Therefore, methods
that assume collective jamming based only on  packing fraction and
local criteria, such as nearest-neighbor coordination
and Voronoi statistics \cite{Ma08}, are incomplete.

\section{Packings of Spheres With a Size Distribution}
\label{poly}

Polydispersity in the size of the particles
constitutes a fundamental feature of the microstructure
of a wide class of dispersions of technological importance,
including those involved in composite solid propellant combustion
\cite{Ke87}, sintering of powders\index{sintered materials} \cite{Ra95},
colloids \cite{Ru89}, transport and mechanical
properties of particulate\index{particulate media}
 composite materials \cite{Ch79}, and flow in packed beds\index{packed beds} \cite{Sc74}.

The spheres generally possess a distribution in radius $R$
characterized by a probability density $f(R)$ that normalizes to unity, i.e.,
\begin{equation}
\int_0^\infty f(R) dR =1.
\end{equation}
The average of any function $w(R)$ is defined by
\begin{equation}
\langle {w(R)} \rangle = \int_{0}^{\infty} w(R) f(R) dR.
\end{equation}
The overall packing fraction $\phi$ of the system is defined as
\begin{equation}
\phi=\rho \langle v_1(R) \rangle,
\label{eta-poly}
\end{equation}
where $\rho$ is the total number density, $v_1(R)$ is given by (\ref{vol-sph}) and $\langle v_1(R) \rangle$ is
the average sphere volume defined by
\begin{equation}
\langle v_1(R) \rangle=\frac{\pi^{d/2}}{\Gamma(1+d/2)}\; \langle R^d
\rangle.
\end{equation}

\begin{figure}
\centerline{\psfig{file=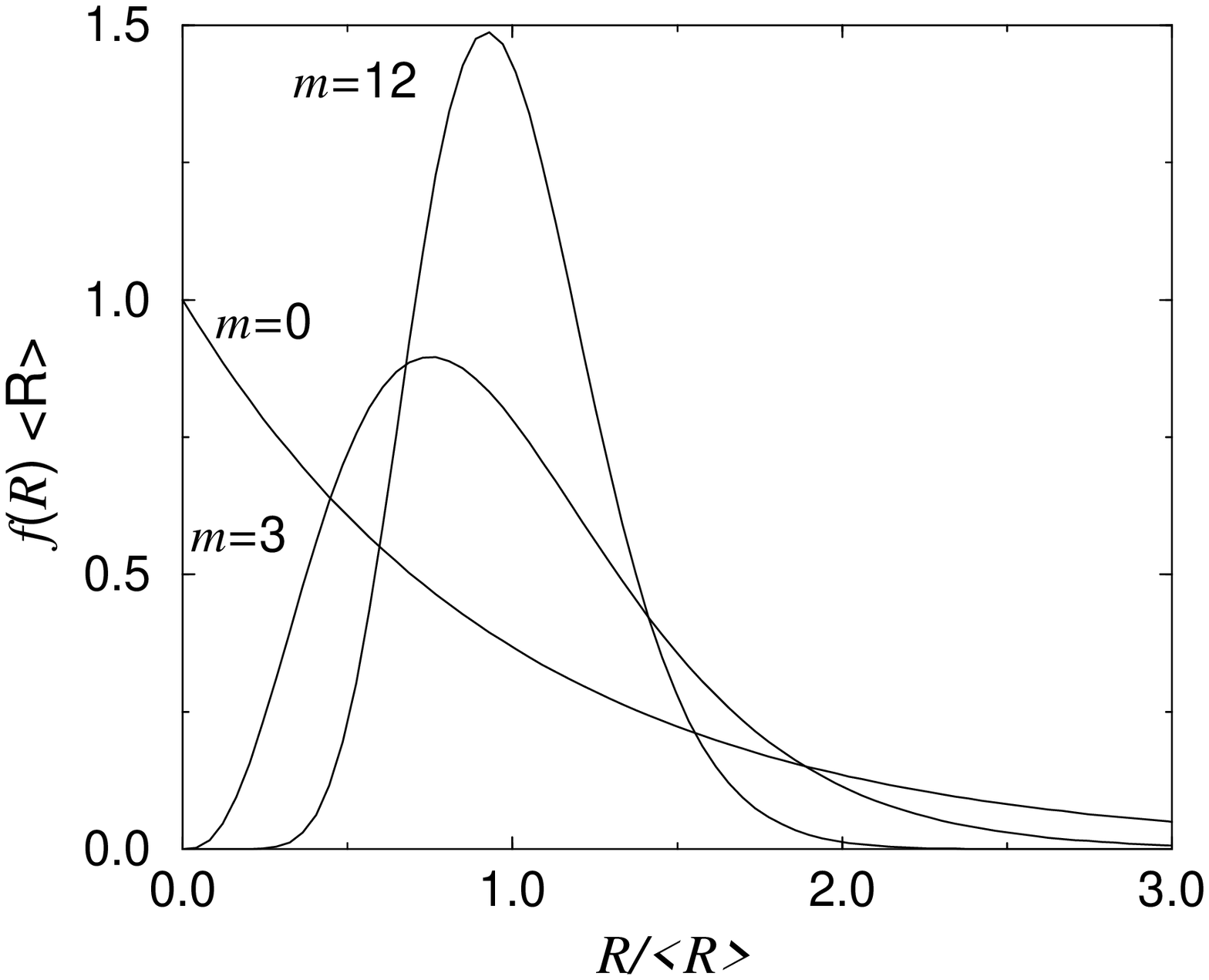,height=2.0in}
\hspace{.15in}\psfig{file=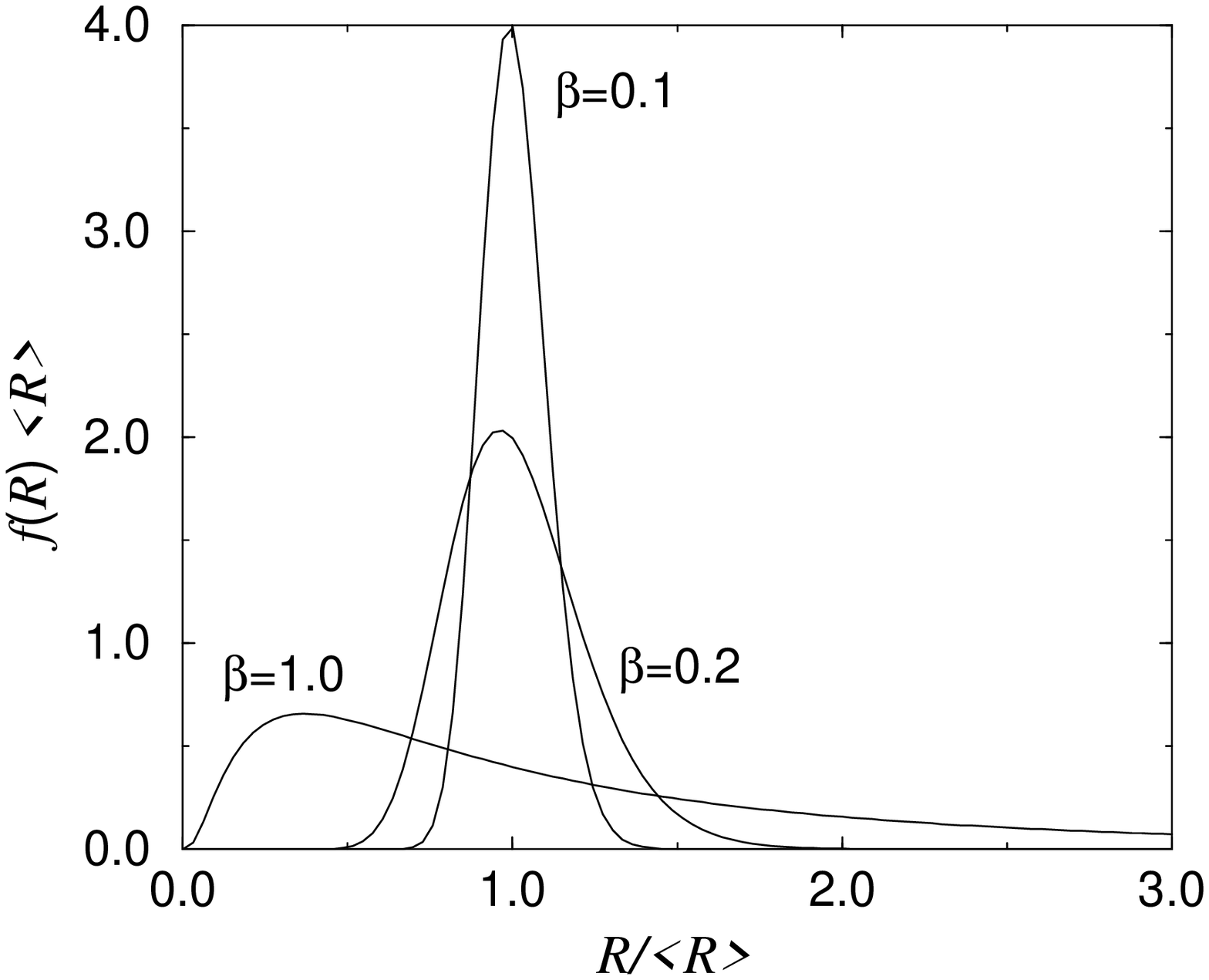,height=2.0in}}
\caption{ Frequently used hard-sphere size distributions. Left panel:
Schulz size distribution for
several values of $m$. Right panel: Log-normal size distribution
for several values of $\beta$. }
\label{size-dist}
\end{figure}

There is a variety of choices for the size distribution
$f(R)$ that deserve consideration. Two continuous probability densities
that have been widely used to characterize physical
phenomena are the Schulz \cite{Sc39} and log-normal \cite{Cr54}
 distributions.  The Schulz  distribution is defined as
\begin{equation}
f(R) = \frac{1}{\Gamma(m+1)} \left(
\frac{m+1}{\langle {R} \rangle} \right)^{m+1} R^{m} \exp \left[
\frac{ - (m + 1)R}{\langle {R}\rangle} \right]. \label{Schulz}
\end{equation}
where $\Gamma(x)$ is the gamma function. When the parameter $m$
is restricted to nonnegative integer values, 
$\Gamma(m+1)=m!$, and the $n$th moment of this distribution is given by
\begin{equation}
\langle{R^{n}\rangle} =  \frac{(m+n)!}{m!}\frac{1}{(m + 1)^{n}}
 \; \langle {R} \rangle^{n} .
\end{equation}
By increasing $m$, the variance decreases, i.e.,
the distribution becomes sharper.  In the monodisperse
limit $m \rightarrow \infty$, $f(R) \rightarrow \delta (R -
\langle{R}\rangle)$. The case $m=0$ gives an exponential
distribution in which many particles have extremely small
radii. By contrast, the log-normal distribution is defined as
\begin{equation}
f(R) = \frac{1}{R \sqrt{2 \pi \beta^{2}}} \exp \left\{ -
\frac{[ \ln (R/\langle R \rangle) ]^{2}}{2 \beta^{2}} \right\},
\end{equation}
where $\beta^2= \langle (\ln R)^2 \rangle -\langle \ln R \rangle^2$.
The quantity $\ln R$ has a normal or Gaussian distribution.
The $n$th moment is given by
\begin{equation}
\langle R^{n} \rangle = \exp (n^{2} \beta^{2}/2)\; \langle R\rangle^{n} .
\end{equation}
As $\beta^{2} \rightarrow 0$, $f(R) \rightarrow \delta (R -
\langle R\rangle)$. Figure \ref{size-dist} shows examples of the Schulz
and log-normal size distributions.

One can obtain corresponding results for  spheres
with $M$ discrete different sizes from the continuous case by letting
\begin{equation}
f(R) = \sum_{i=1}^{M} \frac{\rho_{i}}{\rho} \delta (R -
R_{i}), 
\label{f-discrete}
\end{equation}
where $\rho_{i}$ and $R_{i}$ are number density and radius
of type-$i$ particles, respectively, and $\rho$ is the {\it total number
density}. Therefore, the overall volume fraction using (\ref{eta-poly})
is given by
\begin{equation}
\phi= \sum_{i=1}^M \phi^{(i)}
\end{equation}
where
\begin{equation}
\phi^{(i)}=\rho_i v_1(R_i)
\end{equation}
is the packing fraction of the $i$th component.

Sphere packings with a size distribution   exhibit
intriguing structural features, some
of which are only beginning to be understood. 
It is known, for example,
that a relatively small degree of polydispersity
can  suppress the disorder-order phase
transition seen in monodisperse hard-sphere systems
\cite{Hen96}. Interestingly, equilibrium mixtures of
small and large hard spheres can ``phase separate'' (i.e.,
the small and large spheres demix) at sufficiently
high densities but the precise nature of
such phase transitions  has not yet been established
and is a subject of intense interest; see \cite{Di99} and references
therein.

Our main interest here is in dense polydisperse packings of spheres,
especially jammed ones. 
Very little is rigorously known about the characteristics
of such systems. For example, the maximal overall packing fraction
of even a binary mixture
of hard spheres in $\mathbb{R}^d$, which we denote by $\phi^{(2)}_{\scriptsize \mbox{max}}$, for arbitrary values of the mole fractions and radii $R_1$ and
$R_2$ is unknown, not to mention the determination of the corresponding
structures. However, one can bound $\phi^{(2)}_{\scriptsize \mbox{max}}$ from above and 
below in terms of the maximal packing fraction $\phi^{(1)}_{\scriptsize \mbox{max}}$ for a 
monodisperse sphere packing in the infinite-volume limit
 using the following analysis of \cite{To02a}. It is
clear that $\phi^{(2)}_{\scriptsize \mbox{max}}$
is bounded from below by the maximum packing fraction
$\phi^{(1)}_{\scriptsize \mbox{max}}$. The lower bound $\phi^{(2)}_{\scriptsize \mbox{max}}
\ge \phi^{(1)}_{\scriptsize \mbox{max}}$ is independent of the radii and
corresponds to the case when the two components are completely {\it phase separated}
(demixed),
each at the packing fraction $\phi^{(1)}_{\scriptsize \mbox{max}}$.
Moreover, one can bound $\phi^{(2)}_{\scriptsize \mbox{max}}$
from above in terms of the monodisperse value $\phi^{(1)}_{\scriptsize \mbox{max}}$
for arbitrary values of $R_1$ and $R_2$. Specifically, consider a wide separation
of sizes ($R_1 \ll R_2$) and imagine a {\it sequential process} in which
the larger spheres are first packed at the maximum density
$\phi^{(1)}_{\scriptsize \mbox{max}}$ for a monodisperse packing.
The remaining interstitial space between the larger spheres can now be packed
with the smaller spheres at the packing fraction $\phi^{(1)}_{\scriptsize \mbox{max}}$
provided that $R_1/R_2 \rightarrow 0$. The overall packing fraction in
this limit is given by $1-(1-\phi^{(1)}_{\scriptsize \mbox{max}})^2$,
which is an upper bound for {\it any binary packing}.
Thus,  $\phi^{(2)}_{\scriptsize \mbox{max}} \le 1- (1-\pi/\sqrt{12})^2 
\approx 0.991$ for $d=2$ and $\phi^{(2)}_{\scriptsize \mbox{max}} \le 1- (1-\pi/\sqrt{18})^2
\approx 0.933$ for $d=3$, where $\phi^{(1)}_{\scriptsize \mbox{max}}$ corresponds to the maximal
packing fraction in two or three dimensions, respectively.

The same arguments extend to
systems of $M$ different hard spheres with radii $R_1,R_2, \ldots, R_M$ in $\mathbb{R}^d$ \cite{To02a}.
Specifically, the overall maximal packing fraction $\phi^{(M)}_{\scriptsize \mbox{max}}$ of such
a general mixture in $\mathbb{R}^d$ [where $\phi$ is defined by
(\ref{eta-poly}) with (\ref{f-discrete})] is bounded from above and below by
\begin{equation}
\phi^{(1)}_{\scriptsize \mbox{max}} \le \phi^{(M)}_{\scriptsize
\mbox{max}}
\le 1- (1-\phi^{(1)}_{\scriptsize \mbox{max}})^M.
\label{upper-eta}
\end{equation}
The lower bound corresponds to the case when the $M$ components completely
demix, each at the density $\phi^{(1)}_{\scriptsize \mbox{max}}$.
The upper bound corresponds to the generalization
of the aforementioned ideal sequential packing process for arbitrary
$M$ in which we take the limits $R_1/R_2 \rightarrow 0$, $R_2/R_3 \rightarrow 0,
\cdots, R_{M-1}/R_M \rightarrow 0$. Specific {\it nonsequential} protocols (algorithmic or otherwise)
that can generate structures that  approach the upper bound (\ref{upper-eta}) for arbitrary
values of $M$ are currently unknown and thus the development of such protocols
is an open area of research. We see that in the limit $M\rightarrow \infty$, the
upper bound approaches unity, corresponding to space-filling polydisperse spheres with
an infinitely wide separation in sizes \cite{He90}. 
Furthermore, one can also 
imagine constructing space-filling
polydisperse spheres with a continuous size distribution with sizes ranging to the infinitesimally small \cite{To02a}.

\begin{figure}
\centerline{\psfig{file=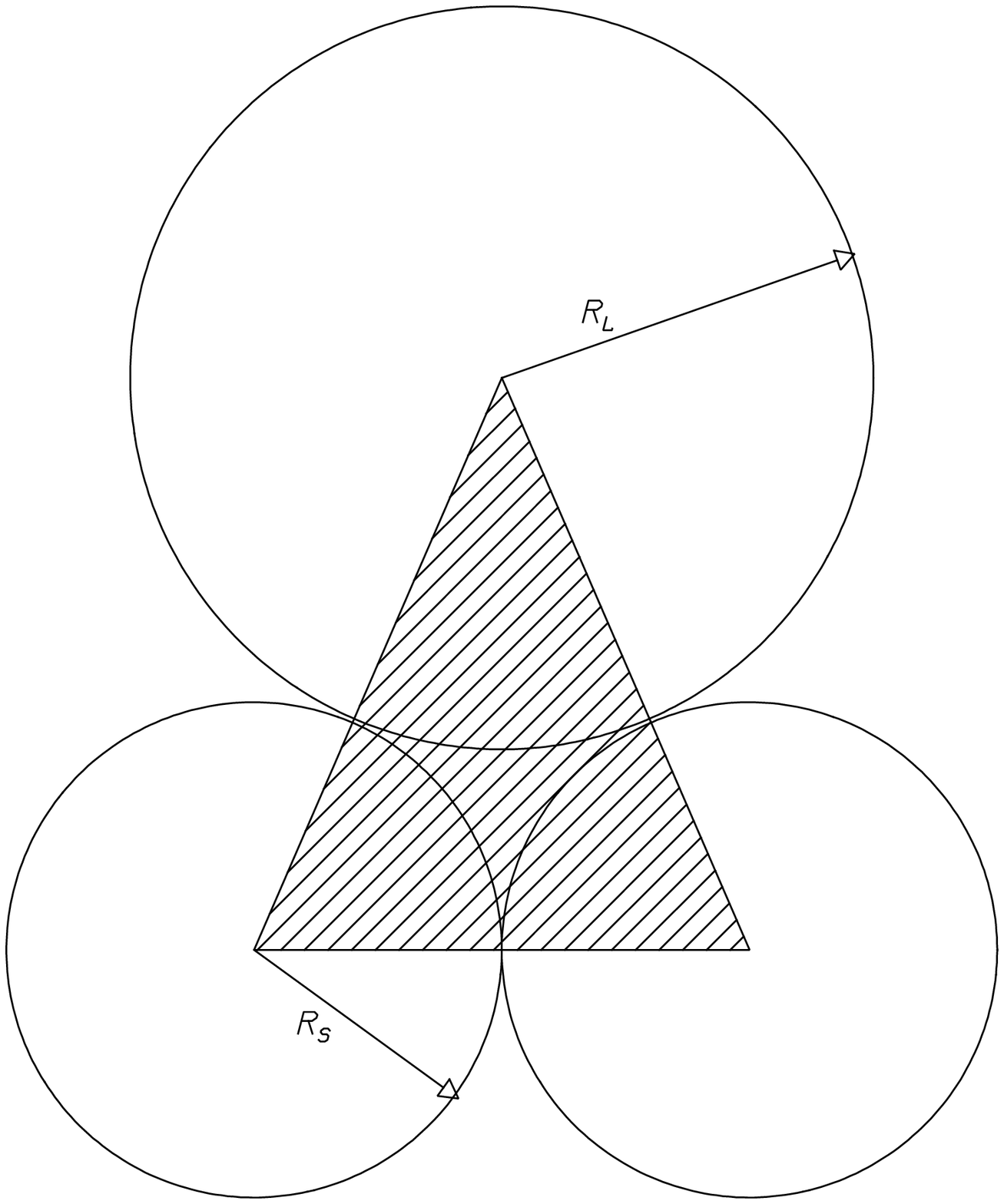,height=3.50in}}
\caption{ One large disk and two small disks in mutual contact
provide the densest local arrangement of binary disks \cite{Fl60}.
The intersection of the shaded triangle with the three disks
yields the local packing fraction $\phi_U=\frac{\displaystyle \pi \alpha^2 + 2(1-\alpha^2)\text{arc}\sin{\left(\frac{\alpha}{1+\alpha}\right)}}
{\displaystyle 2\alpha (1+2\alpha)^{1/2}}$, where  $\alpha = R_{S}/R_{L}$. }
\label{binary}
\end{figure}

Jammed binary packings have received some attention but their
characterization is far from complete. Here we briefly note
work concerned with maximally dense binary packings as well as 
disordered jammed binary packings in two and three dimensions.
Among these cases, we know most about the determination of the maximally dense
binary packings in $\mathbb{R}^2$. Let $R_{S}$ and $R_{L}$
denote the radii of the small and large disks ($R_{S} \leq  R_{L}$), the radii ratio
 $\alpha = R_{S}/R_{L}$ and $x_{S}$ be the number fraction of small disks
in the entire packing. Ideally, it is desired to obtain 
$\phi_{\mbox{\scriptsize max}}$ as a function of $\alpha$ and $x_S$.
In practice, we have a sketchy understanding of the surface
defined by $\phi_{\mbox{\scriptsize max}}(\alpha,x_S)$.
\onlinecite{Fe64} has reported a number of candidate maximally dense packing arrangements
for certain values of the radii ratio in the range $\alpha \geq 0.154701 \ldots$.
Maximally dense binary disk packings have been also investigated to determine the stable
crystal phase diagram of such alloys \cite{Li93}. 
The determination of $\phi_{\mbox{\scriptsize max}}$ for sufficiently small $\alpha$ amounts to finding the
optimal arrangement of the small disks within a {\it tricusp}: the nonconvex cavity between three
close-packed large disks. A particle-growth Monte Carlo algorithm was used to generate
the densest arrangements of small identical  disks (ranging
in number from one through 19)  within such a tricusp  \cite{Uc04a}.
All of these results can be compared to a relatively sharp
upper bound on $\phi_{\mbox{\scriptsize max}}$ given by
\begin{equation}
\phi_{\mbox{\scriptsize max}} \le   \phi_U = \frac{\displaystyle \pi \alpha^2 + 2(1-\alpha^2)\text{arc}\sin{\left(\frac{\alpha}{1+\alpha}\right)}}
{\displaystyle 2\alpha (1+2\alpha)^{1/2}}.
\label{supremum}
\end{equation}
The fraction $\phi_U$ corresponds to the densest local packing arrangement
for three binary disks shown in Fig. \ref{binary} and hence
bounds $\phi_{\mbox{\scriptsize max}}$ from above ~\cite{Fl60}. Inequality 
(\ref{supremum}) also applies to general 
multicomponent packings, where  $\alpha$ is taken to be the ratio of 
the smallest disk radius to the
largest disk radius.

The most comprehensive study of the densest possible packings
of binary spheres in $\mathbb{R}^3$ as well as more general
size-discrete mixtures has been reported in a recent paper by
\onlinecite{Hu08}. These authors generated
candidate maximally dense polydisperse packings 
based on filling the interstices in uniform
three-dimensional tilings of space with spheres
of different sizes. They were able to find for
certain size ratios and compositions a number
of new packings. The reader is referred to \onlinecite{Hu08}
for details and some history on the three-dimensional
problem.

One of the early numerical investigations of disordered jammed packings
of binary disks in $\mathbb{R}^2$ and binary spheres in $\mathbb{R}^3$ employed a ``drop and roll''
procedure~\cite{Vi72}. Such numerical protocols and others \cite{Ok04}, in
which there is a preferred direction in the system, tend to produce
statistically anisotropic packings, which exhibit lower densities than 
than those generated by packing protocols that yield
statistically isotropic packings \cite{Do04a}. It is not clear that
the former packings are collectively jammed. In two dimensions,
one must be especially careful in choosing a sufficiently small
size ratio  in order to avoid the tendency of such
packings to form highly crystalline arrangements.
The LS algorithm has been used successfully to generate
disordered strictly jammed packings of binary disks with
$\phi \approx 0.84$ and $\alpha^{-1}=1.4$  \cite{Do06b}.
By  explicitly constructing an exponential number of jammed packings of binary 
disks with densities spanning the spectrum from the
accepted amorphous glassy state to the phase-separated crystal,
it has been argued \cite{Do06b,Do07c} that there is no 
``ideal glass transition" \cite{Pa05}. The existence
of an ideal glass transition remains a hotly debated
topic of research.

In three dimensions, it was shown by \onlinecite{Sch94}
that increasing polydispersity increases the packing fraction
over the monodisperse value that an amorphous hard-sphere system can possess.
The LS algorithm  \nocite{Lu90} has been extended to
 generate jammed  sphere packings  with a polydispersity 
in size \cite{Ka02c}. It was applied to show that
disordered packings with a wide range of packing fractions
that exceed 0.64 and varying degrees of disorder
can be achieved; see also  \onlinecite{Cha10}. 
 Not surprisingly, the determination of the maximally random jammed (MRJ) state  for
an arbitrary polydisperse sphere packing is a wide open question.
\onlinecite{Cl09} have carried out a beautiful series of experiments to understand
polydisperse random packings of spheres. 
Specifically, they produced  three-dimensional random packings of
frictionless emulsion droplets with a high degree of size polydispersity, and visualize and
characterize them using confocal microscopy.

The aforementioned investigations and the many conundrums
that remain serve to illustrate
the richness of polydisperse jammed packings
but further discussion is beyond the scope of this review.

\section{Packings of Nonspherical Particles}
\label{nonspherical}

Jammed packing characteristics become considerably more complex by allowing for 
nonspherical particle shapes \cite{Wi03,Do04b,Do04d,Ma05,Do05a,Do05b,Do07a,Ro00,Be00,Co06,Ch07,Ch08,To09b,To09c,Gl09,Ji10b,To10b,Ka10,Ch10}. We focus here on the latest developments in this category, 
specifically for particle shapes that are continuous deformations
of a sphere (ellipsoids and superballs) as well as polyhedra.  
Nonsphericity introduces rotational degrees
of freedom not present in sphere packings, and can dramatically
alter the jamming characteristics from those of sphere packings.  
[We note in passing that there has been deep and productive
examination of the equilibrium phase behavior
and transport properties of hard nonspherical particles 
\cite{Fr83,Fr84,Ti93,Ya08}.]

Very recent developments have provided organizing principles to characterize
and classify jammed packings of nonspherical particles in terms
of shape symmetry of the particles \cite{To09b}. We will elaborate
on these principles in Sec.\ref{polyhed} where we discuss 
packings of polyhedra. We will begin the discussion by considering
developments within the last several years on ellipsoid
packings, which has spurred much of the resurgent interest
in dense packings of nonspherical particles.

\subsection{Ellipsoid Packings}
\label{ellip}

One simple generalization of the sphere is an ellipsoid,
the family of which is a continuous deformation of a sphere. A three-dimensional ellipsoid is a centrally
symmetric  body  occupying the region 
\begin{equation}
\left(\frac{x_1}{a}\right)^{2}+  \left(\frac{x_2}{b}\right)^{2}+\left(\frac{x_3}{c}\right)^{2}\le 1,
\end{equation}
where $x_i$ $(i=1,2,3)$ are Cartesian coordinates
and $a$, $b$ and $c$ are the semi-axes of the ellipsoid. Thus,
we see that an ellipsoid is an {\it affine} (linear) transformation of the sphere.
A spheroid is an ellipsoid in which two of the semi-axes are equal, say $a=c$, and 
is  a {\it prolate} (elongated) spheroid  if $b \ge a$ and an {\it oblate} (flattened) spheroid
if $b \le a$.

\begin{figure}[bthp]
\centerline{\includegraphics[height=2.5in,keepaspectratio,clip=]
{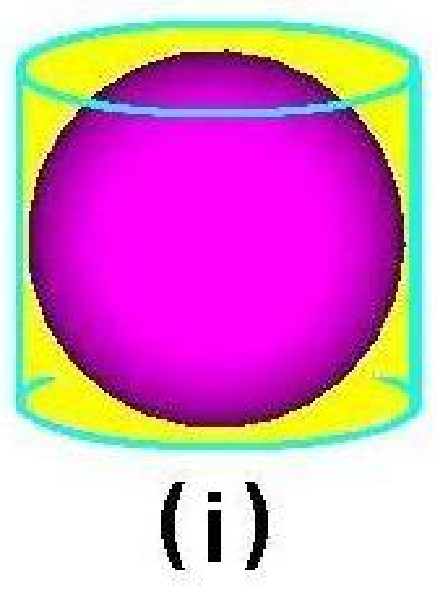} \includegraphics[height=2.5in,keepaspectratio,clip=]
{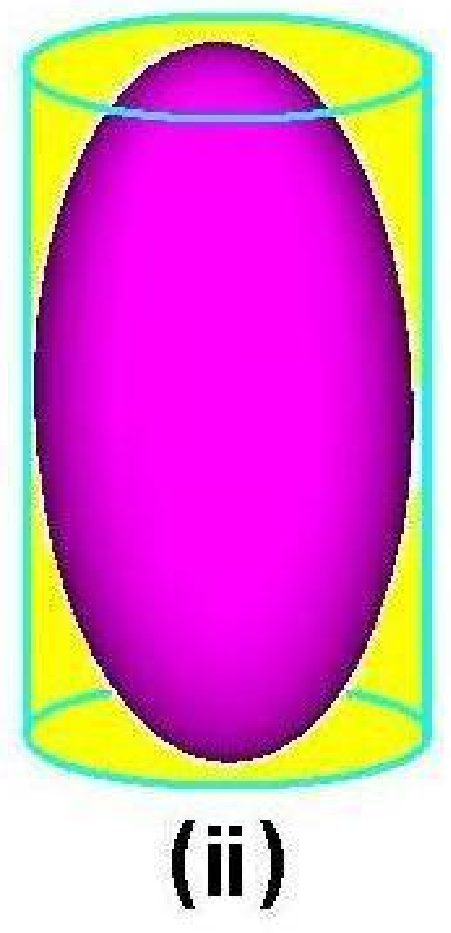} \includegraphics[height=2.5in,keepaspectratio,clip=]
{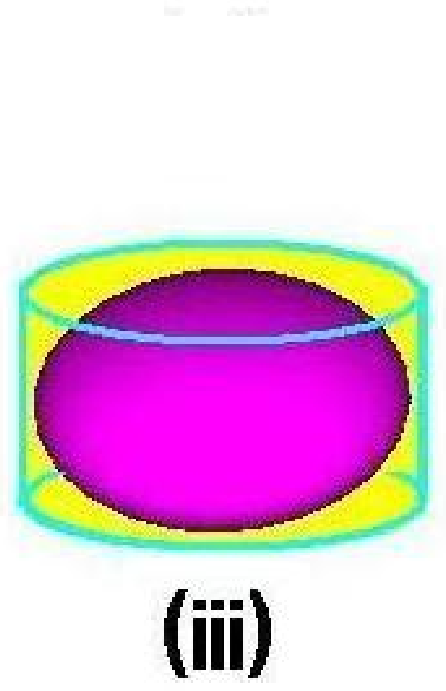}}
\caption{A sphere inscribed within the smallest circular cylinder (i)
undergoes a linear stretch and shrinkage of the space along
the vertical direction leading to a prolate spheroid (ii)
and an oblate spheroid (iii), respectively. This linear
transformation leaves the fraction of space occupied by the
spheroids within the cylinders unchanged from that of the
fraction of the cylinder volume occupied by the sphere, equal
to $2/3$.
}
\label{affine}
\end{figure}

Figure \ref{affine} shows how  prolate and  oblate
spheroids are obtained from a sphere by a linear stretch and shrinkage
of the space along the axis of symmetry, respectively. This figure also illustrates
two other basic points by inscribing the particles within the smallest
circular cylinders. The fraction of space occupied by 
each of the particles within the cylinders is an invariant
equal (due to the affine transformations) to 2/3. This might lead
one to believe that the densest packing of ellipsoids is given
by an affine transformation of one of the densest sphere packings, but
such transformations necessarily lead to ellipsoids that all have exactly
the same orientations. Exploiting the rotational degrees
of freedom so that the ellipsoids are not all required to have
the same orientations turns out to lead to larger packing fractions
than that for maximally dense sphere packings.
Furthermore, because the fraction of space remains the same
in each example shown  in Figure \ref{affine}, the sometimes popular notion
that going to the extreme ``needle-like" limit ($b/a\rightarrow \infty$) or extreme
``disk-like' limit ($b/a\rightarrow 0$) can lead to packing
fractions $\phi$ approaching unity  is misguided.

Experiments on M\&M candies (spheroidal particles) \cite{Do04b,Ma05} as well as numerical
results produced by a modified LS algorithm \cite{Do05a,Do05b}
found MRJ-like packings with packing fractions and mean contact numbers 
that were higher than for spheres. 
This led to a numerical study of the packing fraction $\phi$ and
mean contact number $Z$ as a function of the semi-axes (aspect) ratios.

\begin{figure}[bthp]
\centerline{\includegraphics[height=3.1in,keepaspectratio,clip=]
{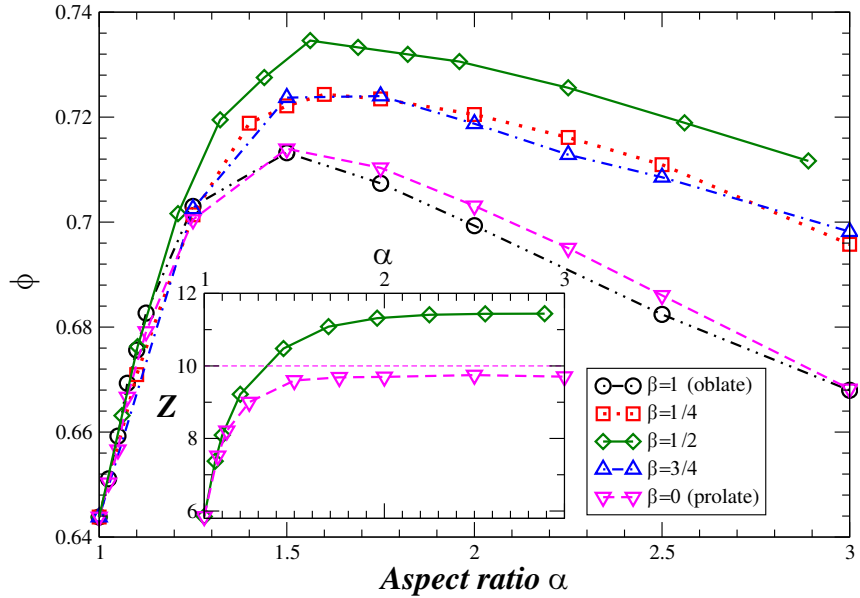}}
\caption{Density $\phi$  versus aspect ratio $\alpha$ for MRJ packings
of 10,000 ellipsoids as obtained in \cite{Do07a}. The semi-axes here are $1, \alpha^\beta, \alpha$. 
The inset shows the mean contact number $Z$ as  a function of $\alpha$.
Neither the spheroid (oblate or prolate) or general ellipsoids
cases attain their isostatic values of $Z=10$ or $Z=12$, respectively.}
\label{10000}
\end{figure}

The results were quite dramatic in several respects.
First, it was shown that  $\phi$ and $Z$, as a function
of aspect ratio, each have a cusp (i.e., non-differentiable) minimum at the sphere point,
and $\phi$ versus aspect ratio possesses a density maximum; see Fig. \ref{10000},
which shows the more refined calculations presented in \cite{Do07a}.
The existence of a cusp at the sphere point runs counter to the prevailing expectation in the literature
that for ``generic" (disordered) jammed frictionless particles  the total number
of (independent) constraints equals the total number of degrees of
freedom $d_f$, implying a mean contact number $Z=2d_f$
($d_{f}=2$ for disks, $d_{f}=3$
for ellipses, $d_{f}=3$ for spheres, $d_{f}=5$ for spheroids, and
$d_{f}=6$ for general ellipsoids). This has been
referred to as the {\it isostatic} conjecture \cite{Al98} 
or {\it isocounting} conjecture \cite{Do07a}.
Since $d_{f}$ increases discontinuously with the introduction of
rotational degrees of freedom as one makes the particles nonspherical,
the isostatic conjecture predicts that ${Z}$ should have
a jump increase at aspect ratio $\alpha=1$ to a value of $Z=12$ for a general
ellipsoid.  Such a discontinuity was not observed by
\onlinecite{Do04b}, rather, it was observed that jammed ellipsoid packings
are \emph{hypostatic}, ${Z}<2d_{f}$, near the sphere point,
and only become nearly  isostatic for large aspect ratios. 
In fact, the isostatic conjecture is only rigorously true 
for amorphous sphere packings after removal of rattlers; generic nonspherical-particle
packings should generally be hypostatic (or sub-isostatic) \cite{Ro00,Do07a}.

Until recently, it was accepted that a sub-isostatic or hypostatic packing of nonspherical 
particles \emph{cannot} be rigid (jammed) due to the existence 
of ``floppy" modes \cite{Al98},
which are unjamming motions (mechanisms) derived within a linear theory
of rigidity, i.e., a first-order analysis in the jamming gap $\delta$
(see Sec. \ref{jam-poly}).  The observation that terms of order higher
than first generally need to be considered was emphasized by \onlinecite{Ro00},
but this analysis was only developed for spheres.
It has recently been rigorously shown that if the curvature of 
nonspherical particles at their contact points are included
in a second-order and higher-order analysis,
then  hypostatic packings of such particles
can indeed be jammed \cite{Do07a}. For example, 
ellipsoid packings are generally not jammed to first
order in $\delta$ but are jammed to second order
in $\delta$ \cite{Do07a} due to curvature deviations
from the sphere.

\begin{figure}[bthp]
\centerline{\includegraphics[height=2.7in,keepaspectratio,clip=]
{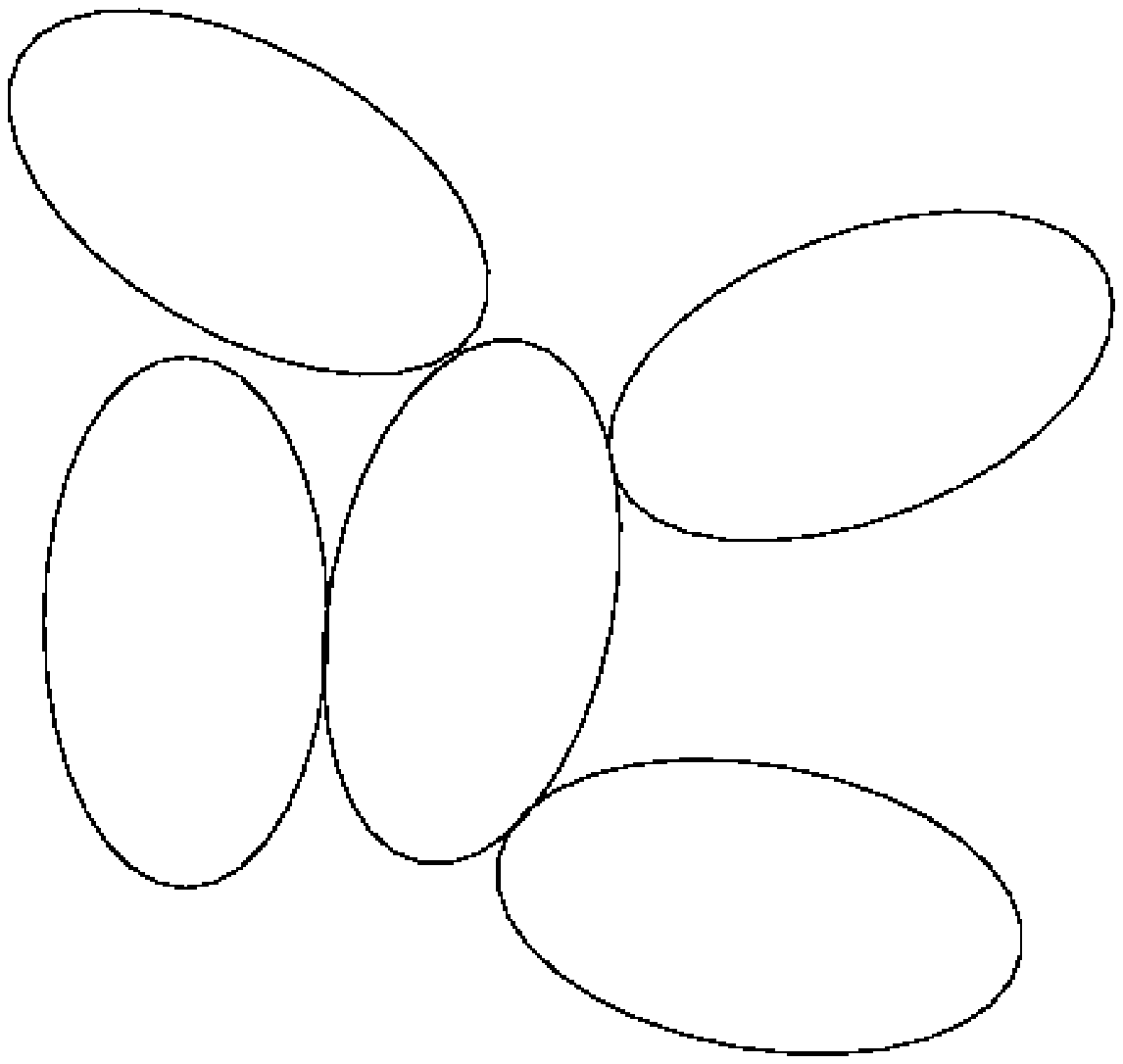} \hspace{0.2in}
\includegraphics[height=2.7in,keepaspectratio,clip=]
{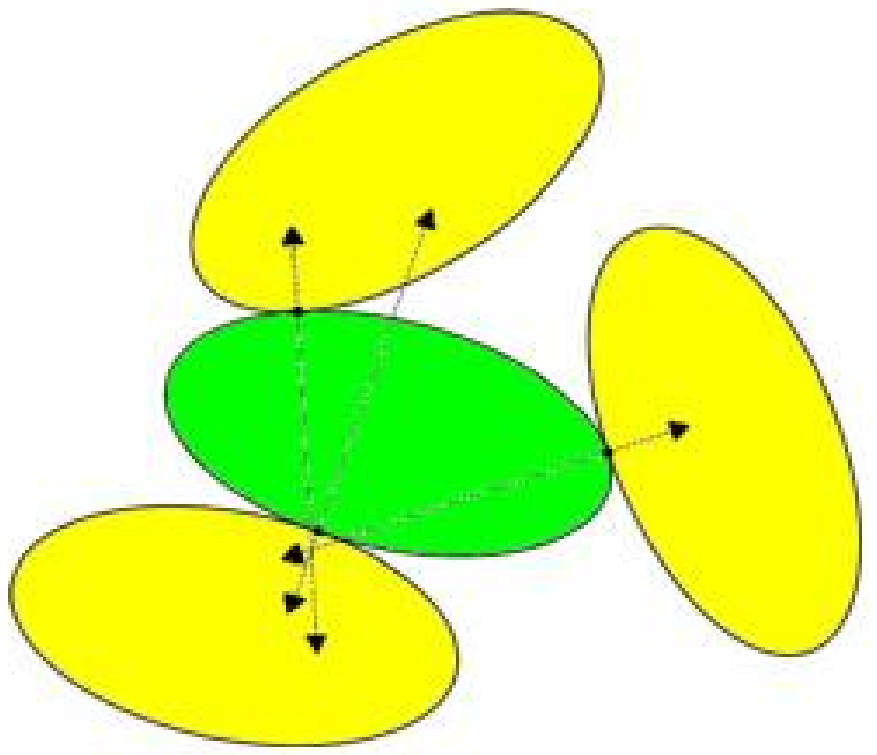}}
\caption{Simple examples of hypoconstrained packings in which all 
particles are fixed, except the central one. Left panel: Generically,
four contacting particles are required to trap the central one. Right panel:
Special correlated configurations only require three contacting
particles to trap the central one. The normal vectors at the points of contact intersect at a common
point, as is necessary to achieve torque balance. }
\label{generic}
\end{figure}

To illustrate how nonspherical jammed packings can  be hypostatic,
Fig. \ref{generic} depicts two simple two-dimensional examples
consisting of a few fixed ellipses and a central particle
that is translationally and rotationally trapped by the fixed particles.
Generically, four contacting particles are required to trap the central one.
However, there are special correlated configurations that only require three contacting
particles to trap the central one. In such instances, the normal vectors at 
the points of contact intersect at a common
point, as is necessary to achieve torque balance. 
At first glance, such configurations might be dismissed as 
probability-zero events. However, it was shown that such {\it nongeneric} 
configurations are degenerate (frequently encountered). 
This ``focusing capacity" toward hypostatic values
of $Z$ applies to large jammed packings of nonspherical particles and
in the case of ellipsoids must be present for sufficiently small aspect ratios
for a variety of realistic packing protocols \cite{Do07a}.
It has been suggested that the degree of nongenericity of the packings
be quantified by determining the fraction of local coordination configurations in which
the central particles have fewer contacting neighbors than the average value $Z$
\cite{Ji10b}.

Having established that curvature deviations from the spherical
reference shape exert a fundamental influence on constraint counting \cite{Do07a},
it is clear that similar effects will emerge when the
hard-particle interactions are replaced by nonspherical particles
interacting with soft short-range repulsive potentials. It immediately
follows that jamming to first and second order in $\delta$ for hard
nonspherical particles, for example,
leads to quadratic and quartic modes in the vibrational 
energy spectrum for packings of  such particles that interact with purely soft repulsive
interactions. The reader is referred to \onlinecite{Mail09}
and \onlinecite{Ze09} for studies of the
latter type for ellipses and ellipsoids, respectively. 

\begin{figure}[bthp]
\centerline{\includegraphics[height=3.3in,keepaspectratio,clip=]
{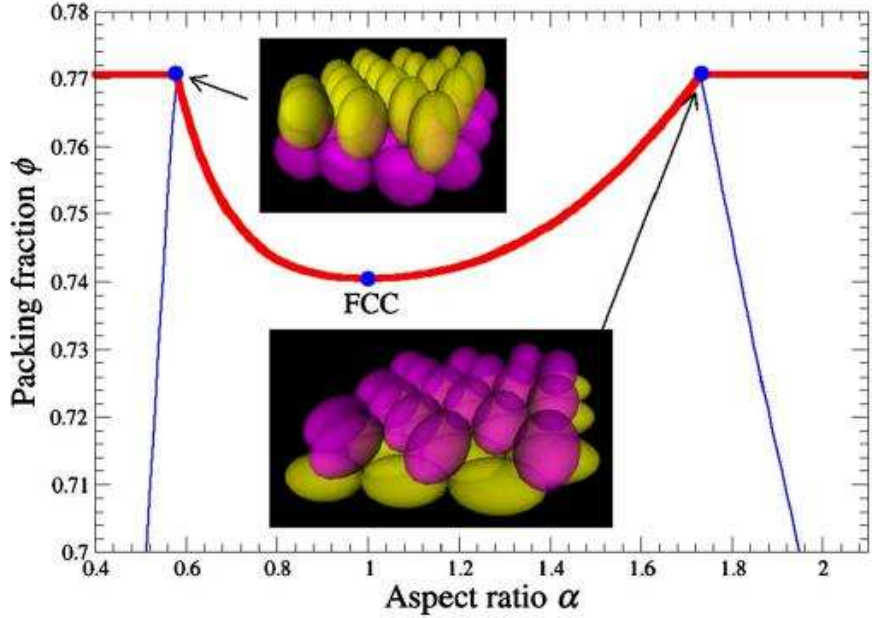}}
\caption{The packing fraction of the ``laminated" non-Bravais lattice packing of
ellipsoids (with a two-particle basis) as a function of
the aspect ratio $\alpha$. The point $\alpha=1$ corresponding
to the face-centered cubic lattice sphere packing is shown, along with the two sharp
maxima in the packing fraction for prolate ellipsoids with $\alpha=\sqrt{3}$
and oblate ellipsoids with $\alpha=1/\sqrt{3}$, as illustrated in
the insets. For both $\alpha < 1/\sqrt{3}$ and $\alpha >\sqrt{3}$, the
packing fractions of the laminated packings drop off precipitously holding
the particle orientations fixed (blue
lines). The presently maximal achievable packing fraction
$\phi=0.7707\ldots$  is highlighted with
a thicker red line, and is constant for $\alpha \le 1/\sqrt{3}$ and $\alpha \ge \sqrt{3}$
because there is an affine stretch by an arbitrary factor  
along a direction in a mirror  plane of the particle directions; see \onlinecite{Do04d}.}
\label{ellipsoids}
\end{figure}

It is noteworthy that in striking contrast with MRJ-like sphere packings, the rattler concentrations 
of the MRJ-like ellipsoid packings appear practically to vanish
outside of some small neighborhood of the sphere point \cite{Do07a}.
It was shown that MRJ-like packings of nearly-spherical ellipsoids 
can be obtained with  $\phi \approx 0.74$, i.e., packing fractions
approaching those of the densest three-dimensional sphere packings \cite{Do04b}.
This suggested that there exist ordered ellipsoid packings
with appreciably higher densities. Indeed, the densest known
ellipsoid packings were subsequently discovered \cite{Do04d}; see Fig. \ref{ellipsoids}.
These represent a new family of non-Bravais lattice packings of ellipsoids with
a packing fraction that always exceeds the density of the 
 densest Bravais lattice packing 
($\phi =0.74048 \ldots$) with a maximal packing fraction of $\phi =0.7707 \ldots$,
for a wide range of aspect ratios ($\alpha \le 1/\sqrt{3}$ and $\alpha \ge \sqrt{3}$). In these densest known packings,  each ellipsoid has 14 contacting neighbors and there
are two particles per fundamental cell.

\begin{figure}[bthp]
\centerline{\includegraphics[height=2.5in,keepaspectratio,clip=]{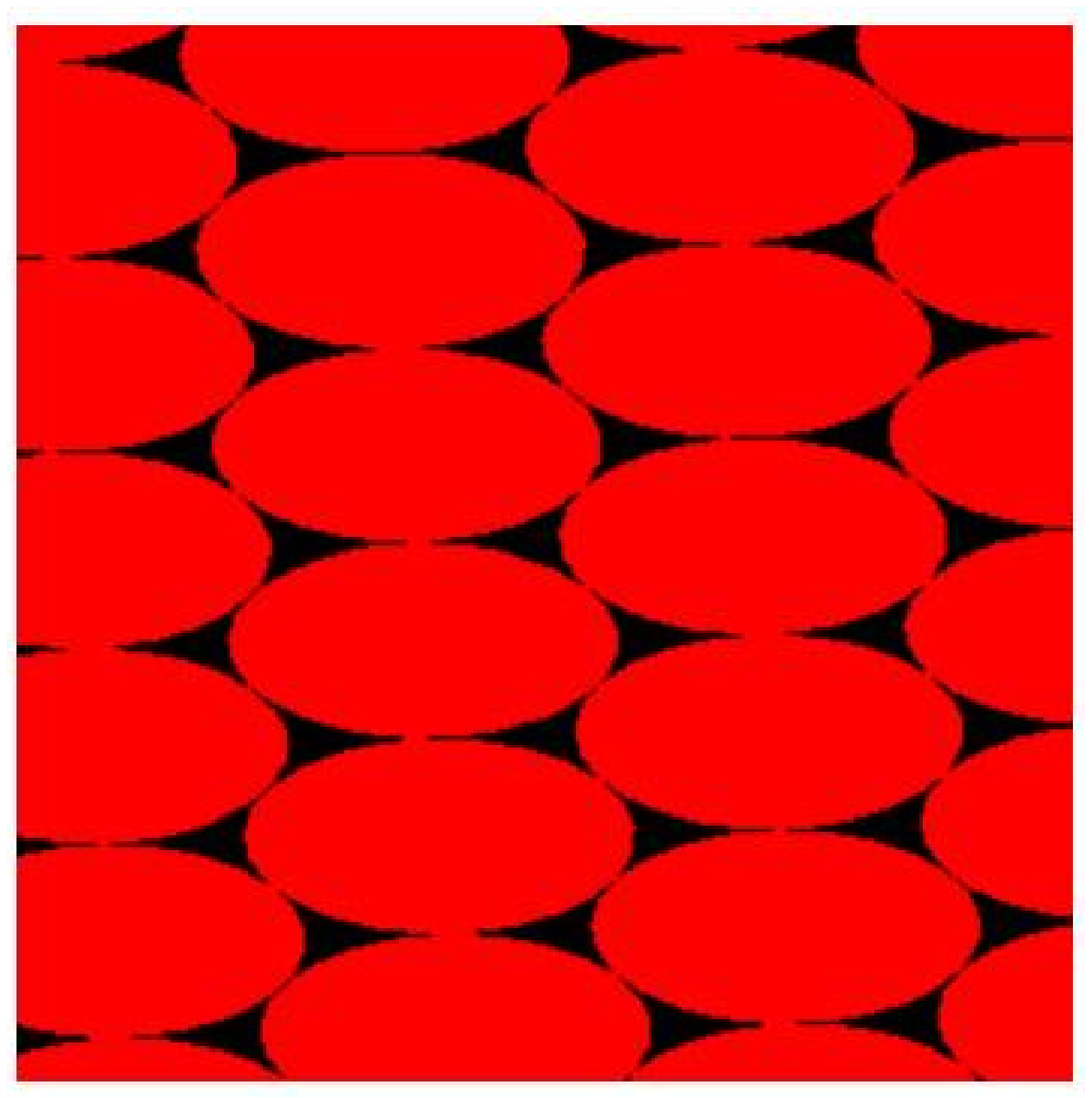}}
\caption{A portion of the densest packing of congruent ellipses, which
is simply an affine transformation of the densest circle packing, i.e.,
the densest triangular lattice of circles. This
maximally dense ellipse packing is not rotationally
jammed for any noncircular shape, but
is strictly translationally jammed.
}
\label{ellipse}
\end{figure}

For identical ellipse packings in $\mathbb{R}^2$, 
the maximally dense arrangement is obtained by an affine
stretching of the optimal triangular-lattice packing 
of circular disks with $\phi_{\mbox{\scriptsize max}}=\pi/\sqrt{12}$,
which leaves $\phi_{\mbox{\scriptsize max}}$ unchanged \cite{Fe64,Do04d}; see Fig. \ref{ellipse}.  This
maximally dense ellipse packing is not rotationally
jammed for any noncircular shape, since it can be sheared continuously without introducing
overlap or changing the density \cite{Do07a}. The packing
is, however, strictly translationally jammed.

\subsection{Superball Packings}
\label{super}

Virtually all systematic investigations of the densest particle packings
have been carried out for convex objects.
A $d$-dimensional \textit{superball} is a centrally
symmetric  body in $d$-dimensional Euclidean space occupying
the region 
\begin{equation}
|x_1|^{2p}+|x_2|^{2p}+\cdots+|x_d|^{2p} \le 1,
\end{equation}
where $x_i$ $(i=1,\ldots,d)$ are Cartesian coordinates
and $p \ge 0$ is the \textit{deformation parameter} (not pressure
as denoted in Sec.~\ref{jam-poly}), which controls
the extent to which the particle shape has deformed from that of a
 $d$-dimensional sphere ($p=1$). Thus, superballs constitute  a large family of
both convex ($p\ge 0.5$) and concave ($0 \le p<0.5$) particles (see Fig.~\ref{superball}).

\begin{figure}
\begin{center}
\includegraphics[height=3.0cm, keepaspectratio]{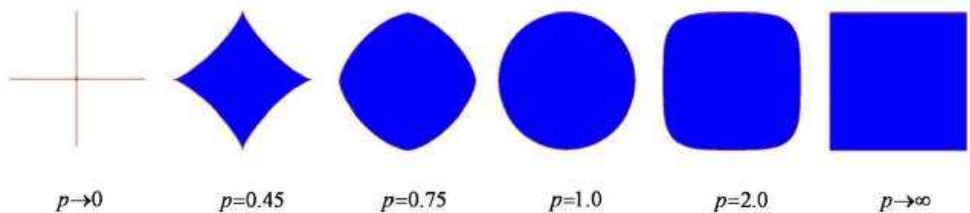}
\end{center}
\caption{(color online) Superdisks with different values of the deformation parameter $p$.}
\label{superdisk}
\end{figure}

 In general, a ``superdisk," the designation in the two-dimensional
case, possesses square
symmetry. As $p$ moves away from unity,
two families of superdisks with square symmetry can be obtained
depending on whether $p < 1$ or $p >1$ 
such that there is a 45-degree rotation with respect to the
``protuberances." When $p<0.5$, the superdisk is concave (see Fig. \ref{superdisk}).
The candidate maximally dense packings  
were recently proposed for all convex and concave shapes \cite{Ji08b}. These are achieved by two different
families of Bravais lattice packings such that $\phi_{\mbox{\scriptsize max}}$
is nonanalytic at the ``circular-disk" point ($p=1$) and increases significantly as
$p$ moves away from unity. 
The broken rotational symmetry of superdisks influences the packing
characteristics in a non-trivial way
that is distinctly different from ellipse  packings.
Recall that for ellipse packings, no improvement over the maximal
circle packing density is possible. For superdisks, one can take advantage of 
the four-fold rotationally symmetric shape of the particle to obtain a  substantial
improvement on the maximal circle packing density.
By contrast, one needs
to use higher-dimensional counterparts of ellipses ($d \ge 3$) in
order to improve on $\phi_{\mbox{\scriptsize max}}$ for spheres. Even for three-dimensional ellipsoids, $\phi_{\mbox{\scriptsize max}}$ increases
smoothly as the aspect ratios of the semi-axes vary from unity \cite{Do04d}, and hence has
no cusp at the sphere point. In fact, congruent
three-dimensional ellipsoid packings
have a cusp-like behavior at the sphere point only when they are randomly
jammed \cite{Do04b}.

\begin{figure}
\begin{center}
\includegraphics[height=3.0cm, keepaspectratio]{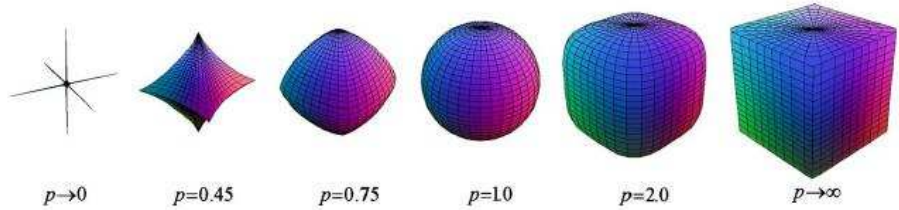}
\end{center}
\caption{(color online). Superballs with different values of the deformation
parameter $p$.} \label{superball}
\end{figure}

Increasing the dimensionality of
the particle imbues the optimal ``superball" packings with  structural characteristics
that are richer than their two-dimensional counterparts \cite{Ji09a}.
For example, in three dimensions, a superball is a perfect sphere at $p=1$,
but can possess two types of shape anisotropy: cube-like shapes 
(three-dimensional analog of the square symmetry of the superdisk)
and octahedron-like shapes, depending on the value of the
deformation parameter $p$ (see Fig.~\ref{superball}).
As $p$ continuously increases
from 1 to $\infty$, we have a family of convex superballs with
cube-like shapes; at the limit $p = \infty$, the superball is a
perfect cube. As $p$ decreases from 1 to 0.5, a family of convex
superballs with octahedron-like shapes are obtained; at $p = 0.5$,
the superball becomes a regular octahedron. When $p<0.5$, the
superball still possesses an octahedron-like shape  but is now concave,
becoming a three-dimensional \textit{``cross"} in the limit $p
\rightarrow 0$. Note that the cube and regular
octahedron (two of the five Platonic polyhedra) have the same group
symmetry (i.e., they have the same 48 space group elements) because they 
are dual to each other. Two polyhedra are dual to each other if the vertices
of one correspond to the faces of the other.

\begin{figure}
\begin{center}
$\begin{array}{c@{\hspace{0.2cm}}c@{\hspace{0.2cm}}c@{\hspace{0.2cm}}c}
\includegraphics[height=4cm, keepaspectratio]{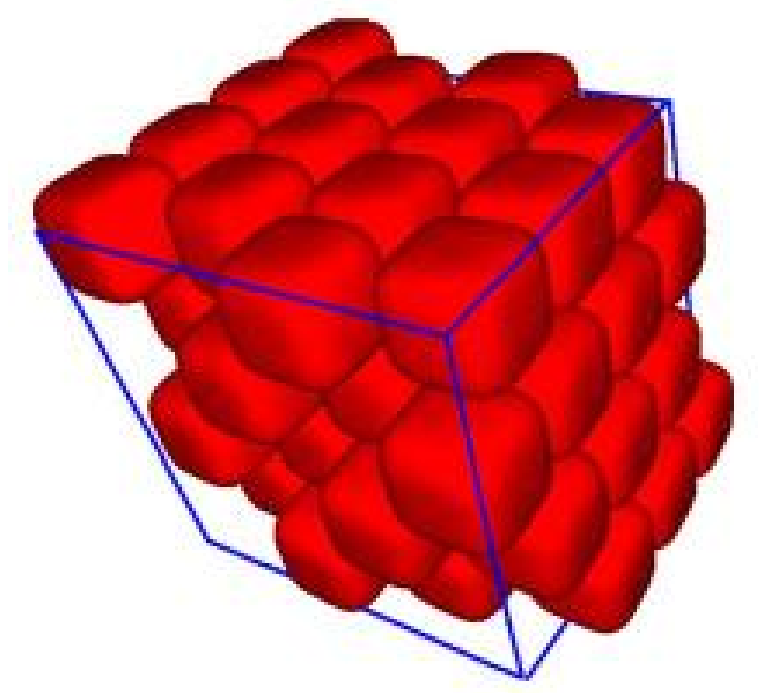} &
\includegraphics[height=4cm, keepaspectratio]{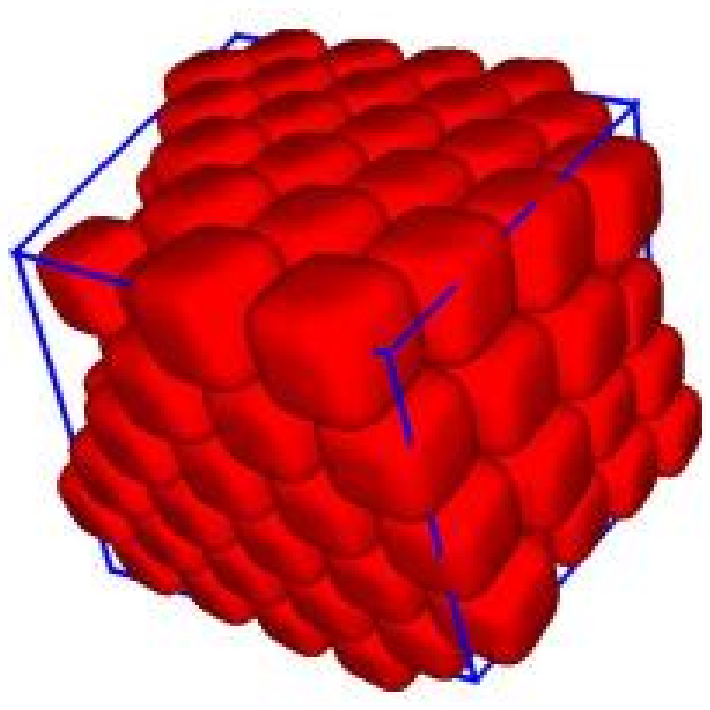} &
\includegraphics[height=4cm, keepaspectratio]{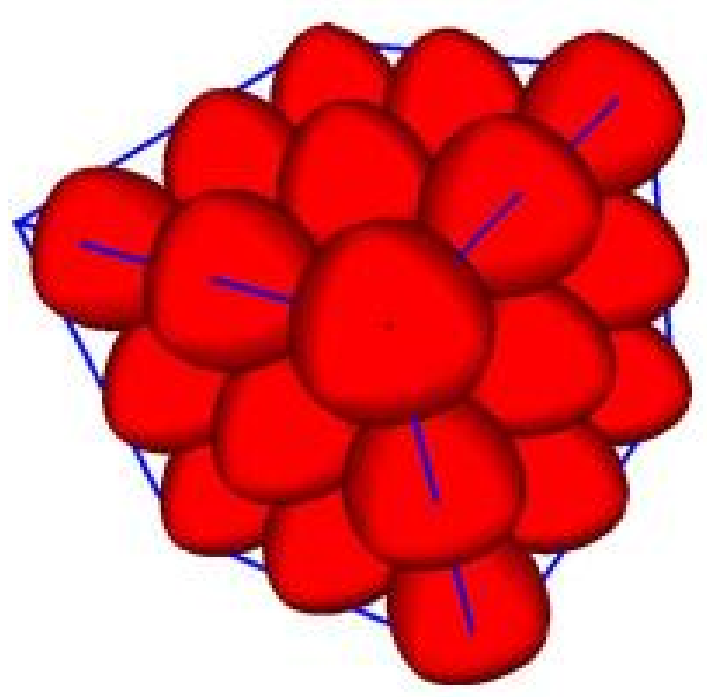} &
\includegraphics[height=4cm, keepaspectratio]{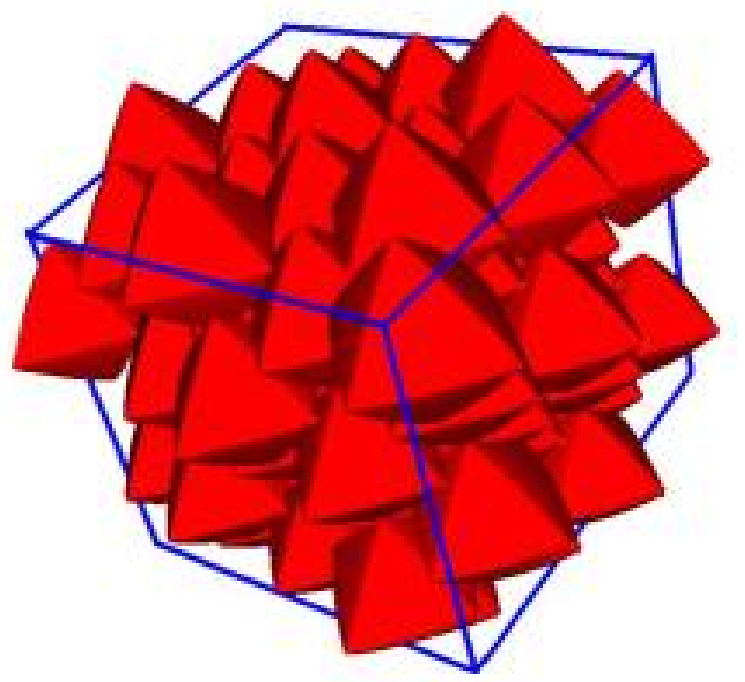} \\
\mbox{(a)} & \mbox{(b)} & \mbox{(c)}  & \mbox{(d)}
\end{array}$
\end{center}
\caption{(color online). Candidate optimal packings of superballs:
(a) The $\mathbb{C}_0$-lattice packing of superballs with $p = 1.8$.
(b) The $\mathbb{C}_1$-lattice packing of superballs with $p = 2.0$.
(c) The $\mathbb{O}_0$-lattice packing of superballs with $p=0.8$.
(d) The $\mathbb{O}_1$-lattice packing of superballs with $p=0.55$.}
\label{fig5}
\end{figure}

\onlinecite{Ji09a} have  obtained analytical constructions for
the densest known superball packings for all convex and concave cases. The candidate maximally dense
packings are certain families of Bravais lattice packings (in which each particle has 12 contacting neighbors)
possessing the global symmetries that are consistent with the symmetries of a superball. Evidence is
provided that these packings are indeed optimal, and Torquato and Jiao \cite{To09c}
have conjectured that the densest packings of all convex superballs
are their densest lattice packings; see Fig. \ref{fig5}. For superballs in the cubic regime ($p>1$), the
candidate optimal packings are achieved by two families of Bravais
lattice packings ($\mathbb{C}_0$ and $\mathbb{C}_1$ lattices) possessing two-fold  and three-fold rotational
symmetry, respectively, which
can both be considered to be continuous deformations of the fcc lattice.
For superballs in the octahedral regime ($0.5<p< 1$), there are also
two families of Bravais lattices ($\mathbb{O}_0$ and $\mathbb{O}_1$ lattices) obtainable from continuous
deformations of the fcc lattice keeping its four-fold rotational
symmetry, and from the densest lattice packing for regular octahedra
\cite{Mi05,Be00}, keeping the translational symmetry of the projected
lattice on the coordinate planes, which are apparently optimal
in the vicinity of the sphere point and the octahedron point,
respectively (see Fig. \ref{fig5}).

\begin{figure} 
$\begin{array}{c}
\includegraphics[height = 6.5cm, keepaspectratio]{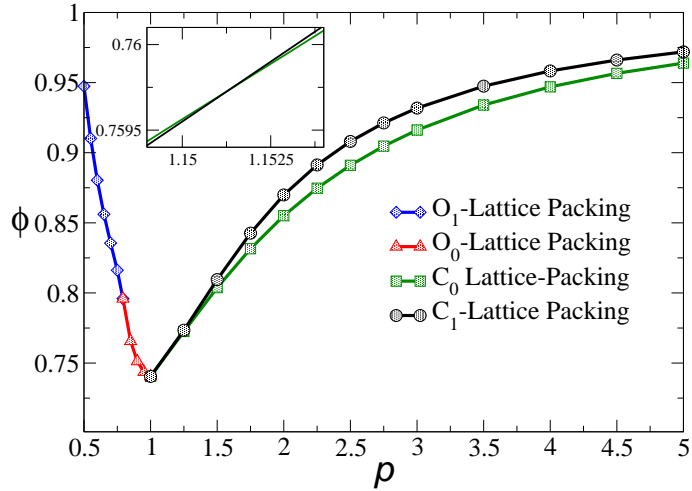}
\end{array}$
\caption{(color online). Density versus  deformation parameter $p$
for the packings of convex superballs \cite{Ji09a}. Insert: Around $p^*_c = 1.1509\ldots$,
the two curves are almost locally parallel to each other.} \label{fig4}
\end{figure}

The proposed maximal packing density $\phi_{\mbox{\scriptsize max}}$ as a function of deformation
parameter $p$ is plotted in Fig.~\ref{fig4}. 
As $p$ increases from unity, the initial increase of $\phi_{\mbox{\scriptsize max}}$
is linear in $(p-1)$ and subsequently $\phi_{\mbox{\scriptsize max}}$
 increases monotonically with $p$ until it reaches unity as the
particle shape becomes more like a cube, which is more efficient
at filling space than a sphere. These characteristics stand in contrast to
those of the densest known ellipsoid packings, achieved by certain crystal
arrangements of congruent spheroids with a two-particle basis,
whose packing density as a function of aspect ratios has zero
initial slope and is bounded from above by a value of
$0.7707\ldots$ \cite{Do04d}. As $p$
decreases from unity, the initial increase of $\phi_{\mbox{\scriptsize max}}$ is
linear in $(1-p)$. Thus, $\phi_{\mbox{\scriptsize max}}$ is a nonanalytic function of
$p$ at $p=1$, which is consistent with conclusions made about
superdisk packings \cite{Ji08b}. However, 
the behavior of $\phi_{\mbox{\scriptsize max}}$ as the superball shape moves off the
sphere point  is distinctly different from that of optimal spheroid
packings, for which $\phi_{\mbox{\scriptsize max}}$ increases smoothly as the aspect
ratios of the semi-axes vary from unity and hence has no cusp at
the sphere point \cite{Do04d}. The density of congruent ellipsoid
packings (not $\phi_{\mbox{\scriptsize max}}$) has a cusp-like behavior at the sphere
point only when the packings are randomly jammed \cite{Do04b}.
The distinction between the two systems results from different
broken rotational symmetries. For spheroids, the continuous
rotational symmetry is only partially broken, i.e., spheroids
still possess one rotationally symmetric axis; and the three
coordinate directions are not equivalent, which facilitates dense
non-Bravais packings. For superballs, the continuous rotational
symmetry of a sphere is completely broken and the three coordinate
directions are equivalently four-fold rotationally symmetric
directions of the particle. Thus, a superball is less symmetric
but more isotropic than an ellipsoid,  a shape characteristic 
which apparently favors  
dense Bravais lattice packings. The broken symmetry of superballs
makes their shapes more efficient in tiling space and thus results
in a larger and faster increase in the packing density as the shape
moves away from the sphere point.

\begin{figure}
$\begin{array}{c}
\includegraphics[height = 6.5cm, keepaspectratio]{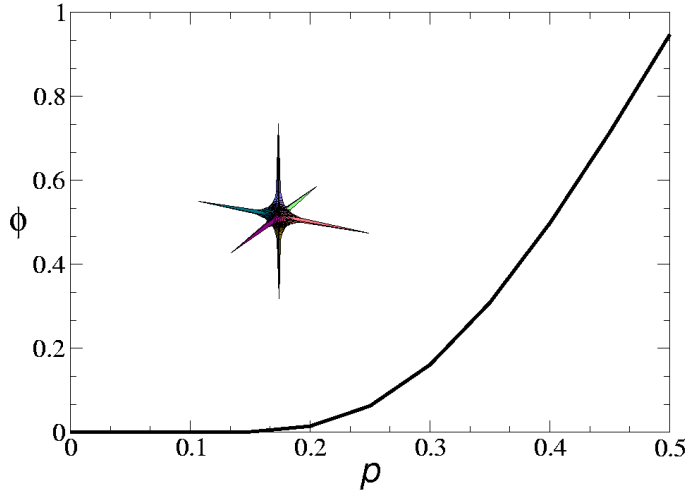}
\end{array}$
\caption{(color online). Density versus  deformation parameter $p$
for the lattice packings of concave superballs \cite{Ji09a}.  Insert: a concave
superball with $p = 0.1$, which will becomes a three-dimensional
cross at the limit $p \rightarrow 0$.} \label{fig12}
\end{figure}

As $p$ decreases from 0.5, the superballs become concave particles,
but they still possess octahedron-like shapes [see Fig.~\ref{superball}(a)]. The lack of
simulation techniques to generate concave superball packings
makes it very difficult to find the optimal packings for the entire
range of concave shapes ($0<p<0.5$). However, based on their conclusions for
convex superball packings, \onlinecite{Ji09a} conjectured that near the octahedron
point, the optimal packings possess similar translational symmetry
to that of the $\mathbb{O}_1$-lattice packing, and based on theoretical considerations 
proposed candidate optimal packings for all concave cases with a density versus $p$ as shown
in Fig.\ref{fig12}.

\begin{figure}
\begin{center}
\includegraphics[height=3.5in, keepaspectratio]{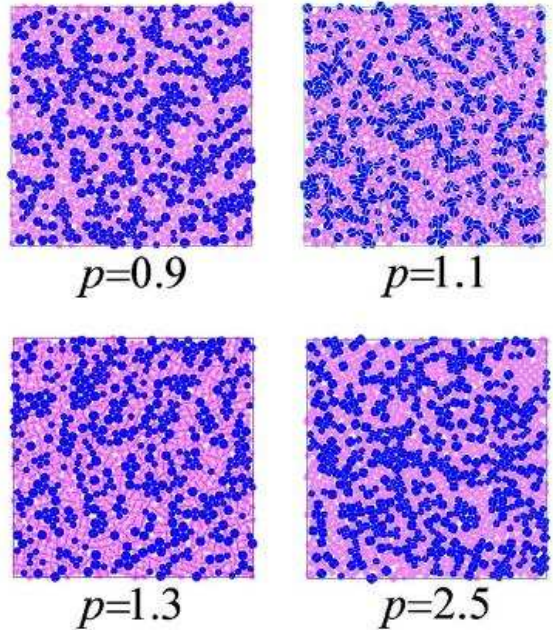} 
\end{center}
\caption{(color online). Local packing structures with more contacts than average (shown in blue)
and those with less contacts than average (shown in pink)
in two-dimensional MRJ superdisk packings for different values of the deformation
parameter $p$, as obtained from \onlinecite{Ji10b}. Note that the degree of nongenericity
decreases as $p$ moves away from its sphere-point value. Here the size ratio (diameter of large superdisks divided
by the diameter of the small superdisks) is $1.4$ and the molar 
ratio (number of large superdisks divided by the number of small superdisks)
is $1/3$. 
}
\label{Generic}
\end{figure}

\onlinecite{Ji10b} have determined the packing fractions of  maximally random jammed (MRJ) packings
of binary superdisks in $\mathbb{R}^2$ and monodispersed superballs in $\mathbb{R}^3$.
They found that the MRJ densities of such packings
increase dramatically and nonanalytically 
as one moves away from the circular-disk and sphere point ($p=1$). Moreover,
these disordered packings were demonstrated to be  hypostatic, i.e., the average number of contacting
neighbors is less than twice the total number of degrees of freedom per particle,
and the packings are mechanically stable.
As a result, the local arrangements of the particles are necessarily nontrivially
correlated to achieve jamming and hence ``nongeneric.''
The degree of ``nongenericity'' of the packings
was quantitatively characterized by determining the fraction of local coordination structures
in which the central particles have fewer contacting neighbors than the average value $Z$.
Figure~\ref{Generic} depicts local packing structures with more contacts than average 
and those with less contacts than average in MRJ binary superdisk packings for different $p$ values.
It was also explicitly shown  that such seemingly ``special'' packing configurations are
counterintuitively not rare. As the anisotropy of the particles increases,
it was shown that the fraction of rattlers decreases while the minimal orientational order (as measured
by the cubatic order metric) increases.
These novel characteristics result from the unique rotational symmetry breaking manner of 
superdisk and superball particles.

\subsection{Polyhedron Packings}
\label{polyhed}

Until some recent developments, very little was known about the densest packings of
polyhedral particles. The difficulty
in obtaining dense packings of polyhedra is related to their complex rotational degrees of freedom and to the
non-smooth nature of their shapes.

\begin{figure}[bthp]
\begin{center}
\includegraphics[width=14.5cm,keepaspectratio]{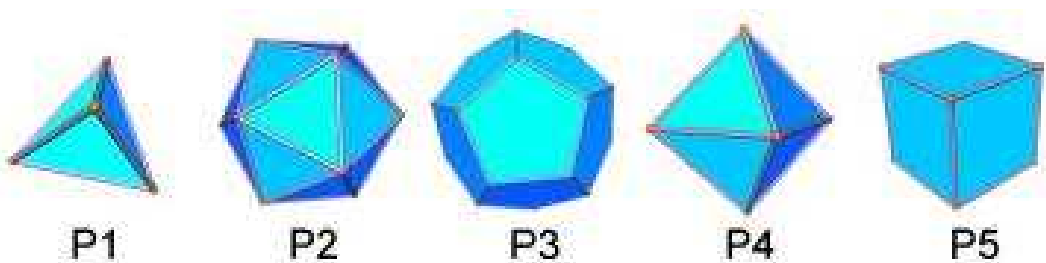} \\
\end{center}
\caption{(color online). The five Platonic solids: tetrahedron (P1), icosahedron (P2),
dodecahedron (P3), octahedron (P4), and cube (P5).}
\label{platonic}
\end{figure}

The Platonic solids  (mentioned in Plato's Timaeus)
are convex polyhedra with faces composed of congruent
convex regular polygons. There are exactly five such solids: the tetrahedron (P1), icosahedron (P2),
dodecahedron (P3), octahedron (P4), and cube (P5) (see Fig. \ref{platonic}) 
[We note in passing that viral capsids often have icosahedral symmetry; see,
for example, \onlinecite{Za04}.]
Here we focus on the problem of the determination of the densest packings of each
of the Platonic solids in three-dimensional Euclidean space $\mathbb{R}^3$,
except for the cube, which is the only Platonic solid that tiles space.

It is useful to highlight some basic geometrical properties
of the Platonic solids that we will employ in subsequent sections of 
this review. The dihedral angle $\theta$ is the interior angle between any two face planes
and is given by
\begin{equation}
\sin{\theta\over 2} = \frac{\cos(\pi/q)}{\sin(\pi/p)},
\end{equation}
where $p$ is the number of sides of each face and $q$ is the number of faces meeting at each vertex \cite{Co73}.
Thus, $\theta$ is $2\sin^{-1}(1/\sqrt{3})$, $2\sin^{-1}(\Phi/\sqrt{3})$, $2\sin^{-1}(\Phi/\sqrt{\Phi^2+1})$,
$2\sin^{-1}(\sqrt{2/3})$, and $\pi/2$, for the tetrahedron, icosahedron, dodecahedron,
octahedron, and cube, respectively, where $\Phi=(1+\sqrt{5})/2$ is the golden ratio.
Since the dihedral angle for the cube is the only one
that is a submultiple of $2\pi$, the cube is the only Platonic solid
that tiles space. It is noteworthy that in addition
to the regular tessellation of $\mathbb{R}^3$ by cubes in the simple cubic lattice
arrangement, there is an infinite
number of other irregular tessellations of space by cubes.
This tiling-degeneracy example vividly illustrates a fundamental
point made by \onlinecite{Ka02d}, namely, packing arrangements of nonoverlapping objects
at some fixed density can exhibit a large variation in their degree
of structural order. We note in passing that there are two regular dodecahedra that 
independently tile
three-dimensional (negatively curved) hyperbolic
space $\mathbb{H}^3$, as well as one cube and one regular icosahedron \cite{Co73};
see Sec. \ref{curved} for additional remarks about packings in curved spaces.

Every polyhedron has a dual polyhedron with faces and vertices interchanged. The dual of each
Platonic solid is another Platonic solid, and therefore
they can be arranged into dual pairs: the tetrahedron is self-dual (i.e., its dual is another
tetrahedron), the icosahedron and
dodecahedron form a dual pair, and the octahedron and cube form a dual pair.

\begin{figure}
\begin{center}
\includegraphics[width=14.5cm,keepaspectratio]{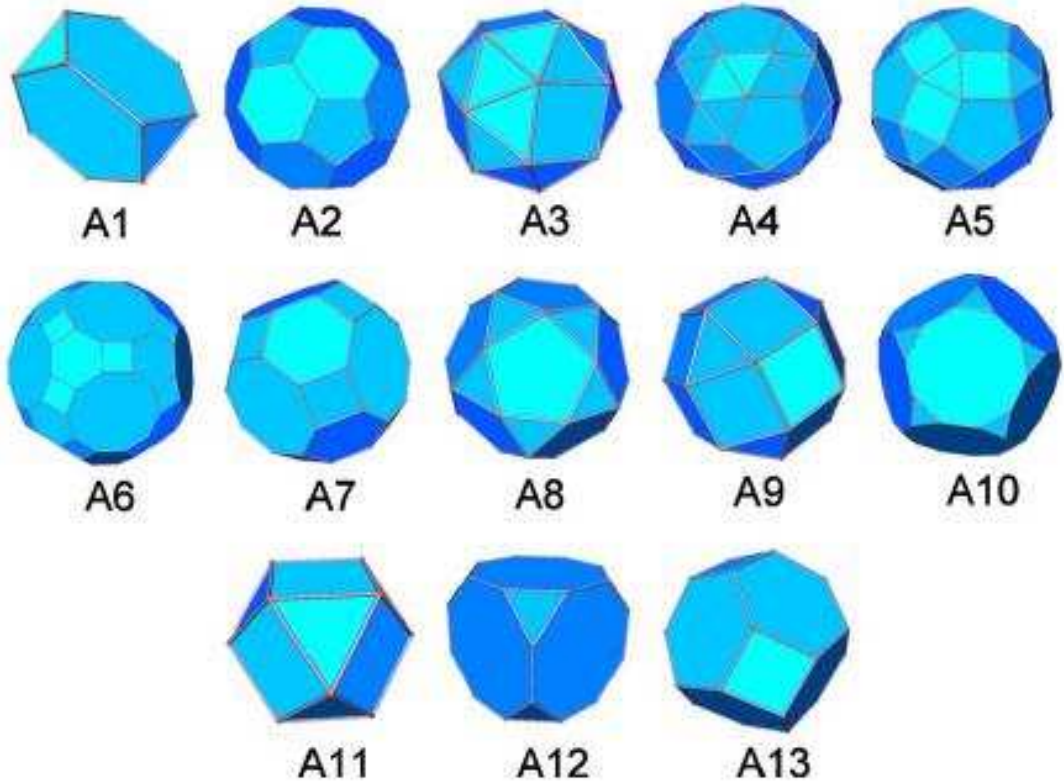} \\
\end{center}
\caption{(color online) The 13 Archimedean solids: truncated tetrahedron (A1),
truncated icosahedron (A2),
snub cube (A3), snub dodecahedron (A4), rhombicosidodecahdron (A5),
truncated icosidodecahedron (A6),
truncated cuboctahedron (A7), icosidodecahedron (A8),  rhombicuboctahedron (A9),
truncated dodecahedron (A10), cuboctahedron (A11),
truncated cube (A12), and truncated octahedron (A13). This typical enumeration
of the Archimedean solids does not count the chiral forms (not shown) of the snub cube (A3)
and snub dodecahedron (A4), which implies that the left-handed and right-handed
forms of each of these pairs lack central symmetry. The remaining 11 Archimedean solids
are non-chiral (i.e., each solid is superposable on its mirror image)
and the only non-centrally symmetric one  among these is the truncated tetrahedron.} \label{archimedean}
\end{figure}

An Archimedean solid is a highly symmetric, semi-regular
convex polyhedron composed of two or more types of regular polygons meeting in
identical vertices. There are thirteen Archimedean solids: truncated tetrahedron (A1),
truncated icosahedron (A2), snub cube (A3), snub dodecahedron (A4), rhombicosidodecahdron (A5),
truncated icosidodecahdron (A6), truncated cuboctahedron (A7), icosidodecahedron (A8),
rhombicuboctahedron (A9), truncated dodecahedron (A10), cuboctahedron (A11),
truncated cube (A12) and truncated octahedron (A13) (see Fig. \ref{archimedean}).
Note that the truncated octahedron is the only Archimedean solid that tiles $\mathbb{R}^3$.
The duals of the Archimedean solids are  an interesting set of 
new polyhedra that are called the Archimedean duals or the Catalan polyhedra,
the faces of which are not regular polygons.
 
Another important observation is that the tetrahedron (P1) and
the truncated tetrahedron (A1) are the
only Platonic and non-chiral Archimedean solids, respectively,
that are not {\it centrally symmetric}. The chiral snub cube and chiral snub dodecahedron
are the only other non-centrally symmetric Archimedean solids.
A particle is centrally symmetric if it has a center $C$ that
bisects every chord through $C$ connecting any two boundary points
of the particle, i.e., the center is a point of inversion symmetry.
 We will see that the central symmetry
of the majority of the Platonic and Archimedean solids (P2 -- P5, A2 -- A13)
distinguish their dense packing arrangements from those of the
non-centrally symmetric ones (P1 and A1)
in a fundamental way.

Tetrahedral tilings of space underlie many different molecular systems \cite{Co06}.
Since regular tetrahedra cannot tile space, it is of
interest to determine the highest density  that such
packings of particles can achieve (one of Hilbert's 18th problem set). It is of interest
to note that the densest Bravais-lattice packing of tetrahedra
(which requires all of the tetrahedra to have the same
orientations) has $\phi =18/49=0.367\ldots$ and each tetrahedron
touches 14 others. Recently, \onlinecite{Co06} showed
that the maximally dense tetrahedron packing cannot be a Bravais lattice
(because dense tetrahedron packings favor face to face contacts) and
found non-Bravais lattice (periodic) packings of regular tetrahedra
with $\phi \approx 0.72$. One such packing is based upon
the filling of ``imaginary" icosahedra with the densest arrangement of 20 tetrahedra
and then arranging the imaginary icosahedra in their densest lattice
packing configuration.
Using ``tetrahedral" dice, \onlinecite{Ch07}
experimentally generated jammed disordered packings of such dice
with $\phi \approx 0.75$; see also \onlinecite{Chaik10}
for a refined version of this work. However, because these dice are not
perfect tetrahedra (vertices and edges are slightly rounded), a
definitive conclusion could not be reached. 
Using physical models and computer algebra system, \onlinecite{Ch08} 
discovered a dense periodic arrangement
of tetrahedra with $\phi=0.7786\ldots$, which exceeds
the density of the densest sphere packing by an appreciable amount.
 
In an attempt to find even denser packings of tetrahedra,
\onlinecite{To09b,To09c} have formulated the problem of generating dense packings of 
polyhedra within an adaptive fundamental cell subject to periodic boundary 
conditions as an optimization problem, which they call the Adaptive 
Shrinking Cell (ASC) scheme. Starting from a variety of initial
unjammed  configurations, this optimization procedure uses 
both a {\it sequential} search of the configurational space of 
the particles and the space of lattices via an adaptive fundamental cell that shrinks
on average to obtain dense packings.  This was used to obtain
a tetrahedron packing consisting of 72 particles per fundamental cell
with packing fraction $\phi =0.782\ldots$ \cite{To09b}.
Using 314 particles per fundamental cell and starting
from an ``equilibrated" low-density liquid configuration,
the same authors were able to improve the packing fraction to $\phi =0.823\ldots$ \cite{To09c}.
This packing arrangement interestingly lacks any long-range
order.  \onlinecite{Gl09} numerically
constructed a periodic packing of tetrahedra made of parallel stacks of ``rings" around
``pentagonal'' dipyramids consisting of 82 particles per fundamental cell
and a density $\phi=0.8503\ldots$. 

\begin{figure}
\centerline{
\includegraphics[  width=5in,keepaspectratio]
{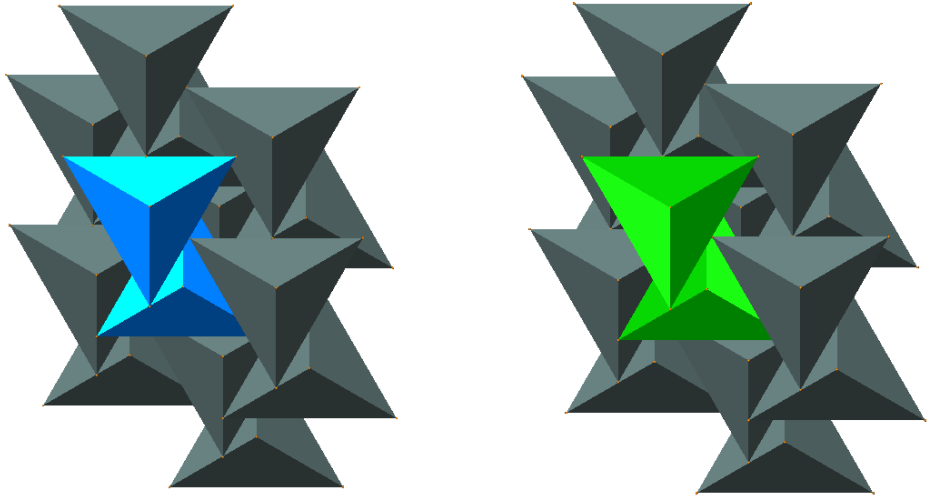}}
\caption{(color online) \footnotesize 
Closely related dense tetrahedron packings. 
Left panel: A portion of one member of the
densest two-parameter family of tetrahedron packings with 4 particles per fundamental cell
and packing fraction $\phi = \frac{12250}{14319}=0.855506\ldots$ \cite{To10a}.
The blue particles (lighter shade) represent two dimers (i.e., 4 tetrahedra) but from this perspective
each dimer appears to be a single tetrahedron. Right panel: A portion of one member of the
densest three-parameter family of tetrahedron packings with 4 particles per fundamental cell
and packing fraction $\phi = \frac{4000}{4671} = 0.856347\ldots$ \cite{Ch10,To10b}.
The green particles (lighter shade) represent two dimers. It is clear that these two packings 
are configurationally almost identical to one another.} 
\label{tetra}
\end{figure}

\onlinecite{Ka10} found a remarkably simple ``uniform" packing of tetrahedra
with high symmetry consisting of only four particles per fundamental
cell (two ``dimers") with packing fraction $\phi=\frac{100}{117}=0.854700\ldots$. 
A {\it uniform} packing has a  symmetry (in this case a point inversion symmetry) 
that takes one tetrahedron to another. A {\it dimer} is composed of a pair of 
regular tetrahedra that exactly share a common face. \onlinecite{To10a} subsequently
presented an analytical formulation to construct a three-parameter
family of  dense uniform dimer packings of tetrahedra again with four particles per fundamental cell.
(A uniform dimer packing of tetrahedra has  a point of inversion symmetry operation
that takes any dimer into another.) 
Making an assumption about one of these parameters
resulted in a two-parameter family, including those with a packing fraction as high as
$\phi=\frac{12250}{14319}=0.855506\ldots$ (see left panel of Fig.~\ref{tetra}).
\onlinecite{Ch10} recognized that 
such an assumption was made in the formulation
of \onlinecite{To10a} and employed a similar formalism
to obtain a three-parameter family of tetrahedron packings,
including the densest known dimer packings of tetrahedra with
the very slightly higher  packing fraction $\phi= \frac{4000}{4671} = 0.856347\ldots$ 
(see right panel of Fig.~\ref{tetra}).
The most  general analytical formulation to date to construct
dense periodic packings of tetrahedra with four particles per fundamental cell
was carried out by \onlinecite{To10b}.
This study  involved a  six-parameter family of dense
tetrahedron packings that includes as special cases 
all of the aforementioned ``dimer" packings of tetrahedra,  including the densest 
known packings with packing fraction $\phi= \frac{4000}{4671} = 0.856347\ldots$. This 
most recent investigation strongly
suggests that the latter set of packings are the densest among
all packings with a {\it four-particle basis}. Whether these packings are the densest
packings of tetrahedra among all packings is an open question 
for reasons given by \onlinecite{To10b}.

\begin{figure}
\centerline{
\includegraphics[  width=1.5in,keepaspectratio]
{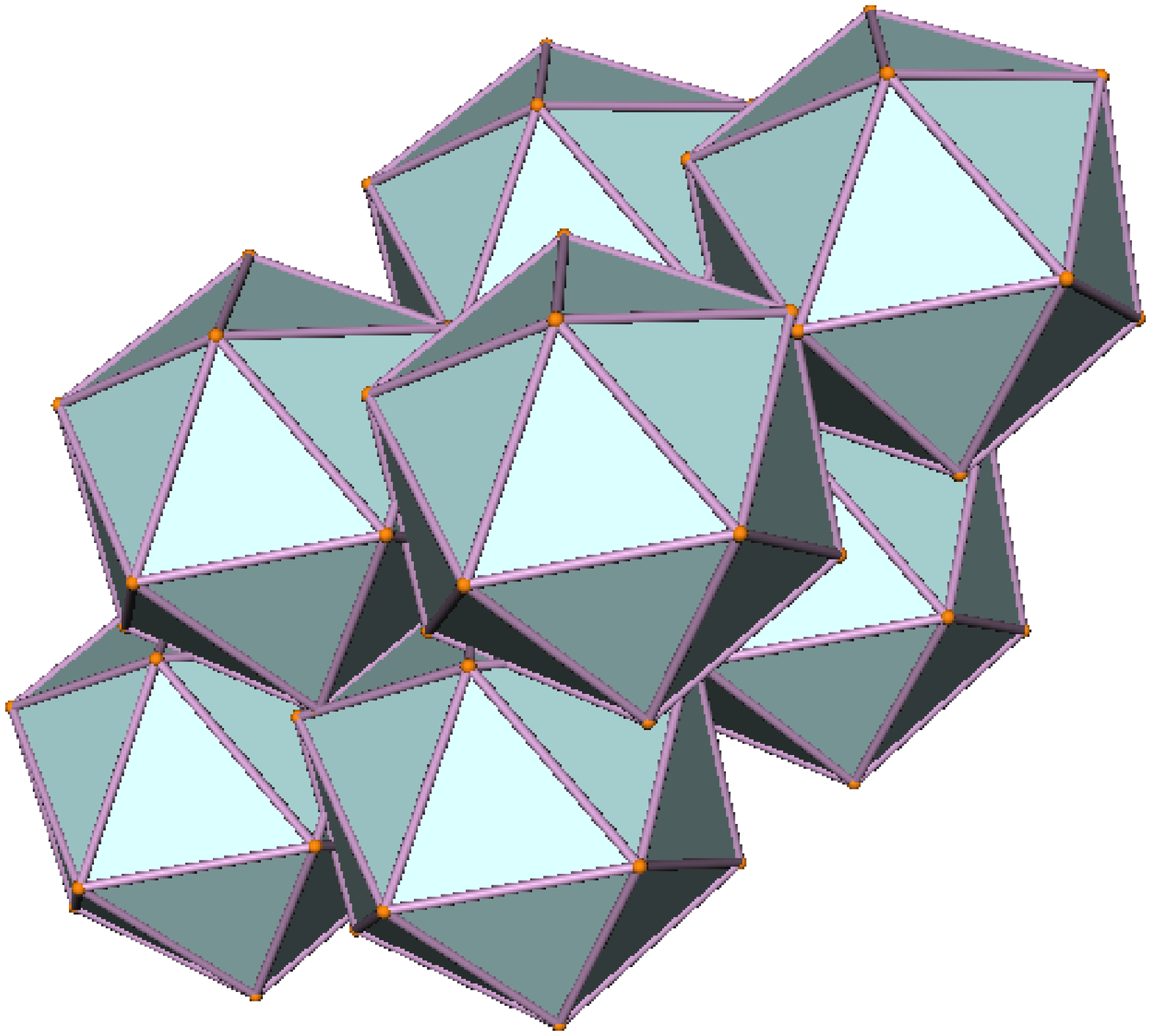}
\hspace{0.16in}
\includegraphics[  width=1.5in,keepaspectratio]
{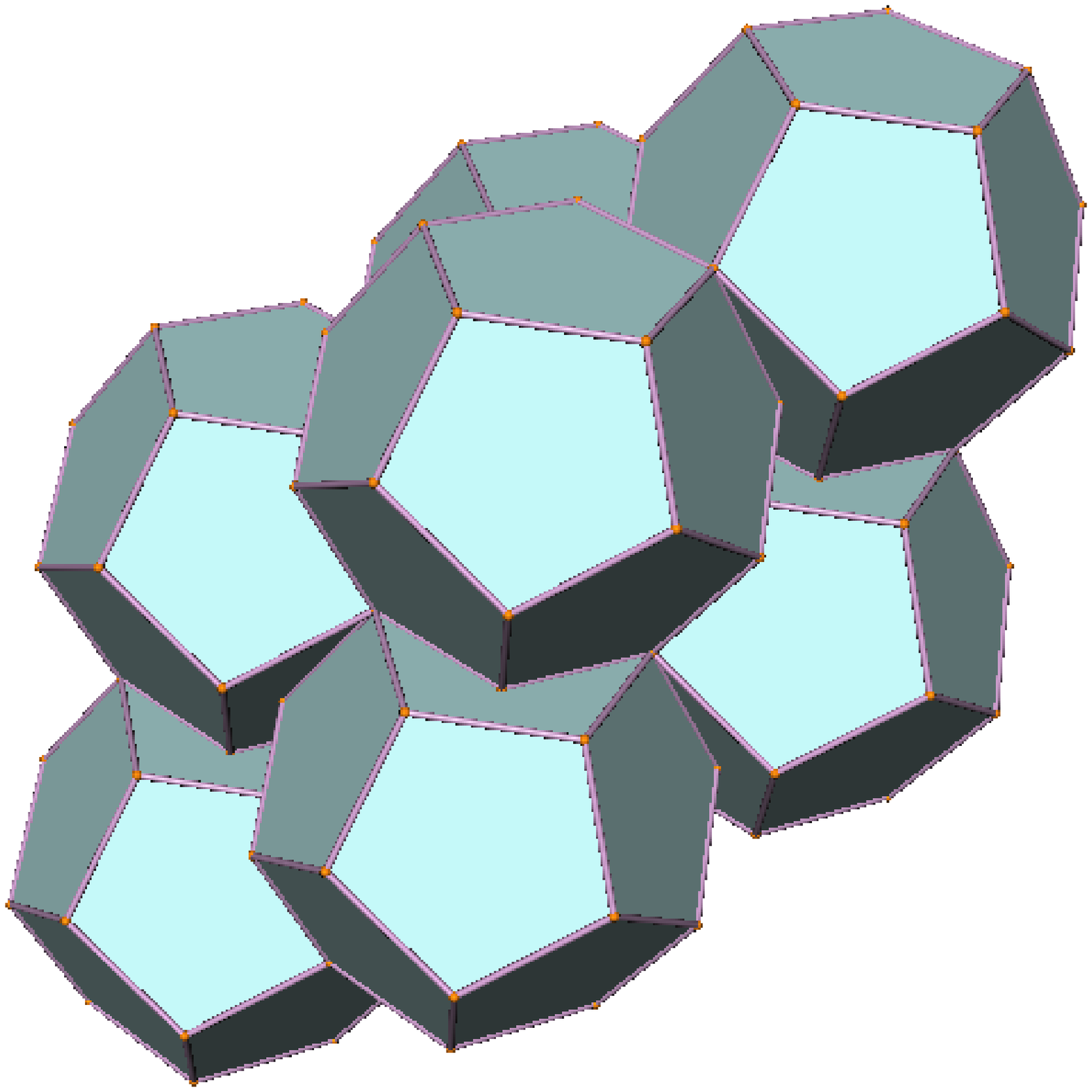}
\hspace{0.16in}
\includegraphics[  width=1.5in,keepaspectratio]
{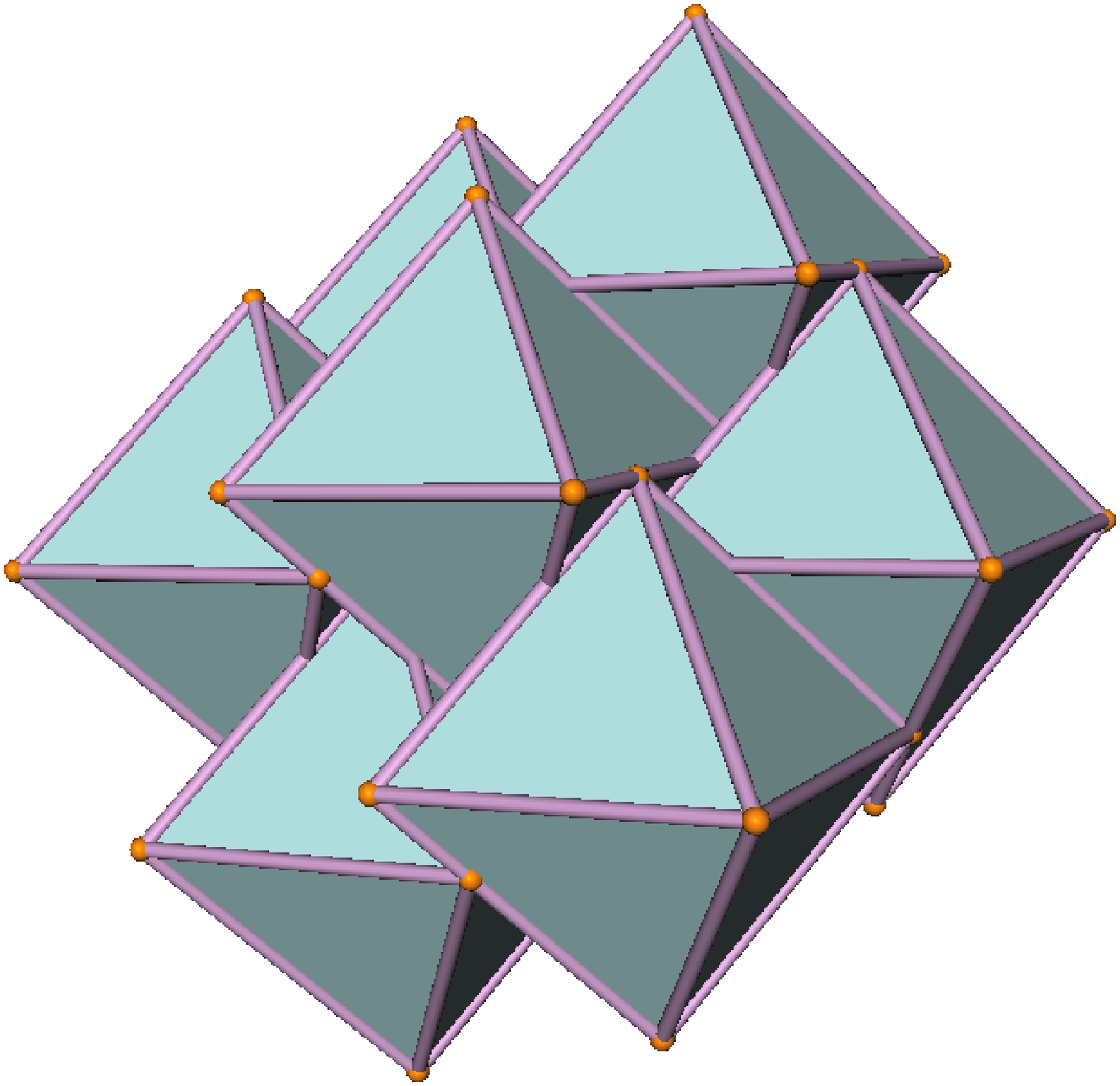}}
\caption{(color online) Portions of the densest lattice packings
of three of the centrally symmetric Platonic solids found by the 
ASC scheme \cite{To09b,To09c}. Left panel: Icosahedron packing
with packing fraction $\phi=0.8363\ldots$. Middle panel:
Dodecahedron packing with packing fraction $\phi=0.9045\ldots$.
Right panel:  Octahedron packing
with packing fraction $\phi=0.9473\ldots$.} 
\label{nontiling}
\end{figure}

Using the ASC scheme and a variety of initial conditions
with multiple particles in the fundamental cell, \cite{To09b,To09c} were also able  to find
the densest known packings of the octahedra, dodecahedra
and icosahedra (three non-tiling Platonic solids) with densities  $0.947\ldots$,
$0.904\ldots$, and $0.836\ldots$, respectively. Unlike the densest
tetrahedron packing, which must be a non-Bravais  lattice
packing, the densest packings of the other non-tiling Platonic
solids found by the algorithm are their previously known optimal (Bravais)
lattice packings \cite{Mi05,Be00}; see Fig. \ref{nontiling}.  These simulation 
results as well as other theoretical considerations, which we briefly 
describe immediately below, led them to general organizing principles
concerning the densest packings of a class of nonspherical particles.

Rigorous upper bounds on the maximal packing fraction $\phi_{\mbox{\scriptsize max}}$
of packings of nonspherical particles of general shape can be used to assess
the packing efficiency of a particular dense packing
of such particles. However, it has been highly challenging to
formulate upper bounds for non-tiling particle
packings that are nontrivially less than unity.  
It has recently been shown  that 
$\phi_{\mbox{\scriptsize max}}$ of a packing of congruent nonspherical particles of
volume $v_{P}$ in $\mathbb{R}^3$ is bounded from above according to
\begin{equation}
\phi_{\mbox{\scriptsize max}}\le \phi_{\mbox{\scriptsize max}}^U= \mbox{min}\left[\frac{v_{P}}{v_{S}}\;
\frac{\pi}{\sqrt{18}},1\right], \label{bound}
\end{equation}
where $v_{S}$ is the volume of the largest sphere that can be inscribed
in the nonspherical particle and $\pi/\sqrt{18}$ is the maximal sphere-packing density
\cite{To09b,To09c}.
The upper bound (\ref{bound}) will be relatively tight for packings of nonspherical
particles provided that the {\it asphericity} $\gamma$
(equal to the ratio of the circumradius to the inradius)  of the
particle is not large.  Since bound (\ref{bound}) cannot generally be sharp (i.e.,
exact) for a non-tiling, nonspherical
particle, any packing  whose density is close to the upper bound
(\ref{bound}) is nearly optimal, if not optimal.
It is noteworthy that a majority of the  centrally symmetric Platonic and Archimedean solids have relatively small asphericities and
explain the corresponding small differences between $\phi_{\mbox{\scriptsize max}}^U$
and the packing fraction of the densest lattice packing  $\phi_{\mbox{\scriptsize max}}^L$
\cite{Mi05,Be00}.

\onlinecite{To09b,To09c} have demonstrated that substantial
face-to-face contacts between any of the centrally symmetric Platonic
and Archimedean solids allow for a  higher packing fraction.
They also showed that central symmetry enables maximal 
face-to-face contacts when particles are {\it aligned}, which is
consistent with the densest packing being
the {\it optimal lattice packing}.

The aforementioned simulation results, upper bound,
and theoretical considerations led to the following three conjectures
concerning the densest packings of polyhedra and other nonspherical
particles in $\mathbb{R}^3$ \cite{To09b,To09c,To10b}:

\noindent
{\bf Conjecture 1}: {\sl The densest packings of the centrally symmetric
Platonic and Archimedean solids are given by their corresponding optimal lattice packings.}
\smallskip

\noindent
{\bf Conjecture 2}: {\sl The densest packing of any convex, congruent polyhedron
without central symmetry generally is not a (Bravais) lattice packing, i.e.,
set of such polyhedra whose optimal packing
is not a lattice is overwhelmingly larger than the set whose optimal
packing is a lattice. 
}
\smallskip

\noindent
{\bf Conjecture 3}: {\sl The densest
packings of congruent, centrally symmetric particles that do not possesses three
equivalent principle axes (e.g., ellipsoids) generally cannot be Bravais lattices.}
\smallskip

Conjecture 1 is the analog of Kepler's sphere conjecture for 
the centrally symmetric Platonic and Archimedean solids. 
Note that the densest known packing of the non-centrally symmetric truncated tetrahedron 
is a non-lattice packing with density at least as high as $23/24 = 0.958333\ldots$ \cite{Co06}.
The arguments leading to Conjecture 1 
also strongly suggest that the densest packings of superballs
are given by their corresponding optimal lattice packings \cite{To09c},
which were proposed by \onlinecite{Ji09a}.

\subsection{Additional Remarks}

It is noteworthy that the densest known packings of all
of the Platonic and Archimedean solids as
well as the densest known packings of superballs \cite{Ji09a}
and ellipsoids \cite{Do04d} in $\mathbb{R}^3$ have packing fractions that exceed
the optimal sphere packing value $\phi^S_{max}=\pi/\sqrt{18}=0.7408\ldots$.
These results are consistent with a conjecture of Ulam
who proposed without any justification [in a private
communication to Martin Gardner \cite{Ga01}] that the optimal
packing fraction for congruent sphere packings is smaller than that for any
other convex body. The sphere is perfectly
isotropic with an asphericity $\gamma$ of unity, and therefore
its rotational degrees of freedom are irrelevant in affecting its packing characteristics.
On the other hand, each of the aforementioned convex nonspherical
particles break the continuous rotational symmetry of the sphere
and thus its broken symmetry  can be exploited to yield the densest possible packings,
which might be expected to exceed $\phi^S_{max}=\pi/\sqrt{18}=0.7408\ldots$ \cite{To09b}.
However, broken rotational symmetry in and of itself
may not be sufficient to satisfy Ulam's conjecture if
the convex particle has a little or no symmetry \cite{To09b}.

Apparently, the two-dimensional analog of Ulam's conjecture
(optimal density of congruent circle packings ($\phi_{\mbox{\scriptsize max}}=\pi/\sqrt{12}=0.906899\ldots$)
is smaller than that for any other convex two-dimensional body) is false. The ``smoothed" octagon 
constructed by \cite{Re34} is conjectured to have
smallest optimal packing fraction ($\phi_{\mbox{\scriptsize max}}=(8-4\sqrt{2}-\ln2)/(2\sqrt{2}-1)=0.902414
\ldots$) among all congruent centrally-symmetric planar
particles.

It will also be interesting to determine whether Conjecture 1
can be extended to other polyhedral packings. The
infinite families of prisms and antiprisms  provide such a
class of packings. A prism is a polyhedron having bases that
are parallel, congruent polygons and sides that are parallelograms.
An antiprism is a polyhedron having bases that are
parallel, congruent polygons and sides that are alternating
bands of triangles.  They are generally much less symmetric
than either the Platonic or Archimedean solids. Moreover,
even the centrally symmetric prisms and antiprisms generally
do not possess three equivalent directions. Thus, it is less
obvious whether Bravais lattices would still provide the optimal
packings for these solids, except for prisms that tile
space, e.g., hexagonal prism or rhombical prisms \cite{To09c}.
\onlinecite{To09c} have also commented on the validity
of Conjecture 1 to polytopes in four and higher dimensions.

\section{Packing Spheres in High-Dimensional  Euclidean Spaces}
\label{high}

\begin{figure}
\begin{center}
\includegraphics[height=2.3in, keepaspectratio]{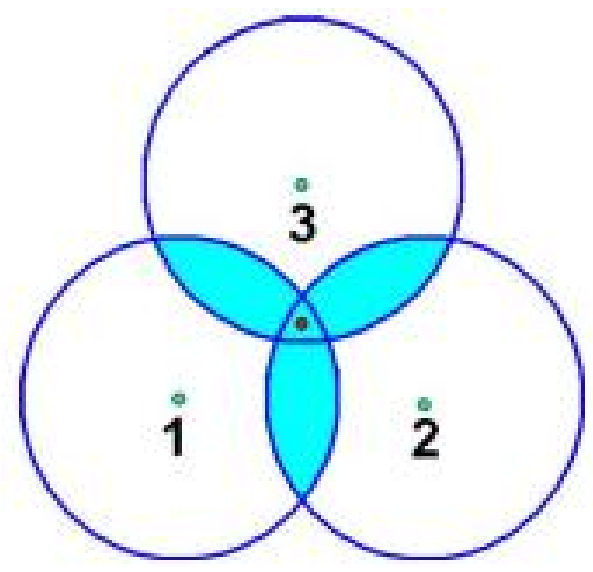}\hspace{0.25in} 
\includegraphics[height=2.3in, keepaspectratio]{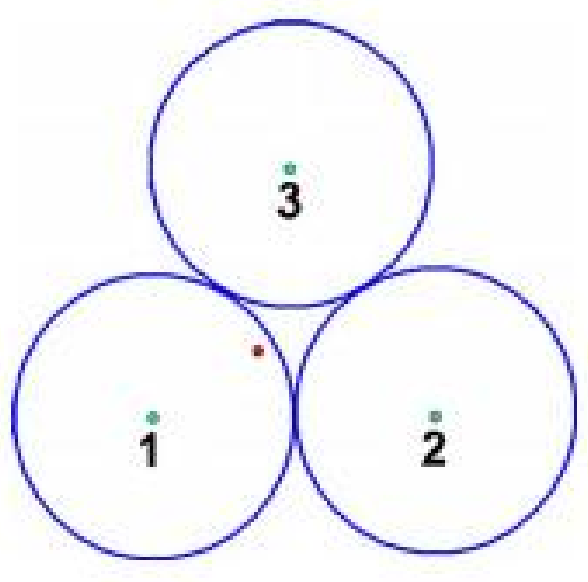}
\caption{(color online) A fundamental problem in communications theory is 
to find the best way to send signals or ``code words"
(amplitudes at $d$ different frequencies) over a noisy channel.
Each code word $\bf s$ corresponds to a coordinate in $\mathbb{R}^d$. The sender
and receiver desire to design a ``code book" that contains a large number of code
words that can be transmitted with maximum reliability given the
noise inherent in any communications channel. Due to the noise in the channel,
the code word $\bf s$ will be received as $\bf r \neq s$ such that
$|{\bf s} -{\bf r}| < a$, where the distance $a$ is  the maximum error 
associated with the channel.
\onlinecite{Sh48} showed that the best way to send code words over a noisy channel
is to design a code book that corresponds to the densest arrangement of spheres 
of radius $2a$ in $\mathbb{R}^d$. To understand this remarkable result,
this figure depicts two choices for the code book in two dimensions.
Left panel: Three code words (depicted as green points) that are circumscribed
by circles of radius $a$ are shown in which the distance between any 
two words is chosen  to be less than $2a$. The received signal (red point) falls
in the region common to all circles and therefore the receiver
cannot determine which of the three code words shown was sent.
Right panel: By choosing a code book in which the code words
are at least a distance $2a$ apart (which corresponds to a sphere packing),  
any ambiguity about the transmitted code word is eliminated. Since
one desires to send many code words per unit volume, 
the best code book corresponds to the densest sphere packing
in $\mathbb{R}^d$ in the limit that the number of code words
tends to infinity.
}
\label{shannon}
\end{center}
\end{figure}

There has been resurgent interest in sphere packings for $d>3$
in both the physical and mathematical sciences 
\cite{Con95,Fr99,Pa00,El00,Pa06,Co02,Co03,To06a,To06b,To06c,Sk06,Ro07,Ad08,Sc08,Co09,Me09,Me09b,
Pa10,Lu10}. 
Remarkably, the optimal way of sending digital signals over noisy channels corresponds
to the densest sphere packing in a high-dimensional space \cite{Sh48,Co93}.
These ``error-correcting" codes underlie a variety of systems in digital
communications and storage, including compact disks, cell phones and the Internet;
see Fig. \ref{shannon}.
Physicists have studied sphere packings in high dimensions to gain insight
into liquid and glassy states of matter as well as
phase behavior in lower dimensions \cite{Fr99,Pa00,Pa06,Sk06,Ro07,Ad08,Me09,Me09b,Pa10}.
Finding the densest packings
in arbitrary dimension is a problem of long-standing interest in discrete geometry \cite{Co93}.
A comprehensive review of this huge
subject is beyond the scope of this article. 
We instead briefly summarize the relevant literature leading to a recent development that
supports the counterintuitive possibility that the densest sphere packings
for sufficiently large $d$ may be disordered \cite{To06b,Sc08}, or
at least possess fundamental cells whose size and structural
complexity increase with $d$.

\begin{figure}
\begin{center}
\includegraphics[height=2.3in, keepaspectratio]{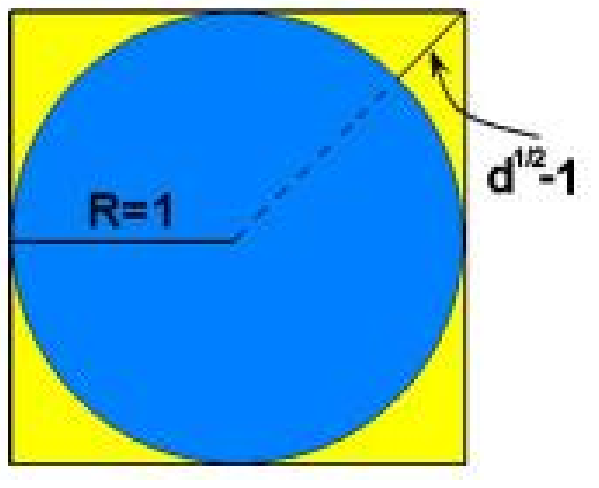}\hspace{0.25in}
\includegraphics[height=2.3in, keepaspectratio]{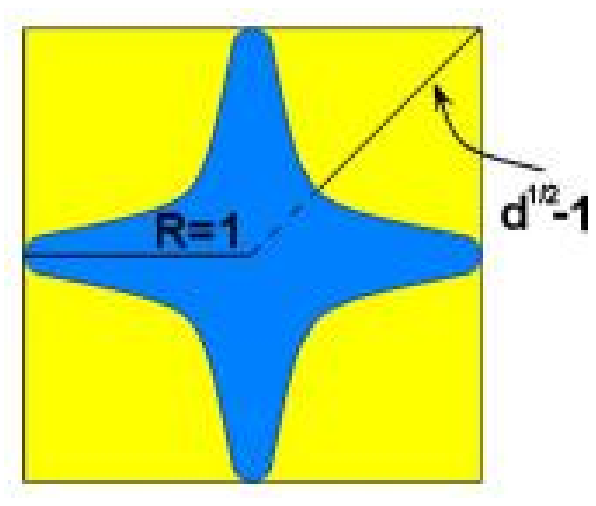}
\caption{(color online) Lattice packings in sufficiently
high dimensions are almost surely {\it unsaturated} because
the ``holes" or space exterior to the spheres dominates $\mathbb{R}^d$ \cite{Co93}.
To get an intuitive feeling for this phenomenon, it is instructive to examine
the hypercubic lattice $\mathbb{Z}^d$ (square lattice for $\mathbb{R}^2$
and the simple cubic lattice for $\mathbb{R}^3$). Left panel: A fundamental
cell of $\mathbb{Z}^d$ represented in two dimensions. Note that the distance
between the point of intersection of the longest diagonal in the hypercube with the hypersphere
boundary and the vertex of the cube along this diagonal is given by 
$\sqrt{d} -1$ for a sphere of unit radius. This means that $\mathbb{Z}^d$
already becomes unsaturated at $d=4$. Placing an additional sphere 
in $\mathbb{Z}^4$ doubles the density of $\mathbb{Z}^4$ and, in particular, yields
the four-dimensional checkerboard lattice packing $D_4$, which is believed
to be the optimal packing in $\mathbb{R}^4$. Right panel:
A schematic ``effective" distorted representation of the hypersphere within the hypercubic
fundamental cell for large $d$, illustrating that the volume content
of the hypersphere relative to the hypercube rapidly diminishes asymptotically. Indeed,
the packing fraction of $\mathbb{Z}^d$ is given by $\phi=\pi^{d/2}/(\Gamma(1+d/2) 2^d)$.
It is the presence of the gamma function, which grows like $(d/2)!$, 
in the denominator that makes $\mathbb{Z}^d$ far from optimal
(except for $d=1$). In fact,
the checkerboard lattice $D_d$ with packing fraction $\phi=\pi^{d/2}/(\Gamma(1+d/2) 2^{(d+2)/2})$ 
becomes suboptimal in relatively low dimensions because it too
becomes dominated by larger and larger holes as $d$ increases.}
\label{holes}
\end{center}
\end{figure}

The sphere packing problem seeks to answer the following
question: Among all packings of congruent spheres in $\mathbb{R}^d$,
what is the maximal packing density $\phi_{\mbox{\scriptsize max}}$
and what are the corresponding arrangements of the spheres \cite{Co93}?
The optimal solutions are known only for the first three space
dimensions \cite{Ha05}. For $4 \le  d \le 9$,
the densest known packings are Bravais lattice
packings  \cite{Co93}. For example, the ``checkerboard" lattice $D_d$, which is
a $d$-dimensional generalization of the fcc lattice
(densest packing in $\mathbb{R}^3$), is believed
to be optimal in $\mathbb{R}^4$ and $\mathbb{R}^5$. 
The remarkably symmetric $E_8$ and Leech lattices in $\mathbb{R}^8$
and $\mathbb{R}^{24}$, respectively, are most
likely the densest packings in these dimensions \cite{Co09}. 
Table \ref{packings} lists the densest known sphere packings in  $\mathbb{R}^d$  for selected $d$.
Interestingly, the non-lattice (periodic) packing $P_{10c}$
(with 40 spheres  per fundamental cell)
is the densest known packing in $\mathbb{R}^{10}$, which is the lowest
dimension in which the best known packing is not a (Bravais) lattice.
It is noteworthy that for sufficiently large $d$,
lattice packings are most likely not the densest
(see Fig.~\ref{holes}),
but it becomes increasingly difficult to find explicit
dense packing constructions as $d$ increases.
Indeed, the problem if finding the shortest
lattice vector in a particular lattice packing (densest
lattice packing) grows super-exponentially
with $d$ and is in the class of NP-hard (non-deterministic polynomial-time hard)problems
\cite{At98}.

\begin{table}[bthp]
\caption{The densest known sphere packings in  $\mathbb{R}^d$  for selected $d$. 
Except for the non-lattice packing $P_{10c}$ in $\mathbb{R}^{10}$, all
of the other densest known packings listed in this table are lattice packings:
$\mathbb{Z}$ is the integer lattice, $A_2$ is the triangular lattice,
$D_d$ is the checkerboard lattice (a generalization of the fcc
lattice), $E_d$ is one the root lattices, and $\Lambda_d$ is the laminated lattice.
The reader is referred to \onlinecite{Co93} for further details. }
\label{packings}
\renewcommand{\baselinestretch}{1.2} \small \normalsize
\centering
\begin{tabular} {|c|c|c|}
\multicolumn{3}{c}{~} \\\hline
 ~Dimension, $d$~ & ~Packing Structure~& ~Packing Fraction, $\phi$~\\ \hline
1 &  $\mathbb{Z}$ & $1$   \\

2 & $A_2$  & $\pi/\sqrt{12}=0.9068\ldots$ \\

3 &  $D_3$  & $\pi/\sqrt{18}=0.7404\ldots$   \\

4 &  $D_4$ &  $\pi^2/16= 0.6168\ldots$\\

5 &  $D_5$ & $2\pi^2/(30\sqrt{2})=0.4652\ldots$ \\

6 &  $E_6$ &  $3\pi^2/(144\sqrt{3})=0.3729\ldots$ \\
7 &  $E_7$ &  $\pi^3/105=0.2952\ldots$ \\
8 &  $E_8$ &  $\pi^4/384=0.2536\ldots$ \\
9 &   $\Lambda_9$  &  $2\pi^4/(945\sqrt{2})=0.1457\ldots$\\
10&   $P_{10c}$ & $\pi^5/3072=  0.09961\ldots$\\
16&  $\Lambda_{16}$ & $\pi^8/645120= 0.01470\ldots$ \\
24&  $\Lambda_{24}$ & $\pi^{12}/479001600=0.001929\ldots$ \\ \hline
\end{tabular}
\end{table}

For large $d$, the best that one can do theoretically
is to devise  upper and lower bounds on 
$\phi_{\mbox{\scriptsize max}}$ \cite{Co93}.
The {\it nonconstructive} lower bound of \onlinecite{Mi05} established the
existence of reasonably dense lattice packings. He found
that the maximal packing fraction $\phi^L_{\mbox{\scriptsize max}}$ among all lattice packings 
for $d \ge 2$ satisfies
\begin{equation}
\phi^L_{\mbox{\scriptsize max}} \ge \frac{\zeta(d)}{2^{d-1}},
\label{mink}
\end{equation}
where $\zeta(d)=\sum_{k=1}^\infty k^{-d}$ is the Riemann zeta function.
Note that for large values of $d$,
the asymptotic behavior of the Minkowski lower bound is controlled by $2^{-d}$.

Since 1905, many extensions and generalizations of (\ref{mink})
have been obtained \cite{Da47,Ball92,Co93,Va09}, but none of these investigations have been able to improve
upon the dominant exponential term $2^{-d}$. It is useful to note that
the  packing fraction of a saturated packing of congruent spheres
in $\mathbb{R}^d$ for all $d$ satisfies 
\begin{equation}
\phi \ge \frac{1}{2^d}.
\label{sat}
\end{equation}
The proof is trivial. A saturated packing of congruent spheres
of unit diameter  and packing fraction 
$\phi$ in $\mathbb{R}^d$ has the property that each point in space lies
within a unit distance from the center of some sphere. Thus, a covering
of the space is achieved if each sphere center is encompassed by a sphere
of unit radius  and the packing fraction of this covering is $2^d \phi \ge 1$.
Thus, the bound (\ref{sat}), which is sometimes called the ``greedy"
lower bound, has the same dominant exponential term as (\ref{mink}).

\begin{figure}[bthp]
\centerline{\includegraphics[height=3.0in,keepaspectratio,clip=]{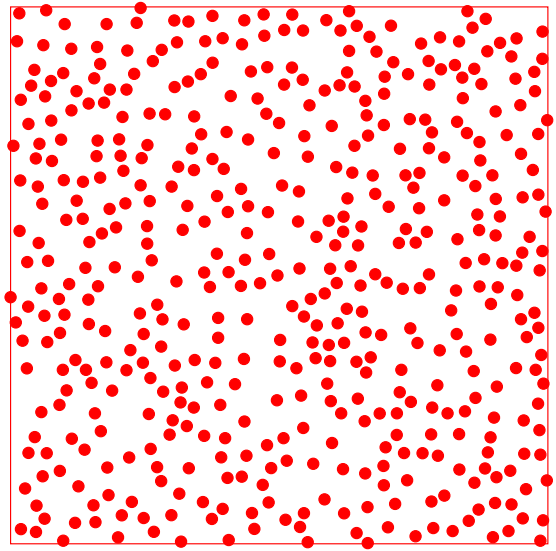}}
\caption{(Color online) \footnotesize The ghost RSA packing model in 
$\mathbb{R}^d$ is a subset of
the Poisson point process and a generalization of the standard RSA process.
The latter is produced by randomly, irreversibly, and sequentially placing 
nonoverlapping objects into a large volume in $\mathbb{R}^d$ that at some
initial time is empty of spheres \cite{Wi66,Fe80,Co88,Vi93}. If an attempt
to add a sphere at some time results in an overlap with an existing sphere in the packing,
the attempt is rejected and further attempts are made until it can
be added without overlapping existing spheres. As the process
continues in time, it becomes more difficult to find available
regions into which the spheres can be added, and eventually
in the {\it saturation} or infinite-time  limit ,
no further additions are possible. In the ghost RSA process, 
a sphere that is rejected is called a ``ghost" sphere.
No additional spheres can be added whether they overlap
an existing sphere or a ghost sphere. The packing fraction
at time $t$ for spheres of unit diameter is given by $\phi(t)=
[1-\exp(-v_1(1) t)]/2^d$, where $v_1(R)$ is the volume
of sphere of radius $R$ [cf. (\ref{vol-sph})]. Thus, we see that
as $t \rightarrow +\infty$, $\phi \rightarrow 2^{-d}$.
This figure shows a configuration of 468 particles
of a ghost RSA packing in a fundamental cell in $\mathbb{R}^2$
at a packing fraction very near its maximal value of 0.25, as adapted
from \onlinecite{Sc08}. 
Note that the packing is clearly unsaturated and there are no contacting particles.}
\label{grsa}
\end{figure}

We know that there exists a disordered but {\it unsaturated} packing construction, known as the ``ghost"
random sequential addition (RSA) packing \cite{To06a}, that achieves the packing $2^{-d}$ 
for any $d$. This packing, depicted in Fig.~\ref{grsa} and described in its caption, 
is a generalization of the standard RSA packing, also described in 
the caption of Fig~\ref{grsa}. It was shown that
all of the $n$-particle correlation functions
of this {\it nonequilibrium} model, in a certain limit, can be obtained analytically for
all allowable densities and in any dimension. This represents the first exactly
solvable disordered sphere-packing model in arbitrary dimension.
(Note that  \onlinecite{Ma86} gave an expression for the pair correlation 
function for this model.)
The existence of this {\it unjammed} disordered packing
strongly suggests that Bravais-lattice packings (which are almost surely
unsaturated for sufficiently large $d$) are far from optimal
for large $d$. Further support for this conclusion
is the fact that the maximal ``saturation" packing fraction of the standard
disordered RSA packing apparently scales as $d \cdot 2^{-d}$ or possibly
$d \cdot \ln(d) \cdot 2^{-d}$ for large $d$ \, \cite{To06d}.
Spheres in both the ghost and standard RSA packings cannot form interparticle
contacts, which appears to be a crucial attribute to
obtain exponential improvement on Minkowski's bound \cite{To06b},
as we discuss below.

The best currently known lower bound on $\phi^L_{\mbox{\scriptsize max}}$ for dimensions
not divisible by four was obtained by  \onlinecite{Ball92}.
He found that 
\begin{equation}
\phi^L_{\mbox{\scriptsize max}} \ge \frac{2(d-1)\zeta(d)}{2^{d}}.
\label{ball}
\end{equation}
For dimensions divisible by four, \onlinecite{Va09} has recently
found the tightest lower bound on the maximal packing fraction
among lattice packings:
\begin{equation}
\phi^L_{\mbox{\scriptsize max}} \ge \frac{6d}{2^{d}e(1-e^{-d})}.
\label{vance}
\end{equation}
Table \ref{table-lower} gives the dominant asymptotic behavior of several
lower bounds on $\phi^L_{\mbox{\scriptsize max}}$ for  large $d$.
Note that the best lower bounds on  $\phi^L_{\mbox{\scriptsize max}}$
improve on Minkowski's bound by a linear factor in $d$ rather
than providing exponential improvement. This suggests
that the packing fraction  of the densest lattice packing
in high $d$  is controlled by the exponential factor $2^{-d}$.

\begin{table}[bthp]
\centering
\caption{ Dominant asymptotic behavior of lower bounds on 
$\phi_{\mbox{\scriptsize max}}^L$
for sphere packings for  large $d$. The bound due to \onlinecite{Va09}
is applicable for dimensions divisible by four. The best
lower bounds do not provide exponential improvement
on Minkowski's lower bound; they instead only improve
on the latter by a factor linear in $d$. \label{table-lower}}
\begin{tabular}{c|c}
\multicolumn{2}{c}{~}\\ \hline\hline
$(2)  2^{-d}$  &  \mbox{Minkowski (1905)} \\ \hline

$[\ln (\sqrt{2}) d] 2^{-d}$  &   \mbox{Davenport and Rogers (1947)}  \\ \hline

$(2 d)2^{-d}$  &   \mbox{Ball (1992)}   \\ \hline

$(6 d/e)2^{-d}$  &   \mbox{Vance (2009)}   \\ \hline\hline
\end{tabular}
\end{table}

\begin{table}[bthp]
\centering
\caption{ Dominant asymptotic behavior of upper bounds on $\phi_{\mbox{\scriptsize max}}$ for sphere packings
for large $d$.\label{upper}}
\begin{tabular}{c|c}
\multicolumn{2}{c}{~}\\ \hline\hline
$(d/2)2^{-0.5d}$  &  \mbox{Blichfeldt (1929)} \\ \hline

  $(d/e)2^{-0.5d}$  &   \mbox{Rogers (1958)}  \\ \hline

  $2^{-0.5990d}$  &   \mbox{Kabatiansky and Levenshtein (1979)}   \\ \hline\hline
\end{tabular}
\end{table}

Nontrivial upper bounds on the
maximal packing fraction  $\phi_{\mbox{\scriptsize max}}$ for any sphere
packing in $\mathbb{R}^d$ have been derived. 
\onlinecite{Bl29} showed that the maximal packing fraction for all $d$ satisfies
$\phi_{\mbox{\scriptsize max}} \le (d/2+1)2^{-d/2}.$
This upper bound was improved by  \onlinecite{Ro58,Ro64}
by an analysis of the Voronoi cells. For large $d$, Rogers'
upper bound asymptotically becomes $d \cdot 2^{-d/2}/e$. 
\onlinecite{Ka78} found an even stronger bound, which in the limit $d\rightarrow \infty$
yields $\phi_{\mbox{\scriptsize max}} \le 2^{-0.5990d}$. All of these
upper bounds prove that the maximal packing fraction tends to zero in the limit $d \rightarrow \infty$.
This rather counterintuitive high-dimensional property of sphere
packings  can be understood by recognizing that almost
all of the volume of a $d$-dimensional sphere for large $d$ is concentrated
near the sphere surface. For example, the volume contained
with such a sphere up to 99\% of it radius is $(99/100)^d$,
which tends to zero exponentially fast. Thus, in high-dimensional
sphere packings (densest or not), almost all of the volume is occupied by the
void space (space exterior to the spheres), which is to be contrasted
with the densest sphere packings in low dimensions in which
the volume contained within the spheres dominates over the void volume.

\onlinecite{Co03} obtained and computed linear programming upper bounds,
which provided improvement over Rogers' upper bounds for dimensions
4 through 36, but it is not yet known whether they improve
upon the Kabatiaksky-Levenshtein upper bounds for large $d$. 
 \onlinecite{Co09} used these techniques
to prove that the Leech lattice is the unique densest lattice in $\mathbb{R}^{24}$. They
also proved that no sphere packing in
$\mathbb{R}^{24}$ can exceed the density of the Leech lattice
by a factor of more than $1+1.65 \times 10^{-30}$, 
and gave a new proof that $E_8$ is
the unique densest lattice in $\mathbb{R}^8$.
Table \ref{upper} provides the dominant asymptotic behavior of several
upper bounds on $\phi_{\mbox{\scriptsize max}}$ for  large $d$.
Note that the best upper and lower bounds on $\phi_{\mbox{\scriptsize max}}$ differ by an exponential
factor as $d \rightarrow  \infty$.

Since 1905, many extensions and generalizations of Minkowski's bound
have been derived \cite{Co93}, but none of them have  improved
upon the dominant exponential term $2^{-d}$. 
\onlinecite{To06b} used a conjecture concerning the existence
of disordered sphere packings  and a $g_2$-invariant optimization procedure 
that maximizes $\phi$
associated with a radial  ``test" pair correlation function $g_2(r)$
to provide the putative exponential improvement on Minkowski's 100-year-old bound
on  $\phi_{\mbox{\scriptsize max}}$. Specifically, a {\it $g_2$-invariant process} \cite{To02c} is one in which 
the functional form of a ``test" pair correlation $g_2({\bf r})$ 
function remains invariant as density varies, for all ${\bf r}$, over the range of
packing fractions
\begin{equation}
0 \le \phi \le \phi_*.
\end{equation}
The terminal packing fraction $\phi_*$ is the maximum achievable density
for the $g_2$-invariant process subject to satisfaction of certain
nonnegativity conditions on pair correlations. For any test $g_2(r)$ that  is a function
of radial distance $r \equiv |\bf r|$ associated with  a packing, 
i.e., $g_2(r)= 0$ for $r < D$, they maximized the corresponding
packing fraction,
\begin{equation}
\phi_* \equiv \lim_{max} \phi,
\end{equation}
subject to satisfying the following two necessary conditions:
\begin{equation}
g_2(r) \ge 0 \qquad \mbox{for all} \; r,
\label{cond1}
\end{equation}
and
\begin{equation}
S(k) \ge 0 \quad \mbox{for all} \; k.
\label{cond2}
\end{equation}
Condition (\ref{cond2}) is a necessary condition
for the existence of any point process [cf. (\ref{factor})]. 
When there exist sphere packings with a $g_2$
satisfying these conditions in the  interval $[0,\phi_*]$,
then one has the lower bound on the maximal packing fraction
given by
\begin{equation}
\phi_{\mbox{\scriptsize max}} \ge \phi_*
\end{equation}

\onlinecite{To06b} conjectured that
a test function $g_2(r)$ is a pair correlation function
of a translationally invariant {\it disordered} sphere packing 
in $\mathbb{R}^d$ for  $0 \le \phi \le \phi_*$
for sufficiently large $d$  if and only if
the conditions (\ref{cond1}) and (\ref{cond2})  are satisfied.  
There is mounting evidence to support this conjecture.
First, they identified a decorrelation principle, which states that unconstrained
correlations in disordered sphere packings vanish asymptotically in high dimensions
and that the $g_n$ for any $n \ge 3$ can be inferred entirely (up to small
errors) from a knowledge
of $\rho$ and $g_2$.  This decorrelation principle, among other results, provides
justification for the conjecture of \onlinecite{To06b},
and is vividly exhibited by the exactly solvable
ghost RSA packing process \cite{To06a} as well as by computer simulations
in high dimensions of the maximally random jammed state \cite{Sk06} and the
standard RSA packing  \cite{To06c}. Second, other necessary
conditions on $g_2$ \cite{Cos04,To06b,Ho09} appear to only have relevance in very low dimensions.
Third, one can recover the form of known rigorous bounds [cf. (\ref{mink}) and (\ref{ball})]
for specific test $g_2$'s when the conjecture is invoked. Finally, in these two instances,
configurations of disordered sphere packings on the torus
have been numerically constructed with such $g_2$ in
low dimensions for densities up to the terminal packing fraction \cite{Cr03,Uc06a}.

Using a particular test pair correlation corresponding to a disordered sphere
packing, \onlinecite{To06b} found a conjectural lower bound on $\phi_{\mbox{\scriptsize max}}$ 
that is controlled by $2^{-(0.77865\ldots) d}$
and the associated lower bound on the average contact (kissing) number $Z$ 
is controlled by $2^{(0.22134\ldots)d}$
(a highly overconstrained situation).
These results counterintuitively suggest that the densest packings as
$d$ increases without bound may exhibit increasingly complex
fundamental cells, or even become disordered at some sufficiently large $d$ rather than periodic.
The latter possibility would imply
the existence of disordered classical ground states for some continuous potentials.
\onlinecite{Sc08} demonstrated  that there is a wide class of test
functions (corresponding to disordered packings) that lead to precisely the same putative exponential improvement 
on Minkowski's lower bound and therefore the asymptotic form $2^{-(0.77865\ldots) d}$ 
is much more general and robust than previously
surmised.

Interestingly, the optimization problem defined above is the {\it dual}
of the infinite-dimensional linear program (LP) devised by \onlinecite{Co02}
to obtain upper bounds on the maximal packing fraction; see  \onlinecite{Co03}
for a proof.
In particular, let $f(r)$ be a radial function in $\mathbb{R}^d$ such that
\begin{eqnarray}
f(r)& \le& 0 \quad \mbox{for}\quad  
r \ge D, \nonumber \\{\tilde f}(k) &\ge& 0 \quad \mbox{for all} \; k,
\label{cohn-elkies1}
\end{eqnarray}
where ${\tilde f}(k)$ is the Fourier transform of $f(r)$.
Then the number density $\rho$ is bounded from above by
\begin{equation}\min \frac{f(0)}{2^d{\tilde f}(0)}.
\label{cohn-elkies2}
\end{equation}
The radial function $f(r)$ can be physically interpreted to be a
{\it pair potential}. 
The fact that its Fourier transform must be
nonnegative for all $k$ is a well-known stability condition
for many-particle systems with pairwise interactions \cite{Ru99}. We see that whereas
the LP problem specified by (\ref{cond1}) and (\ref{cond2}) 
utilizes information about pair correlations, its dual program ({\ref{cohn-elkies1})
and ({\ref{cohn-elkies2})  uses information about
pair interactions. As noted by \onlinecite{To06b},  even if  there does
not exist a sphere packing with $g_2$ satisfying conditions
(\ref{cond1}), (\ref{cond2}) and the hard-core constraint
on $g_2$, the terminal packing fraction $\phi_*$ can never exceed the
Cohn-Elkies upper bound. Every LP has a dual program  and when
an optimal solution exists, there is no {\it duality gap} between
the upper bound and lower bound formulations.
Recently, \onlinecite{Co07b}  proved that there is no duality
gap.


\section{Remarks on Packing Problems in Non-Euclidean Spaces}
\label{curved}

Particle packing problems in non-Euclidean (curved) spaces have  been
the focus of research in a variety of fields, including
physics \cite{Bo06,Ka07b}, biology \cite{Ta30,Go67,Pr90,To02d,Za04}, communications theory \cite{Co93},
and geometry \cite{Co93,Ha04,Co07}. Although a comprehensive overview
of this topic is beyond the scope of this review, we highlight here
some of the developments for the interested reader in spaces
with constant positive and negative curvatures.
We will limit the discussion to packing spheres on the positively curved
unit sphere $S^{d-1} \subset \mathbb{R}^d$ and in negatively curved hyperbolic space $\mathbb{H}^d$.

The {\it kissing} (or contact) number $\tau$ is the number of spheres of unit radius
that can simultaneously touch a unit sphere  $S^{d-1}$ \cite{Co93}. The kissing
number problem asks for the maximal kissing number $\tau_{max}$ in $\mathbb{R}^d$.
The determination of the maximal kissing number
in $\mathbb{R}^3$ spurred a famous debate between Issac Newton and
David Gregory in 1694. The former correctly thought the answer 
was 12, but the latter wrongly believed that 13 unit spheres could simultaneously
contact another unit sphere. The optimal kissing number 
$\tau_{max}$ in dimensions greater than three is only known for 
$\mathbb{R}^4$ \cite{Mu08}, $\mathbb{R}^8$ and $\mathbb{R}^{24}$ 
\cite{Lev79,Od79}. Table \ref{kissing} lists the largest  known
kissing numbers  in selected dimensions.

\begin{table}[bthp]
\caption{The largest known kissing numbers for identical spheres in 
$\mathbb{R}^d$  for selected $d$. 
Except for $\mathbb{R}^9$ and $\mathbb{R}^{10}$,
the largest  known kissing numbers listed in this table 
are those found in the densest lattice packings listed in Table \ref{packings}.
$P_{9a}$ is a non-lattice packing with an average
kissing number of $235$ $3/5$ but with a maximum kissing
number of 306. $P_{10b}$ is a non-lattice packing with an average
kissing number of $340$ $1/3$ but with a maximum kissing
number of 500.
The reader is referred to \onlinecite{Co93} for further details. }
\label{kissing}
\renewcommand{\baselinestretch}{1.2} \small \normalsize
\centering
\begin{tabular} {|c|c|c|}
\multicolumn{3}{c}{~} \\\hline
 ~Dimension, $d$~ & ~Packing Structure~& ~Kissing Number, $\tau$~\\ \hline
1 &  $Z$ & $2$   \\

2 & $A_2$  & $6$ \\

3 &  $D_3$  & $12$   \\

4 &  $D_4$ & $24$  \\

5 &  $D_5$  & $40$  \\

6 &  $E_6$ & $72$  \\
7 &  $E_7$ & $126$   \\
8 &  $E_8$ & $240$  \\
9 &   $P_{9a}$  & 306 \\
10&   $P_{10b}$ & 500\\
16&  $\Lambda_{16}$ & 4,320 \\
24&  $\Lambda_{24}$ & 196,560 \\
 \hline
\end{tabular}
\end{table}

In geometry and coding theory, a {\it spherical code} with parameters ($d,N,t$) is a set of $N$ 
points on the unit sphere $S^{d-1}$  such that no two distinct points
in that set have inner product greater than or equal to $t$, i.e.,
the angles between them are all at least $\cos^{-1} t$. The fundamental
problem is to maximize $N$ for a given value of $t$, or equivalently to minimize $t$
given $N$ [sometimes called the Tammes problem, which was  motivated by an application
in botany \cite{Ta30}].
One of the first rigorous studies of spherical codes was presented
by \onlinecite{Sc51}. \onlinecite{Del77} introduced much of the most important mathematical machinery
to understand spherical codes and designs. One natural generalization of the best way to distribute points on
$S^{d-1}$ (or $\mathbb{R}^d$)  
is the energy minimization problem: given some potential function depending on the 
pairwise distances between points, how should the points be arranged so as to minimize 
the total energy (or what are the ground-state configurations)? 
The original Thomson problem of ``spherical crystallography" seeks the ground states of 
electron shells interacting via the Coulomb potential; but it is also
profitable to study ground states of particles interacting with other potentials
on $S^{d-1}$ \cite{Bo06}. \onlinecite{Co07} have introduced the beautiful idea of a 
{\it universally optimal configuration}, a unique configuration that minimizes
a class of potentials. In particular, they proved that for any fixed number of points $N$
on $S^{d-1}$ there is a  universally optimal configuration that minimizes all completely monotonic potential functions 
(e.g., all inverse power laws). 

The optimal spherical code problem is related to the densest local packing
(DLP) problem in $\mathbb{R}^d$\cite{Ho10b}, which involves the placement of $N$ 
nonoverlapping spheres of unit diameter near an additional
fixed unit-diameter sphere such that the greatest radius $R$ from the center of
the fixed sphere to the centers of any of the $N$ surrounding spheres is minimized.
Let us recast the optimal spherical code problem as the placement of the centers of $N$ nonoverlapping spheres of unit
diameter onto the surface of a sphere of radius $R$ such that $R$ is minimized. 
It is has been proved that for any $d$, all solutions for $R$ between unity and the golden ratio
$\tau=(1+\sqrt{5})/2$ to the optimal spherical code problem for $N$ spheres are also solutions to the corresponding
DLP problem \cite{Ho10a}. It follows that for any packing of nonoverlapping spheres
of unit diameter, a spherical region of radius $R$ less than or equal to  $\tau$ centered on an
arbitrary sphere center cannot enclose a number of sphere centers greater than one
more than the number that than can be placed on the spherical region's surface.

We saw in Sec.\ref{mrj} that monodisperse circle (circular disk) 
packings in $\mathbb{R}^2$ have a great tendency 
to crystallize at high densities due to a lack of geometrical frustration.
The hyperbolic plane $\mathbb{H}^2$ (for a particular constant negative curvature,
which measures the deviation from the flat Euclidean plane) provides a two-dimensional space 
in which global crystalline order in dense circle packings is frustrated, and thus affords
a means to use circle packings to understand fundamental features of  simple liquids,
disordered jammed states and glasses. \onlinecite{Ka07b} formulated
an expression for the equation of state for disordered hard disks in $\mathbb{H}^2$ 
and compared it to corresponding results obtained from molecular dynamics simulations.
\onlinecite{Ka07b} derived a generalization of the virial equation in $\mathbb{H}^2$
relating the pressure to the pair correlation function and  developed the appropriate setting 
for extending integral-equation approaches of liquid-state theory.
For a discussion of the mathematical subtleties associated with finding the densest
packings of identical $d$-dimensional spheres in $\mathbb{H}^d$, the 
reader is referred to \onlinecite{Ra03}.

\section{Challenges and Open Questions}
\label{open}

The geometric-structure approach advanced and explored in this review provides a comprehensive 
methodology to analyze and compare jammed disk and sphere packings across their infinitely rich variety. This approach also highlights aspects
of present ignorance, thus generating many challenges and open questions for future investigation.  Even
for identical spheres, detailed characterization of jammed structures across the simple two-dimensional ($\phi$-$\psi$)  order maps  outlined in Sec. 
\ref{order} is still very incomplete. A partial list
of open and challenging questions in the case
of sphere packings includes the following:

\begin{enumerate}
\item Are the strictly jammed ``tunneled" crystals \cite{To07} the family 
of lowest density collectively jammed packings under periodic boundary conditions?  

\item How can the extremal jammed packings that inhabit the upper and lower boundaries of 
occupied regions of each of those order maps be unambiguously identified?  

\item What would be the shapes of
analogous occupied regions if the two-parameter versions illustrated in Fig. \ref{maps} were
 to be generalized to three or more parameters?  

\item To what extent can the rattler concentration in collectively
or strictly sphere packings be treated as an independent variable?  What is the upper limit to attainable rattler
concentrations under periodic boundary conditions?

\item What relations can be established between order metrics and geometry
of the corresponding configurational-space polytopes?

\item Can upper and lower bounds be established for the number
of collectively and/or strictly jammed states for $N$ spheres?

\item Upon extending the geometric-structure approach to Euclidean dimensions
greater than three, do crystalline arrangements with arbitrarily large unit cells 
or even disordered jammed packings ever provide the highest attainable densities?
\end{enumerate}

     Jamming characteristics of nonspherical and even non-convex hard particles is an area of research that is
still largely undeveloped and therefore deserves intense research attention.   Many of the
same open questions identified above for sphere packings are equally relevant to packings of nonspherical particles.
An incomplete list of open and challenging questions for such particle packings includes the following:

\begin{enumerate}
\item What are the appropriate generalizations of the jamming categories for packings of nonspherical particles?

\item Can one devise incisive order metrics for packings of nonspherical  particles 
as well as  a wide class of many-particle systems (e.g., molecular, biological, cosmological,
and ecological structures)?

\item Can sufficient progress be made to answer the first two questions
that would lead to useful order maps? If so, how do the basic features
of order maps depend on the shape of the particle? What are 
the lowest density jammed states?

\item Why does the deformation of spheres into ellipsoids have the effect of strongly inhibiting (if not completely
preventing) the appearance of rattlers in disordered ellipsoid packings for a wide range
of aspect ratios? Does this same inhibitory phenomenon extend to disordered
jammed packings of other nonspherical shapes?

\item For non-tiling nonspherical particles, can an upper bound on the maximal packing fraction
be derived that is always strictly less than unity? (Note that upper bound (\ref{bound}) is
strictly less than unity provided that the asphericity is sufficiently small.)

\end{enumerate}

 Jamming characteristics of particle packings in non-Euclidean spaces is an area of research that is
still largely undeveloped and therefore deserves research attention. This is true
even for simple particle shapes, such as spheres. For example, we know very little
about the jamming categories of spheres in curved spaces. Does curvature
facilitate jamming or not? Does it induce ordering or disordering? 
These are just a few of the many challenging issues that 
merit further investigation.

     In view of the wide interest in packing problems that the research community displays there is reason to be
optimistic that substantial conceptual advances are forthcoming.  It will be fascinating to see how future review
articles covering this subject document those advances.

\begin{acknowledgments}
The authors are grateful to Yang Jiao for producing  many of the figures and 
generating the data for Fig. \ref{histo}. We thank Yang Jiao, Henry Cohn 
and  Aleksandar Donev for useful discussions. This work was supported by the Division of
Mathematical Sciences at the National Science Foundation
under Award No. DMS-0804431 and by the MRSEC Program
of the National Science Foundation under Award No.
DMR-0820341.
\end{acknowledgments}



\end{document}